\pdfoutput=1
\documentclass[a4paper,11pt,english,DIV=12,parskip=half-,headsepline,footsepline=false,
  bibliography=totoc,listof=totoc,listof=nochaptergap,numbers=noenddot,captions=tableheading]{scrreprt}

\pdftrailerid{}
\input{glyphtounicode}
\pdfglyphtounicode{ohm}{2126}%
\usepackage[T1]{fontenc}
\usepackage{lmodern}
\usepackage[ngerman,main=english]{babel}
\usepackage{microtype}
\usepackage{amsmath,amssymb}
\usepackage{upgreek}
\usepackage{textcomp}
\usepackage{siunitx}
\usepackage{graphicx}
\usepackage{wrapfig}
\usepackage[section]{placeins}%
\usepackage{array,booktabs,longtable,multirow}
\newcolumntype{B}[1]{>{\raggedright\arraybackslash\hangindent=1em\hangafter=1\makebox[1em][l]{\textbullet}}p{#1}}
\newcommand{\proscons}[3]{%
  \begin{table}[ht]
  \centering
  \caption{#1}\label{#2}
  {\small\renewcommand{\arraystretch}{1.2}%
  \begin{tabular*}{\textwidth}{@{}B{0.47\textwidth}@{\extracolsep{\fill}}B{0.47\textwidth}@{}}
  \toprule
  \multicolumn{1}{@{}l}{\textbf{Advantages}} & \multicolumn{1}{l@{}}{\textbf{Disadvantages}} \\
  \midrule
  #3
  \bottomrule
  \end{tabular*}}
  \end{table}}
\usepackage{enumitem}
\usepackage[dvipsnames]{xcolor}
\usepackage{tikz}
\usetikzlibrary{arrows.meta,calc,positioning,shapes.misc}
\usepackage{pgfplots}
\pgfplotsset{compat=1.18}
\usepgfplotslibrary{groupplots}
\pgfplotsset{thesisplot/.style={
  font=\small, tick label style={font=\footnotesize}, label style={font=\small},
  grid=major, major grid style={black!15}, axis line style={black!70},
  tick align=outside, tick pos=left, every axis plot/.append style={line width=0.6pt}}}
\colorlet{plotline}{black}%
\colorlet{plotfill}{black!45}%
\colorlet{plothighlight}{black!10}%
\colorlet{thgrey}{black!50}%
\usepackage{listings}
\usepackage{newfloat}
\usepackage[format=plain,font=small,labelfont=bf,labelsep=colon,justification=centering,skip=6pt]{caption}
\usepackage{tablefootnote}
\usepackage[headsepline]{scrlayer-scrpage}
\usepackage{xurl}
\usepackage[hidelinks,pdftitle={Direct Electromechanical Actuation of the Shift Drum in Non-Synchronized Motorcycle Transmissions: From Actuator Sizing and Electronics Design to Closed-Loop Position Control - A Gear Change in Less Than 80 ms on a Gearbox Test Rig},pdfauthor={Johannes Stockhammer},
  pdfkeywords={automated manual transmission, motorcycle, shift drum actuator, shift-by-wire, worm gear, self-locking, position control, closed-loop control, electronics design, shift time, gearbox test rig}]{hyperref}

\counterwithout{figure}{chapter}
\counterwithout{table}{chapter}
\counterwithout{equation}{chapter}
\counterwithout{footnote}{chapter}
\DeclareFloatingEnvironment[fileext=loa,name=Appendix,placement=htbp]{appendixfloat}
\counterwithout{appendixfloat}{chapter}
\makeatletter
\renewcommand\@biblabel[1]{[\ifnum#1<100 0\fi\ifnum#1<10 0\fi#1]}
\makeatother
\renewcaptionname{english}{\bibname}{References}
\renewcaptionname{english}{\contentsname}{Table of Contents}

\clearpairofpagestyles
\ihead{\includegraphics[height=7mm]{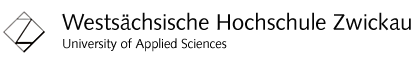}}
\ohead{\includegraphics[height=7mm]{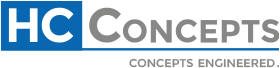}}
\ofoot*{\pagemark}

\KOMAoptions{headsepline}

\DeclareRobustCommand{\nocc}{\\{\itshape Not covered by the CC license}}

\setlist{itemsep=1pt,topsep=3pt}

\begin{document}

\begin{titlepage}
\begin{center}
\includegraphics[height=12mm]{figures/logo_whz}\hfill\includegraphics[height=12mm]{figures/logo_hcc}\par
\vspace*{4mm}
{\Large\bfseries Westsächsische Hochschule Zwickau\par}
\vspace{2mm}
{\large Faculty of Electrical Engineering\par}
\vspace{10mm}
{\large Diploma Thesis\par}
\vspace{9mm}
{\LARGE\bfseries\sffamily Direct Electromechanical Actuation\\ of the Shift Drum in Non-Synchronized\\ Motorcycle Transmissions\par}
\vspace{5mm}
{\large From Actuator Sizing and Electronics Design to Closed-Loop Position Control:\\ A Gear Change in Less Than \qty{80}{\milli\second} on a Gearbox Test Rig\par}
\vspace{9mm}
{\large for the attainment of the academic degree\\ Diplom-Ingenieur (FH)\\ in the degree program Automotive Electronics\par}
\vspace{8mm}
{\large submitted by\par}
\vspace{1mm}
{\Large\bfseries Johannes Stockhammer\par}
\vfill
{\large\textbf{Supervisors:}\\ Prof. Dr.-Ing. Matthias Würfel\\ Prof. Dr.-Ing. Mirko Bodach\par}
\vspace{6mm}
{\large In cooperation with HC-Concepts Engineering GmbH\par}
\vspace{6mm}
{\large Zwickau, 2019\\ \textit{Revised English edition, 2026}\par}
\end{center}
\end{titlepage}

\thispagestyle{empty}
\begingroup\small
\vspace*{\fill}
\textbf{Edition.} This document is the revised English edition (2026) of the diploma thesis ``Entwickeln eines automatisierten Schaltgetriebes für Motorräder'', submitted in 2019 at the Westsächsische Hochschule Zwickau (University of Applied Sciences), Germany. For this edition, errors were corrected, an abstract and a summary of the results were added, figures were redrawn and the layout was revised.

\textbf{Citation.} Stockhammer, J.: Direct Electromechanical Actuation of the Shift Drum in Non-Synchronized Motorcycle Transmissions. Diploma thesis, Westsächsische Hochschule Zwickau, 2019. Revised English edition, 2026.

\textbf{License.} This publication is licensed under the Creative Commons license Attribution-NonCommercial-ShareAlike 4.0 International (CC BY-NC-SA 4.0): \url{https://creativecommons.org/licenses/by-nc-sa/4.0/}

Excluded from this license are third-party contents, the logos of Westsächsische Hochschule Zwickau and HC-Concepts Engineering GmbH in the page header, and all contents marked as not covered by the CC license.

\textbf{Figures.} Unless stated otherwise, all figures, photographs and tables are the author's own work. Figures marked ``based on'' follow the cited source.

\textbf{Data and code.} The firmware of the shift controller and the measured shift times are published at \url{https://doi.org/10.5281/zenodo.23002655} under the licenses stated there.

\textbf{Trademarks.} Brand and product names are used for identification only and may be trademarks of their owners.

\textbf{Tools.} AI tools were used for the translation, for language editing and for preparing the schematic drawings and charts of this edition, based on the author's own data and specifications.

\textbf{Disclaimer.} This work documents a proof of concept on a gearbox test rig from 2019 and is not a construction manual. The shift system is not tested or approved for use in road vehicles. Statements on products and the market reflect the state of 2019.

\endgroup
\clearpage

\pagenumbering{roman}

\chapter*{Acknowledgements}\addcontentsline{toc}{chapter}{Acknowledgements}\label{sec:1}

The diploma thesis was prepared in cooperation with the company HC-Concepts Engineering GmbH in Nußdorf am Haunsberg. Looking back at my work in the company, I remember above all the very good working atmosphere and the appreciation shown by my team colleagues. I remember a very demanding time. Nevertheless, I am proud that I was allowed to realize such a new development in the course of my diploma thesis. I am therefore grateful for the great freedom in realizing my ideas. Beyond the project work, I was able to gain important experience while working with many different people.

My special thanks go above all to Mr. Ing. Claus Holweg, who gave the impetus for realizing the project idea. Without him, the implementation of the diploma thesis would not have been possible. I especially thank Mr. Ronald Kuttler and the entire team of HC-Concepts for the excellent cooperation and support.

On the part of the Westsächsische Hochschule Zwickau, special thanks go to my supervisors Prof. Dr.-Ing. Matthias Würfel and Prof. Dr.-Ing. Mirko Bodach for their repeated active support. Many thanks for the uncomplicated scheduling of appointments whenever I had questions!

Special thanks also go to my friends and my family, who actively supported me during my studies, even though the time for joint activities fell somewhat short while I was preparing my diploma thesis.

Many thanks to all of you!

\chapter*{Abstract}\addcontentsline{toc}{chapter}{Abstract}\label{sec:2}

This thesis presents the direct electromechanical actuation of the shift drum in a non-synchronized motorcycle transmission. A DC motor drives the shift drum through a self-locking worm gear, so the entire foot shift mechanism is eliminated. As a reference, 84 manual gear changes of a large-displacement series-production motorcycle were measured on a chassis dynamometer; the mean shift time is \qty{80}{\milli\second} for upshifts and \qty{92}{\milli\second} for downshifts. The system design covers the complete chain from the mechanics to the software. The actuator motor is sized from inertia, friction and gear ratio. A dedicated shift controller is developed with supply filtering, signal conditioning for the position sensor and the incremental encoder, an H-bridge motor driver and a custom circuit board based on a SAMD21 microcontroller. The embedded firmware implements position control of the shift drum with a proportional controller, without inner speed or current loops. On a gearbox test rig, a gear change is completed in less than \qty{80}{\milli\second} without overshoot in a single measurement, below the mean of the manual reference shifts. The analysis reveals a trade-off in the choice of the worm gear ratio: at a ratio of 30, about \qty{93}{\percent} of the motor torque is needed to accelerate the motor shaft and the worm. A ratio of about 8 halves the required torque, but the worm gear then loses its self-locking. The measured shift times are published as an open dataset.

\textbf{Keywords:} shift-by-wire, automated manual transmission, motorcycle, shift drum actuator, worm gear, self-locking, inertia matching, embedded position control, gearbox test rig, shift time

\begin{otherlanguage}{ngerman}
\chapter*{Kurzfassung}\addcontentsline{toc}{chapter}{Kurzfassung}\label{sec:3}
{\bfseries Direkte elektromechanische Betätigung der Schaltwalze in nicht synchronisierten Motorradgetrieben -- Von der Auslegung des Stellmotors und der Elektronikentwicklung bis zur Lageregelung: Ein Gangwechsel in weniger als \qty{80}{\milli\second} am Getriebemodell\par}

Die Arbeit behandelt die direkte elektromechanische Betätigung der Schaltwalze in einem nicht synchronisierten Motorradgetriebe. Ein Gleichstrommotor treibt die Schaltwalze über ein selbsthemmendes Schneckengetriebe an, die gesamte Fußschaltmechanik entfällt. Als Referenz werden 84 manuelle Gangwechsel eines hubraumstarken Serienmotorrads am Rollenprüfstand gemessen; die mittlere Schaltzeit beträgt \qty{80}{\milli\second} beim Hochschalten und \qty{92}{\milli\second} beim Runterschalten. Die Systementwicklung umfasst die gesamte Kette von der Mechanik bis zur Software. Der Stellmotor wird aus Massenträgheit, Reibung und Übersetzung ausgelegt. Ein eigenes Schaltsteuergerät wird mit Versorgungsfilter, Signalaufbereitung für Positionssensor und Inkrementalgeber, H-Brücken-Motortreiber und einer eigenen Leiterplatte auf Basis eines SAMD21-Mikrocontrollers entwickelt. Die Firmware realisiert die Lageregelung der Schaltwalze mit einem P-Regler ohne unterlagerte Drehzahl- und Stromregelung. Am Getriebemodell ist ein Gangwechsel in einer Einzelmessung in weniger als \qty{80}{\milli\second} ohne Überschwingen abgeschlossen und liegt damit unter dem Mittelwert der manuellen Referenzschaltungen. Die Auswertung zeigt einen Zielkonflikt bei der Wahl der Schneckenübersetzung: Bei einer Übersetzung von 30 entfallen rund \qty{93}{\percent} des Motordrehmoments auf die Beschleunigung von Motorwelle und Schnecke. Eine Übersetzung von etwa 8 halbiert das notwendige Drehmoment, die Selbsthemmung geht dabei jedoch verloren. Die gemessenen Schaltzeiten sind als offener Datensatz veröffentlicht.

\textbf{Schlagwörter:} Shift-by-Wire, automatisiertes Schaltgetriebe, Motorrad, Schaltwalze, Schneckengetriebe, Selbsthemmung, Trägheitsanpassung, eingebettete Lageregelung, Getriebemodell, Schaltzeit
\end{otherlanguage}

\tableofcontents
\clearpage
\pagenumbering{arabic}

\chapter{Introduction}\label{sec:4}

For some years now, electronics has also been finding its way into two-wheelers. The well-known rule ``Two-wheelers lag 10 years behind developments in the automotive sector'' has long ceased to apply. It has become apparent that the solutions from the automotive sector cannot be adopted directly for motorcycles. A two-wheeler places entirely new requirements on the technology. The way in which new technology is introduced, however, is analogous to the automotive sector. First, the high-priced top models of the manufacturers are equipped with new technology and tested. The customers of these motorcycles demand the latest technical innovations. For customers of low-priced motorcycles, a good price-performance ratio and rugged technology are the main priorities. If expensive systems from the upper class are to be implemented here, this places very special requirements on the development. On the one hand, the systems must offer a similar range of functions; on the other hand, they must be very low-priced and rugged. This confronts engineers with new challenges again and again.

From the point of view of customers and mechanics, the two-wheeler sector is also changing. Two-wheelers have always been a popular object for hobby mechanics and tinkerers. Many problems and faults are visible thanks to the easily accessible mechanics and can be fixed with simple tools. With the introduction of electronics, the overall system becomes increasingly complex, and faults are no longer directly recognizable. Changing a small part has an effect on the entire vehicle. Without suitable fault diagnosis equipment and special tools, the possible repair measures are limited. Thus, many mourn the ``old'' technology and perceive new electronic components as unnecessary or unmanageable.

It makes sense to take a critical look at new technology, but at the same time electrification is at the core of the present day. Often it is better to take a second look. In many cases, the technology turns out to be quite different from what was previously assumed.

\section{Motivation}\label{sec:5}

The more than 100-year history of motor vehicles is characterized by technical masterpieces and new technologies. It is therefore a great experience to carry on a small part of this history oneself. New laws and standards force manufacturers to meet ever stricter emission standards in the two-wheeler sector as well, and thus fuel consumption is also increasingly moving into focus. At the same time, customers take a high level of riding comfort and certain equipment features for granted. In order to meet all these criteria, gearboxes are required whose shift points, and thus the engine operating point, can be selected depending on the specification by the engine control unit. Many large-displacement motorcycles already have automatic or automated gearboxes, but for high-volume production motorcycles in the low-price segment, such a system is not yet available on the market. With the introduction of European emission legislation in India, there is a particularly great demand here.

\section{Tasks and Objectives of the Diploma Thesis}\label{sec:6}

The task of the diploma thesis is the development of the electronic components for controlling an automated gearbox. The development of the gearbox prototypes and the mechanical components is carried out by a team from HC-Concepts.

First of all, the framework conditions are analyzed, all requirements are presented in a requirements specification and evaluated according to relevance and feasibility. Once the final requirement items have been determined, an overall concept is designed. This concept is divided into three subareas: mechanics, electronics and software. As a first step, the development of the mechanics is carried out in parallel with the development of the hardware. Subsequently, the controlled system is analyzed and a control concept for position control is designed. The development of the control software takes place with the aid of the constructed gearbox test rig. The combustion engine is replaced by an electric motor. The masses of the vehicle are represented by model rotating masses.

The aim of the diploma thesis is the development of a gearbox test rig to demonstrate the project idea as a ``proof of concept''. To save time and costs, it is realized as an open gearbox test rig without oil filling. Among other things, the development department's main focus is the evaluation of the effort, the costs and the advantages and disadvantages compared to conventional systems. If the tests are completed successfully, the system is to be used in a production motorcycle.

The thesis investigates whether the shift drum of a conventional, non-synchronized motorcycle gearbox can be driven by an electric motor with a self-locking worm gear in such a way that gear changes take place in a similar time as with manual actuation. In detail, the following points are considered:

\begin{itemize}
\item Determination of the shift times of manual gear changes as a reference for the design
\item Influence of inertia, friction and gear ratio on the required motor torque
\item Feasibility of a simple position control without subordinate speed and current control
\end{itemize}

\subsection{General Conditions}\label{sec:7}

The project of the company HC-Concepts represents the automation of a conventional 6-speed gearbox for motorcycles in series production. For manufacturers to be able to comply with the increasingly strict emission legislation, a fully automated gearbox in a cost-effective design is a great advantage.

The planned field of application is large quantities of light motorcycles on the South and Southeast Asian market. If series use is successful, the system is to be used for motorcycles in Europe in the future as well. Especially the rising population in large Indian cities can afford comfortable motorcycles as a means of transport due to the economic upswing. The motorcycle no longer serves purely for locomotion; it is increasingly becoming a fun and lifestyle product. Customers strive for technical refinements and developments from Europe.

Due to the high traffic volume in large cities, smog pollution is becoming an increasing health problem for the population. One starting point of policy is, among other things, the reduction of two-stroke mopeds with insufficient exhaust gas cleaning. Above all, the ten largest cities of India are affected by smog pollution. In the three largest cities, Mumbai, New Delhi and Kolkata, there is an urgent need for action. Thus, Indian policymakers have adopted the European emissions legislation. Indian vehicle manufacturers are primarily confronted with additional costs and development effort. For European vehicle manufacturers, it is the opportunity to open up new markets through the transfer of knowledge and know-how.

\subsection{Requirements}\label{sec:8}

The following requirements are placed on the new development:

\begin{enumerate}
\item Maximum weight of all components on the vehicle three~kilograms
\item Execution of the shifting movement by an electric motor
\item Reduction of mechanical components compared to manual shift actuation
\item Average shift time of \qty{100}{\milli\second}
\item Development of control electronics that can also be used as an add-on control unit for retrofit solutions
\item Maintenance-free within the service life of the vehicle (Table~\ref{tab:1})
\item Simplest possible control strategy with overload protection, if possible without additional speed and current control
\end{enumerate}

\chapter{State of the Art}\label{sec:9}

Various systems of automatic and continuously variable transmissions exist on the market, with the corresponding advantages and disadvantages for the planned use. Both series and retrofit solutions are available.

Gearboxes have a direct influence on the driving and emission behavior of motorcycles. Above all, the fuel consumption and the CO\textsubscript{2} emissions depending on it are directly related to the operating point of the engine. However, HC, CO and NO\textsubscript{X} also depend on the operating point and thus on the gear ratio. Often only stationary operating points are discussed, but the dynamic operating behavior, which is important for test cycles, is also decisively influenced by a gearbox. The gearbox therefore matters for the development of new vehicles and engines. In this chapter, various systems already available on the market are presented.

In addition to commercial systems, research also deals with automated manual transmissions. For passenger cars, modeling and control of the shifting operation have been investigated extensively \cite{Gli06}, \cite{Luc07}. The focus there is on the actuation of the clutch and the shift actuator as well as on synchronization. For motorcycles, studies on the automation and quality of gear changes in sport motorcycles are available \cite{Gia14}. Furthermore, a dual clutch transmission for large-displacement motorcycles has been brought into series production \cite{Wat11}. The design of electric-motor shift actuators is described, among others, for gearboxes of electric vehicles \cite{Hu16}. The direct electric-motor drive of the shift drum of a non-synchronized motorcycle gearbox via a self-locking gear is not covered in the literature considered. However, this design is of particular interest for cost-effective motorcycles in large quantities. This is where the present thesis comes in.

\section{Continuously Variable Transmissions}\label{sec:10}

For scooters, the term ``Variomatic'' has become established in common usage as a description for this transmission. Technically correct, the transmission can be described as a continuously variable belt drive transmission or as a CVT. The control of these transmissions can be electrohydraulic or mechanical by means of centrifugal weights. In two-wheelers, almost exclusively mechanical controls are used due to the simple and sturdy technology. An electrohydraulic design allows more precise control of the gear ratio during driving. This type of control is used above all in passenger cars.

Continuously variable transmissions are used above all where a fixed engine operating point is important and great importance is attached to comfort. The rider has to neither shift manually nor operate the clutch; the driving force is simply controlled via the throttle grip. These transmissions are widespread in mopeds and scooters with low to medium engine power, since the operating point can be placed in the elastic range of the engine. Due to the narrow speed band of small 2-stroke engines, comfort also matters here.

\textbf{Function:} By means of a continuously variable ratio, the operating point can be varied over a wide range, limited by the possible ratio spread of the transmission. Ranges and characteristic curves for different requirements can be defined within the power hyperbola of the combustion engine. The transmission control unit contains these maps and controls the operating points as defined in the development process. Many control units ``learn'' the rider's driving style and then optimize the shifting and operating behavior of the engine. In the two-wheeler sector, continuously variable transmissions are mostly designed as belt drives. Due to the flexing work of the belt, the efficiency is considerably lower than that of gear transmissions. The selection of a consumption-optimized engine operating point is thus almost canceled out again~\cite[pp.~134\,f.]{Har07}.

The engine map in Figure~\ref{fig:1}\,a shows three possible control curves: Depending on the driving situation, consumption-oriented or performance-oriented control may be necessary. For everyday operation, a compromise between both control strategies is selected. With an automated transmission, the control moves dynamically between the consumption-optimized and the performance-optimized curve within the everyday operating range~\cite[pp.~134\,f.]{Har07}.

\begin{figure}[tbp]
\begin{minipage}[t]{0.48\textwidth}
\textbf{a)}\par\medskip
\centering
\begin{tikzpicture}[y=0.72cm, font=\small, >={Stealth[length=2mm]}]%
  \def\fullload{(0.6,2.3) .. controls (1.4,3.6) and (2.4,4.05) .. (3.5,4.05)
                .. controls (4.6,4.05) and (5.5,3.8) .. (6.2,3.3)}
  \begin{scope}
    \clip (0.6,0) -- \fullload -- (6.2,0) -- cycle;
    \foreach \c in {5,9,14}
      \draw[densely dotted, domain=0.6:6.4, samples=40] plot (\x,{\c/\x});
    \foreach \a/\b in {0.45/0.25, 0.95/0.5, 1.5/0.8}
      \draw[thin, rotate around={12:(2.8,3.05)}] (2.8,3.05) ellipse ({\a} and {\b});
  \end{scope}
  \draw[thick] \fullload;
  \fill (2.8,3.05) circle (1.3pt);
  \node[fill=white, inner sep=0.5pt, above=1.5pt] at (2.8,3.05) {$b_{\mathrm{e,min}}$};
  \node at (2.6,4.35) {Full-load curve};
  \node[fill=white, inner sep=1pt] at (3.6,1.75) {$P=\mathrm{const.}$};

  \coordinate (pmax) at (5.6,3.62);
  \draw[very thick] (0.85,0) -- (0.85,2.3) .. controls (1.6,3.05) .. (2.8,3.15)
    .. controls (4.0,3.25) .. (pmax);
  \draw[thick, dash pattern=on 5pt off 2pt on 1pt off 2pt]
    (0.85,0.25) -- (3.6,0.7) -- (5.1,1.4) -- (pmax);
  \draw[thick, dashed] (1.6,0) -- (1.6,2.2) -- (pmax);

  \draw[->] (0,0) -- (6.6,0) node[below left] {Engine speed $n$};
  \draw[->] (0,0) -- (0,4.6);
  \node[rotate=90, above] at (0,2.3) {Engine torque $T$};

  \begin{scope}[shift={(0.2,-1.05)}, y=1cm]
    \draw[very thick] (0,0) -- (0.6,0) node[right] {consumption-oriented};
    \draw[thick, dash pattern=on 5pt off 2pt on 1pt off 2pt] (0,-0.42) -- (0.6,-0.42)
      node[right] {performance-oriented};
    \draw[thick, dashed] (0,-0.84) -- (0.6,-0.84) node[right] {compromise};
  \end{scope}
\end{tikzpicture}
\end{minipage}\hfill
\begin{minipage}[t]{0.48\textwidth}
\textbf{b)}\par\medskip
\centering
\begin{tikzpicture}[font=\small, >={Stealth[length=2mm]}]
  \def\rp{0.45}\def\rs{1.2}\def\d{3.6}
  \pgfmathsetmacro{\phi}{asin((\rs-\rp)/\d)}
  \coordinate (P) at (0,0);
  \coordinate (S) at (\d,0);
  \draw[thick] (P) circle (1.0);
  \draw[thick] (S) circle (1.45);
  \draw[very thick]
    ($(P)+(90+\phi:\rp)$) arc[start angle=90+\phi, end angle=270-\phi, radius=\rp]
    -- ($(S)+(270-\phi:\rs)$) arc[start angle=-90-\phi, end angle=90+\phi, radius=\rs]
    -- cycle;
  \foreach \c in {P,S} {
    \draw (\c) circle (0.09);
    \draw ($(\c)+(-0.15,0)$) -- ++(0.3,0) ($(\c)+(0,-0.15)$) -- ++(0,0.3);
  }
  \draw[->] (P) -- ++(-120:\rp) coordinate[pos=0.5] (rm);
  \draw[thin] (rm) -- (-1.15,-1.05) node[below left=-1pt] {$r_{\mathrm{prim}}$};
  \draw[->] (S) -- ++(-40:\rs) node[pos=0.5, above right, fill=white, inner sep=1pt] {$r_{\mathrm{sec}}$};
  \node[align=center, above] at (0,1.0) {Primary pulley\\(crankshaft)};
  \node[align=center, above] at (\d,1.45) {Secondary pulley\\(output)};
  \node at (1.45,-1.2) {V-belt};
\end{tikzpicture}
\end{minipage}
\caption[Engine map and CVT belt drive]{Engine map and CVT. a) Engine map (based on~\protect\cite[p.~135]{Har07}), b) CVT belt drive\nocc}
\label{fig:1}
\end{figure}
The function is as follows: At idle, the largest possible ratio is present. The effective lever arm $r_{\mathrm{prim}}$ (primary) is at its smallest, $r_{\mathrm{sec}}$ (secondary) at its largest. The ratio is thus determined by:
\[ i=\frac{{r}_{\mathrm{sec}}}{{r}_{\mathrm{prim}}} \]
If the engine speed is now increased, centrifugal weights press the two primary conical pulleys together. Since the belt has a fixed length, the secondary conical pulleys are pushed apart. The ratio decreases. Above a certain speed, the maximum lever arm at the primary conical pulley is reached. The ratio cannot decrease any further. Due to the operating behavior of the combustion engine, a clutch is required for starting off.

\looseness=-1 \textbf{Description of advantages and disadvantages:} CVT belt drives are used above all when a fixed engine operating point is required due to low engine power. Electro-hydraulic control systems for precise control of the ratio are too complex and too expensive for two-wheelers of this category.

Table~\ref{tab:proscvt} summarizes the advantages and disadvantages of the transmission.

\proscons{Advantages and disadvantages of CVT belt drives}{tab:proscvt}{
Constant operating points over a wide speed range & Poor efficiency due to slip and flexing work of the belt \\
Low-cost and simple design if the mechanical control is sufficient & Regular replacement of the belt required \\
High riding comfort & High inertia due to the conical pulleys \\
No additional operation by the rider required & ``Rubber band effect'' during sporty riding \\
}

\section{Yamaha Chip Controlled-Shift}\label{sec:11}

In 2006, Yamaha introduced a sport tourer with the semi-automatic shift system YCC-S with the model FJR1300AS. The ``AS'' version with YCC-S was temporarily withdrawn from the market and has been available again since 2013~\cite{Mot19}.

The shift system supports and relieves the rider in everyday traffic situations. The omission of the clutch lever increases riding comfort in city traffic and in traffic jams. Although the system changes gears automatically at the rider's request, a pure automatic mode with fixed shift maps does not exist. Depending on personal preference, the request to upshift or downshift can be made either by foot or by a switch on the handlebar. The MCU\phantomsection\label{bm:Ref14713289} controls the optimal clutch actuation~\cite[p.~1]{FJR19}.

\textbf{Function:} Technically, the gearbox can be described as an automated gearbox with electronic clutch actuation. The gearbox is a sequential 5-speed gearbox in which neutral is not located between 1st and 2nd gear as usual, but is arranged in first place before 1st gear~\cite[p.~2, Fig.~2]{FJR19}. A shift actuator is mounted on the gearbox itself, which turns the shift shaft via an actuating mechanism. The foot shift lever has no mechanical connection to the shift drum; the shift actuation is carried out electromechanically (Appendix~\ref{app:1}). The clutch disengagement is controlled by the MCU during a gear change. The clutch actuation depends on the engine operating point and the engine speed. Shift requests are checked for plausibility, and critical requests are not executed, e.g. downshifting into excessively high speeds or upshifting into excessively low speeds. During operation, the entire electronic shift system carries out self-tests. If a fault occurs, the system is switched off and the fault state is signaled to the rider. Gear changes are no longer possible in the event of a fault~\cite[p.~1]{FJR19}.

\textbf{Description of advantages and disadvantages:} With the YCC-S in the FJR1300AS model, Yamaha has implemented a comfortable shift system for large-displacement touring motorcycles. The design is based on a conventional manual transmission. Table~\ref{tab:prosyccs} summarizes the advantages and disadvantages.

\proscons{Advantages and disadvantages of the Yamaha YCC-S}{tab:prosyccs}{
Solid and sturdy design & No reduction of moving parts compared to manual shift actuation \\
Small package size compared to dual clutch transmissions & Too complex for motorcycles in the lower price segment \\
Widely used, hydraulically actuated clutches can be employed & Large number of controls for shift actuation \\
}

\section{ShiftFX - Electronic Shift Transmission}\label{sec:12}

\looseness=-1 The Canadian company Biperformance Development Corporation has developed a novel transmission control system for motorcycles with a conventional manual transmission. The product is available in the aftermarket. As with Yamaha's YCC-S, the clutch is actuated automatically; however, the clutch lever for manual actuation in certain situations is retained. The foot shift lever is omitted. There is a semi-automatic and a fully automatic mode. In fully automatic mode, no shift interventions by the rider are required. In semi-automatic mode, the rider selects the gears with switches on the handlebar. If desired, gear changes take place in less than \qty{50}{\milli\second}. Through precisely controlled clutch actuation and speed synchronization between engine and transmission, the system reduces transmission and clutch wear. In addition, semi-automatic gear changes are very easy for the rider and increase comfort on long journeys. The system is also available without automatic clutch actuation; in this version, gear changes are clearly noticeable~\cite{Ele19}.

\textbf{Function:} The mechanical gear changes are essentially carried out by two components: rotation of the shift drum via the actuator motor and an electrohydraulically actuated clutch. The electronic control is performed by the sensors already present on the motorcycle via a dedicated transmission control unit~\cite{Ele19}.

\textbf{Description of advantages and disadvantages:} The actuator motor is mechanically connected to the shift drum through a two-stage gear ratio. By means of this gear ratio, a torque of \qty{3.5}{\newton\metre} can be applied to the shift drum. A detent mechanism is used to lock a gear step in place~\cite{Ele19}. The detent forces must be overcome with every gear change. Thus, the actuator motor must be dimensioned larger than the rotation of the shift drum requires. Although the automatic gear changes eliminate the mechanical components of the foot shift lever, they do not eliminate the detent of the gear steps. The two-stage reduction gearbox is added. The complete design including the control unit weighs four kilograms~\cite{Shi19}.

OEMs demand a design that is as simple and cost-effective as possible. Additional gearboxes and detent mechanisms, which in turn are maintenance-intensive for the end user, usually offer neither cost nor competitive advantages. Should the system be used in a series application, essential components must be adapted. Table~\ref{tab:prosshiftfx} summarizes the advantages and disadvantages.

\proscons{Advantages and disadvantages of ShiftFX}{tab:prosshiftfx}{
Well-thought-out design of the overall system with many sensors & Detent forces must be overcome with every gear change \\
Automatic clutch actuation and intervention in the engine control for smooth gear changes & Multi-stage gearbox required to drive the shift drum \\
Elimination of the foot shift lever and the associated mechanism & Complex design of the mechanism \\
}

\section{Evaluation of the Systems}\label{sec:13}

At the beginning of a new development, it is important to evaluate the current state of the art with regard to the required criteria. Parts of systems that have proven themselves over many years can often be incorporated into the new development.

The use of a CVT combines many advantages with simple and sturdy technology. Above all, riding comfort and cost-effective implementation are in the foreground. The operating characteristic of the engine can be set excellently with a variable gear ratio -- exactly as required in the task definition. However, this is cancelled out again by the poor efficiency of the transmission. A belt-driven CVT is therefore not usable for the intended purpose.

The automated shift system Yamaha YCC-S: It is evident that automated shift systems are already available on the market in small quantities. A fully automatic mode does not yet exist, however -- but it can easily be realized electronically. The mechanism for carrying out the gear changes is, however, too complex for the intended purpose. The goal is to carry out the gear changes with as few moving parts as possible. The entire mechanism for manual gear changes is to be eliminated. YCC-S is an in-house development of Yamaha; using the system requires long-term license negotiations. Should the system be used, adaptations are necessary. If the rights are not held by HC-Concepts, further developments are not possible without restriction. Thus, such a design is not suitable for the intended purpose.

ShiftFX offers an excellently integrated solution for motorcycles with a conventional manual gearbox. The technology has been in use for several years and has proven itself. For the mechanical rotation of the shift drum, however, an additional gearbox with a detent is required. In the end product, production and maintenance are thus confronted with higher costs. In addition, Biperformance Development Corporation holds the patents for the actuator motor in combination with the gearbox for driving the shift drum. The licenses for using the system would have to be acquired. Due to the required additional mechanisms and the patent rights, this system will not be used in this project~\cite{Shi19}.

Ideas for simplifying the ShiftFX system have already been presented in the development team of HC-Concepts. Above all, the number of parts and the production costs are to be reduced. The system is designed for the aftermarket; the components are developed with the greatest possible flexibility. By specializing in specific engine/gearbox combinations, essential properties of the system can be optimized.

\chapter{Electrical and Mechanical Components}\label{sec:14}

This chapter describes the hardware of the shift system. Figure~\ref{fig:2} shows the overall concept on the gearbox test rig. The gearbox test rig and its drive are presented in Sections~\ref{sec:16} and~\ref{sec:17}, followed by the shift drum position sensor (Section~\ref{sec:18}) and the sizing of the actuator motor (Section~\ref{sec:19}).

\par\noindent\begin{minipage}{\textwidth}
\centering
\input{figures/tikz/concept}\par
\captionof{figure}{System concept on the gearbox test rig\nocc}
\label{fig:2}
\end{minipage}
\clearpage

\section{Gearbox Test Rig}\label{sec:16}

For the development, a production-ready manual gearbox from the most recent development work of the company HC-Concepts is used. The internal components of the gearbox are used unchanged, which reduces the development effort. However, mechanical changes to the shift drum actuation require a redesign of the housing. The goal is to obtain a gearbox without oil filling and with direct access to the individual mechanical components. Gear changes must be visible from the outside. Furthermore, individual components must be replaceable within a short time.

The self-locking worm gear, as an essential part of the automation, is mounted outside the housing. Changes such as new gear wheels, bearings or electric motors can thus be carried out within a few minutes.

For mounting the shift drum position sensor and the actuator motor, adapters are designed and manufactured in the in-house 3D printing department. In later series production, the gearbox will be redesigned so that the worm gear and the position sensor can be mounted directly.

The gearbox is a single-stage, sequential, non-synchronized 6-speed gearbox. Table~\ref{tab:1} lists the parameters used in the design calculations.

\par\medskip\noindent\begin{minipage}{\textwidth}
\centering
\captionof{table}{Data of the gearbox test rig (source:~\protect\cite[p.~1]{Ron19})}\label{tab:1}
{\small
\begin{tabular}{lc}
\toprule
\textbf{Parameter} & \textbf{Value} \\
\midrule
Rated speed & \qty{3600}{\per\minute} \\
Rated service life & \qty{60000}{\kilo\metre} \\
Moment of inertia of the shift drum & \qty{120e-6}{\kilogram\metre\squared} \\
Mass of the heaviest shift fork (1-3, 2-4) & \qty{0.063}{\kilogram} \\
Mass of the heaviest gear wheel (1-3-4) & \qty{0.236}{\kilogram} \\
Shift travel of the shift forks & \qty{5}{\milli\metre} \\
Shift travel on the shift drum & \ang{27} \\
Gear spacing on the shift drum & \ang{60} (\ang{30} to neutral) \\
Gear ratio of the worm gear & 30 \\
Efficiency of the worm gear & \qty{42}{\percent} \\
Gear ratio of the electric gearbox drive & 2.4 \\
\bottomrule
\end{tabular}}
\end{minipage}\par\medskip

\section{Gearbox Drive}\label{sec:17}

In the motor vehicle, the gearbox input shaft is driven by the crankshaft of the combustion engine via the primary drive. The gearbox test rig is to be operated at speeds similar to those in the later vehicle. A belt drive is required to adapt the speed of the electric motor. In addition, it is possible to control the operating voltage of the drive motor with a PWM controller (Figure~\ref{fig:3}). As an equivalent to the ignition interruption of the combustion engine, a relay is used in the gearbox test rig to switch off the electric motor during the gear change (relay in Figure~\ref{fig:2}). The rotating masses of the combustion engine and of the rear wheel including the chain drive could only be replicated as a model. The load changes therefore do not behave exactly as in the actual vehicle. This circumstance must be taken into account during development.

\begin{wrapfigure}[21]{r}{0.37\textwidth}
\vspace*{-\intextsep}\vspace*{-1.3pt}%
\centering
\begin{tikzpicture}[font=\scriptsize, x=1cm, y=1cm,
    key/.style={draw, thick, rounded corners=2pt, minimum width=1.08cm, minimum height=0.78cm,
                align=center, inner sep=2.5pt}]
  \draw[thick, rounded corners=3pt] (0,0) rectangle (5.6,3.0);
  \draw[thick] (0.4,1.75) rectangle (4.4,2.7);
  \foreach \i in {0,...,3} {
    \begin{scope}[shift={(0.75+0.9*\i,1.88)}]
      \draw (0,0) rectangle (0.45,0.7);
      \draw (0,0.35) -- (0.45,0.35);
    \end{scope}
  }
  \node[font=\small] at (4.95,2.2) {\%};
  \draw (0.4,1.42) -- (5.2,1.42);
  \node[key] at (0.97,0.72) {MENU};
  \node[key] at (2.19,0.72) {$\blacktriangle$};
  \node[key] at (3.41,0.72) {$\blacktriangledown$};
  \node[key] at (4.63,0.72) {RUN/\\[-1pt]STOP};
\end{tikzpicture}\par
\captionof{figure}{PWM controller of the gearbox drive\nocc}
\label{fig:3}
\vspace{\intextsep}%
\captionof{table}{Friction torques of the gearbox drive}\label{tab:2}
{\small
\begin{tabular}{>{\centering\arraybackslash}p{\dimexpr 0.22\textwidth-2\tabcolsep\relax}>{\centering\arraybackslash}p{\dimexpr 0.13\textwidth-2\tabcolsep\relax}}
\toprule
\textbf{Gear} & \textbf{Torque} \\
\midrule
1 & \qty{98}{\milli\newton\metre} \\
N (output blocked) & \qty{92}{\milli\newton\metre} \\
2 & \qty{84}{\milli\newton\metre} \\
3 & \qty{104}{\milli\newton\metre} \\
4 & \qty{106}{\milli\newton\metre} \\
5 & \qty{129}{\milli\newton\metre} \\
6 & \qty{152}{\milli\newton\metre} \\
\bottomrule
\end{tabular}}
\end{wrapfigure}
For the dimensioning of the gearbox drive, the torque required to overcome the friction must be determined at the gearbox input shaft. The torque is measured with variable weights on a thread around the circumference. A single measurement is carried out for each gear. In real driving operation, the gearbox position neutral is engaged above all when the vehicle is stationary. The output shaft of the gearbox must be blocked when determining the friction torque.

\textbf{Design of the gearbox drive:} A torque of \qty{152}{\milli\newton\metre} is required to drive the gearbox input shaft in 6th gear (Table~\ref{tab:2}). The input speed of the gearbox should lie in the medium engine speed range. With a maximum gearbox input speed of \qty{3600}{\per\minute}, the design in the medium speed range is based on \qty{1000}{\per\minute} to \qty{2000}{\per\minute} (Table~\ref{tab:1}). According to the type plate, the electric motor delivers \qty{100}{\milli\newton\metre} at \qty{3000}{\per\minute}; a gear ratio between the motor and the gearbox input shaft is required. The belt drive has a gear ratio of 2.4 (Table~\ref{tab:1}). Thus, the motor speed is reduced by a factor of 2.4 and the motor torque is increased by a factor of 2.4.

\section{Shift Drum Position Sensor}\label{sec:18}

For the gearbox test rig, parts from current developments are used. The rotary potentiometer as position sensor is therefore predetermined. The arrangement of the shift drum as well as of the input and output shaft was adapted to the gearbox test rig accordingly in comparison to the series gearbox. This results in an opposite direction of rotation of the shift drum and of the potentiometer compared to the manufacturer's specification. The potentiometer must be mounted either on the right or on the left side according to the direction of rotation of the shift drum. A change of the mounting position without changing the direction of rotation of the shift drum results in mirrored voltage values compared to the manufacturer. The position values of the individual gears must therefore be determined manually. To ensure correct function, the gear positions are measured in the installed state; Table~\ref{tab:4} lists the measured voltages and the ADC values used as gear positions in the firmware. Table~\ref{tab:3} summarizes the technical data of the position sensor, Figure~\ref{fig:4}\,a shows its design.

\textbf{Determination of the gear positions:} The gear positions are determined directly on the gearbox test rig (Figure~\ref{fig:2}). The shift drum is turned to the position of the respective gear, and the ADC value is transmitted from the microcontroller to the PC via the serial interface. The actual voltage at the signal output of the potentiometer is measured with the multimeter. In addition, it was determined that a deviation of \ang{\pm 5} from the required angular position of the shift drum can be regarded as uncritical. With the 12-bit resolution of the ADC in the microcontroller, this corresponds to a deviation of $\pm 57$ ADC values. Due to the opposite direction of rotation of the potentiometer, the gear positions according to the manufacturer do not match the actual voltage values. The firmware therefore uses the gear positions determined on the gearbox test rig (Table~\ref{tab:4}, Figure~\ref{fig:32}). The ADC values are mirrored (4095 minus the conversion result, Section~\ref{sec:39}), so that they rise when shifting up.

\par\medskip\noindent\begin{minipage}{\textwidth}
\begin{minipage}[t]{0.48\textwidth}
\textbf{a)}\par\medskip
\centering
\begin{tikzpicture}[font=\scriptsize, >={Stealth[length=1.8mm]}]
  \draw[thick, fill=white, rounded corners=6pt] (-0.15,1.45) rectangle (0.7,2.55);
  \draw[thick, fill=white, rounded corners=6pt] (2.9,1.45) rectangle (3.75,2.55);
  \draw[thick] (0.2,2.0) circle (0.2) (3.4,2.0) circle (0.2);
  \draw[thick, fill=white, rounded corners=8pt] (0.5,0.75) rectangle (3.1,3.25);
  \draw[thick] (1.8,2.0) circle (0.85);
  \draw[thick] (1.8,2.0) circle (0.5);
  \draw[thick] (1.5,1.93) rectangle (2.1,2.07);
  \draw[thick, fill=white] (1.15,0.15) rectangle (2.45,0.75);
  \foreach \x in {1.3,1.7,2.1} \draw (\x,0.25) rectangle ++(0.2,0.4);

  \draw[thick] (4.6,1.5) rectangle (5.4,2.5);
  \draw[thick] (5.4,1.8) rectangle (5.75,2.2);
  \draw[dash dot, thin] (4.4,2.0) -- (6.0,2.0);
  \draw[->] (6.0,1.55) arc[start angle=-60, end angle=60, radius=0.52] node[above right=-2pt] {$\varphi$};

  \draw[thin] (1.8,2.0) -- (1.8,3.55) node[above] {Rotor with driver};
  \node[anchor=south west, inner sep=1pt] at (3.2,2.6) {Mounting};
  \node[anchor=north west, align=left, inner sep=1pt] at (2.62,0.6) {Connector\\(+5\,V, signal, GND)};%
  \node[anchor=north] at (5.1,1.35) {Side view};
\end{tikzpicture}
\end{minipage}\hfill
\begin{minipage}[t]{0.48\textwidth}
\textbf{b)}\par\medskip
\centering
\begin{tikzpicture}[font=\scriptsize]
  \draw[thick] (1.0,3.0) -- (4.4,3.0) (1.0,1.0) -- (4.4,1.0);
  \draw[thick] (4.4,3.0) arc[start angle=90, end angle=-90, x radius=0.25, y radius=1.0];
  \draw[thick] (4.6,2.7) -- (5.4,2.7) (4.6,1.3) -- (5.4,1.3);
  \draw[thick] (5.4,2.7) arc[start angle=90, end angle=-90, x radius=0.2, y radius=0.7];
  \draw[thick, fill=white] (1.0,2.0) ellipse (0.32 and 1.0);
  \draw[thick] (1.0,2.0) ellipse (0.16 and 0.5);
  \foreach \a in {50,130,230,310} \draw ($(1.0,2.0)+(\a:0.24 and 0.76)$) circle (0.035);
  \draw[thick, fill=white] (0.1,1.85) -- (1.0,1.85) arc[start angle=-90, end angle=90, x radius=0.05, y radius=0.15]
    -- (0.1,2.15);
  \draw[thick, fill=white] (0.1,2.0) ellipse (0.05 and 0.15);
  \draw[thick] (3.5,1.0) -- (3.5,0.6) -- (3.25,0.15) (3.7,1.0) -- (3.7,0.6) -- (3.45,0.15);
  \draw[thick] (4.95,1.3) -- (4.95,0.75) (5.15,1.3) -- (5.15,0.75);
  \draw[thick, fill=white] (4.6,0.15) rectangle (5.5,0.75);
  \foreach \x in {4.68,4.85,5.02,5.19,5.36} \draw (\x,0.35) rectangle ++(0.1,0.2);

  \node[anchor=north] at (0.35,1.78) {Shaft};
  \node[anchor=south] at (2.7,3.05) {DC motor};
  \node[anchor=south] at (5.2,3.05) {Incremental encoder};
\end{tikzpicture}
\end{minipage}\par
\captionof{figure}[Potentiometer and actuator motor]{Potentiometer and actuator motor. a) Potentiometer, b) actuator motor\nocc}
\label{fig:4}
\end{minipage}

\par\bigskip\noindent\begin{minipage}{\textwidth}
\begin{minipage}[b]{0.48\textwidth}
\centering
\captionof{table}{Technical data of the potentiometer}\label{tab:3}
{\small
\begin{tabular*}{\linewidth}{@{}l@{\extracolsep{\fill}}c@{}}
\toprule
\textbf{Parameter} & \textbf{Value} \\
\midrule
Sensing principle & Hall effect \\ \addlinespace[5pt]
Supply voltage & \qty{5}{\volt} \\ \addlinespace[5pt]
Output voltage & \qty{0.5}{\volt} to \qty{4.4}{\volt} \\ \addlinespace[5pt]
Position accuracy & \ang{\pm 2} \\ \addlinespace[5pt]
Electrical angle of rotation & approx.\ \ang{330} \\
\bottomrule
\end{tabular*}}
\end{minipage}\hfill
\begin{minipage}[b]{0.48\textwidth}
\centering
\captionof{table}{Gear positions: measured voltage and ADC value in the firmware}\label{tab:4}
{\small
\begin{tabular*}{\linewidth}{@{}c@{\extracolsep{\fill}}cc@{}}
\toprule
\textbf{Position} & \textbf{Voltage} & \textbf{ADC value} \\
\midrule
1st gear & \qty{4.480}{\volt} & 700 \\
Neutral & \qty{4.090}{\volt} & 1000 \\
2nd gear & \qty{3.680}{\volt} & 1280 \\
3rd gear & \qty{2.905}{\volt} & 1840 \\
4th gear & \qty{2.151}{\volt} & 2480 \\
5th gear & \qty{1.345}{\volt} & 3100 \\
6th gear & \qty{0.569}{\volt} & 3700 \\
\bottomrule
\end{tabular*}}
\end{minipage}
\end{minipage}

\section{Actuator Motor}\label{sec:19}

The electric motor maxon DCX32L GB KL 12V was selected for the development tests (Figure~\ref{fig:4}\,b)~\cite{B7A19}. As the calculation of the dynamic behavior shows (Figure~\ref{fig:8} and the following calculation), a highly dynamic motor with a low moment of inertia is required to drive the shift drum. During the gear changes, the motor is briefly operated at high torque. In the test setup, an incremental encoder is required for the precise measurement of the dynamic behavior; angular velocity or speed, relative position and angular acceleration can be recorded via this encoder. In the completed project, this encoder is omitted.

The calculation of the position control requires knowledge of the torque required at the shift drum to overcome the individual friction forces. For an optimal design, the friction of each guide of a shift track is determined separately. For this purpose, the shift track on the shift drum is positioned between two gears, while a thread with weights hangs on the circumference of the shift drum. The weight is increased until the tangential force required to move the shift fork is applied. The friction torque can be calculated from the radius and the weight. The measurement results are listed as torque values in Table~\ref{tab:5}.

If the shift forks are removed, a torque of \qty{15}{\milli\newton\metre} is measured at the motor shaft (actuator motor) to turn the shift drum.

\par\medskip\noindent\begin{minipage}{\textwidth}
\centering
\captionof{table}{Friction torques of the shift drum}\label{tab:5}
{\small
\begin{tabular}{>{\centering\arraybackslash}p{\dimexpr 0.146\textwidth-2\tabcolsep\relax}>{\centering\arraybackslash}p{\dimexpr 0.219\textwidth-2\tabcolsep\relax}>{\centering\arraybackslash}p{\dimexpr 0.236\textwidth-2\tabcolsep\relax}>{\centering\arraybackslash}p{\dimexpr 0.238\textwidth-2\tabcolsep\relax}}
\toprule
\textbf{Guide} & \textbf{Shift track 1/3} & \textbf{Shift track 5/6} & \textbf{Shift track 2/4} \\
\midrule
A & \qty{105}{\milli\newton\metre} & \qty{55}{\milli\newton\metre} & \qty{80}{\milli\newton\metre} \\
B & \qty{126}{\milli\newton\metre} & \qty{75}{\milli\newton\metre} & \qty{96}{\milli\newton\metre} \\
C & \qty{130}{\milli\newton\metre} & not present & \qty{78}{\milli\newton\metre} \\
D & not present & not present & \qty{102}{\milli\newton\metre} \\
\bottomrule
\end{tabular}}
\end{minipage}

\subsection{Measurement of the Shift Times}\label{sec:20}

\begin{wrapfigure}{r}{0.45\textwidth}
\vspace*{-\intextsep}\vspace*{-0.9pt}%
\centering
\includegraphics[width=0.43\textwidth]{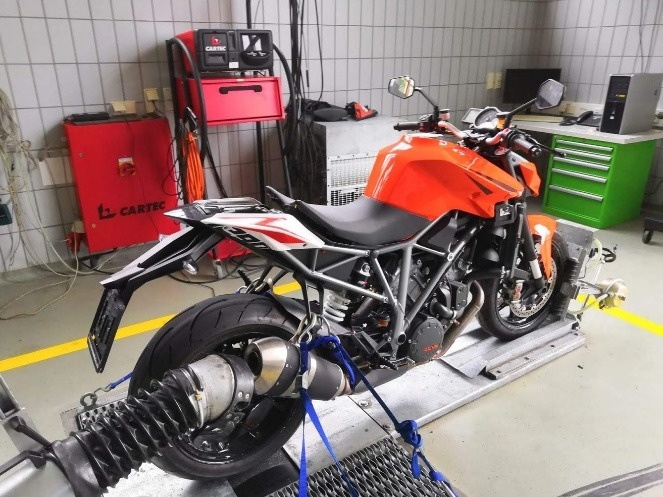}
\caption{Motorcycle chassis dynamometer}
\label{fig:5}
\end{wrapfigure}
The shift times of the automated system should be similar to those of a manual gear change. Knowledge of the shift times is necessary for the following calculation; therefore, the shift times for manual gear changes are determined during a run on the test bench (Figure~\ref{fig:5}). A large-displacement series-production motorcycle with a non-synchronized six-speed gearbox is used as the test motorcycle. For the test bench run, all rider assistance systems are deactivated, as these would falsify the measurement results on the test bench. A shift drum position sensor is also installed by the manufacturer. However, the potentiometer installed as standard on the vehicle does not correspond to the potentiometer used in the project. The voltage values of the gear positions are different.

The voltage value at the shift drum position sensor (potentiometer) is recorded with an oscilloscope, and the shift time is determined from it. The signal line to the engine control unit must not be disconnected. On this model, a fault in the gear detection system activates protective functions in the engine control unit and sets a speed limit in test bench operation. Once disconnected, the shift drum position sensor must be re-taught with the manufacturer's diagnostic tool.

To obtain measurement results that are as realistic as possible, several different driving conditions are tested:

\begin{itemize}
\item Accelerating from low engine speeds or vehicle speeds
\item Accelerating at high engine speeds / vehicle speeds
\item Downshifting when braking hard into a corner
\item Relaxed riding style
\item Sporty riding style
\item Shifting with and without clutch
\item Starting on inclines / under high load
\end{itemize}
Of course, a test run on the test bench never corresponds \qty{100}{\percent} to the daily conditions on the road. The personal riding style has a large influence. However, the measured values are sufficiently accurate for the calculation of the actuator motor. As an example, the gear change from 1st to 2nd gear is considered in more detail. Figure~\ref{fig:6} shows the rotation angle of the shift drum, calculated from the voltage at the potentiometer.

\begin{figure}[ht]
\centering
\begin{tikzpicture}[marker/.style={draw, circle, fill=white, inner sep=1pt, font=\scriptsize}]
\begin{axis}[
    thesisplot,
    width=0.97\textwidth, height=5.0cm,
    xmin=0, xmax=475, xtick={0,100,...,400},
    ymin=-5, ymax=78, ytick={0,20,...,60},
    xlabel={$t$ in \unit{\milli\second}}, ylabel={$\varphi_{\mathrm{drum}}$ in \unit{\degree}},
    clip=false,
  ]
  \addplot[plotline] table[x=t_ms, y=phi_deg] {figures/data/gear_change_test_bench.dat};
  \draw[thin] (axis cs:35,11.9) -- (axis cs:35,1);
  \draw[thin] (axis cs:170,19.2) -- (axis cs:170,8);
  \draw[thin] (axis cs:277,56.6) -- (axis cs:290,48);
  \draw[thin] (axis cs:325.9,73.3) -- (axis cs:322,64);
  \draw[thin] (axis cs:425.1,69.9) -- (axis cs:425.1,61.5);
  \node[marker] at (axis cs:35,11.9) {1};
  \node[marker] at (axis cs:170,19.2) {2};
  \node[marker] at (axis cs:277,56.6) {3};
  \node[marker] at (axis cs:325.9,73.3) {4};
  \node[marker] at (axis cs:425.1,69.9) {5};
\end{axis}
\end{tikzpicture}\par
\caption{Oscillogram of a gear change on the test bench\nocc}
\label{fig:6}
\end{figure}
Different states can be read from the curve:

\begin{itemize}
\item Position 1: Rotation angle \ang{0} - 1st gear is engaged \\
The shift lever is not actuated.
\item Position 2: Rotation angle \ang{7} - 1st gear is engaged \\
The shift lever is pushed upward with the foot. The free travel (\ang{5}) is taken up. The applied force deforms the shift fork and turns the shift drum by a further \ang{2}. Only when the load is removed (by operating the clutch) can the gear wheel be moved.
\item Position 3: Rotation angle \ang{46} - Gearbox is load-free \\
Tooth-on-tooth position of the shift dogs. Due to the speed difference, some time elapses before the shift dogs can slide into one another. The foot force strains the mechanism by approx. \ang{1}.
\item Position 4: Rotation angle \ang{64} - 2nd gear is engaged \\
The dogs slide into one another. Due to the strain in the mechanism, the shift drum overshoots its end position.
\item Position 5: Rotation angle \ang{60} - 2nd gear is engaged \\
The shift drum has reached its end position. The foot shift lever is no longer actuated and the detent of the shift star has engaged.
\end{itemize}
\par\noindent\ignorespaces%
\begin{wraptable}{r}{0.57\textwidth}
\vspace*{-\intextsep}\vspace*{-2.5pt}%
\centering
\caption{Statistical parameters of the manual shift times}\label{tab:shifttimes}
{\small\setlength{\tabcolsep}{4pt}%
\begin{tabular}{lcc}
\toprule
\textbf{Parameter} & \textbf{Upshifting} & \textbf{Downshifting} \\
\midrule
Number of measurements & 42 & 42 \\
Arithmetic mean & \qty{80.1}{\milli\second} & \qty{92.4}{\milli\second} \\
Median & \qty{71.5}{\milli\second} & \qty{80}{\milli\second} \\
Minimum & \qty{26}{\milli\second} & \qty{21}{\milli\second} \\
Maximum & \qty{300}{\milli\second} & \qty{257}{\milli\second} \\
Standard deviation & \qty{47.9}{\milli\second} & \qty{57.4}{\milli\second} \\
\bottomrule
\end{tabular}}
\end{wraptable}
\textbf{Evaluation of the measurement results:} The measured shift times are now entered into the CAS system as a list (Appendix~\ref{app:4}). A total of 84~measured values were collected, 42~for upshifting and 42~for downshifting. Table~\ref{tab:shifttimes} summarizes the most important statistical parameters of the measurement series. The mean upshift time is \qty{80.1}{\milli\second}, the mean downshift time \qty{92.4}{\milli\second}.

With standard deviations of \qty{47.9}{\milli\second} and \qty{57.4}{\milli\second}, the scatter of the shift times is large. The \qty{95}{\percent} confidence interval of the mean is \qty{80.1 \pm 14.5}{\milli\second} for upshifting and \qty{92.4 \pm 17.4}{\milli\second} for downshifting. \qty{83}{\percent} of the upshifts and \qty{69}{\percent} of the downshifts are completed after \qty{100}{\milli\second} or earlier. The target of \qty{100}{\milli\second} for the automated system is thus within the range of manual gear changes.

The individual values are listed in Appendix~\ref{app:4} and are additionally freely available as a data set in CSV format~\cite{Sto26}.

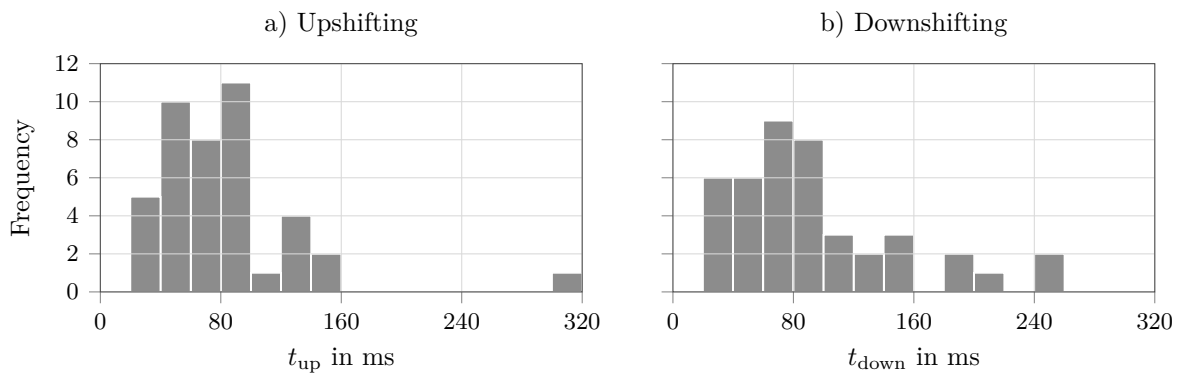
\begin{figure}[hbp]%
\centering
\begin{tikzpicture}
\begin{groupplot}[
    thesisplot, grid=none, ymajorgrids,
    group style={group size=2 by 1, horizontal sep=1.2cm, y descriptions at=edge left},
    width=0.505\textwidth, height=4.6cm,%
    ymin=0, ymax=12, ytick={0,2,...,12},
    ylabel={Frequency},
    ybar interval, area style,
    xmin=0, xmax=320, xtick={0,80,...,320}, xticklabels={0,80,160,240,320},
    x tick label as interval=false,
  ]
  \nextgroupplot[xlabel={$t_{\mathrm{up}}$ in \unit{\milli\second}}, title={a) Upshifting}]
    \addplot[draw=white, fill=plotfill] coordinates {
      (20,5) (40,10) (60,8) (80,11) (100,1) (120,4) (140,2) (160,0) (180,0) (200,0)
      (220,0) (240,0) (260,0) (280,0) (300,1) (320,0)};
  \nextgroupplot[xlabel={$t_{\mathrm{down}}$ in \unit{\milli\second}}, title={b) Downshifting}]
    \addplot[draw=white, fill=plotfill] coordinates {
      (20,6) (40,6) (60,9) (80,8) (100,3) (120,2) (140,3) (160,0) (180,2) (200,1)
      (220,0) (240,2) (260,0)};
\end{groupplot}
\end{tikzpicture}\par
\caption[Histograms of the 84 measured manual shift times]{Histograms of the 84 measured manual shift times (bin width \qty{20}{\milli\second}). a) Upshifting, b) downshifting\nocc}
\label{fig:7}
\end{figure}
\textbf{Upshifting:} In the histogram (Figure~\ref{fig:7}\,a), most upshifts occur between \qty{80}{\milli\second} and \qty{100}{\milli\second}. The spread of the measurement results lies between \qty{20}{\milli\second} and \qty{320}{\milli\second}. The bar between \qty{300}{\milli\second} and \qty{320}{\milli\second} is to be interpreted as a single outlier. All other measured values are plausible. The distribution function of the measured values is right-skewed.

\textbf{Downshifting:} In the histogram, most gear changes cluster between \qty{60}{\milli\second} and \qty{80}{\milli\second}. The spread lies between \qty{20}{\milli\second} and \qty{260}{\milli\second} and is somewhat smaller compared to the upshift measurement series. No outliers are visible. The measured values are plausible. The distribution function of the measured values is likewise right-skewed.

\subsection{Dimensioning and Design}\label{sec:21}

For dimensioning, the torque and speed profile during the gear changes is used. The shift times are decisive for the magnitude of the torque and acceleration values; from these, the load on the worm gear and on the drive components is determined. Due to the high impulse load caused by the gear changes, continuous operation must not be assumed for the design of the components. The short-term load with high speed and acceleration values increases wear many times over compared to steady-state operation. For determining the service life, not only the operating time but also the shift frequency is relevant.

The selected electric motor has excellent technical data; the disadvantage is the price. The aim is to find out which parameters are decisive for the gear changes. For series production, an inexpensive electric motor can subsequently be selected with precise knowledge of the requirements.

\textbf{Calculation of the motor rotation angle during the gear change:} The motor rotation angle changes with the gear ratio of the worm gear. The gear ratio can therefore be adapted to the technical data of the motor. If the motor has a high torque, the gear ratio can be reduced. In contrast, for an electric motor with a low moment of inertia and high rated speeds, the gear ratio is increased. In the first design of the gearbox test rig, the gear ratio of the worm gear was set to 30 (Table~\ref{tab:1}):
\begin{equation}
{i}_{\mathrm{worm}}=30
\label{eq:3}
\end{equation}
With the gear spacing of the shift drum, $\varphi_{\mathrm{gear}}=\ang{60}=\pi/3\,\mathrm{rad}$ (Table~\ref{tab:1}), the motor rotation angle is obtained as follows:
\begin{equation}
{\varphi }_{\mathrm{motor}}={\varphi }_{\mathrm{gear}}\cdot {i}_{\mathrm{worm}}=10\cdot \pi \,\mathrm{rad}
\label{eq:4}
\end{equation}
\textbf{Mean motor speed during the gear change:} The arithmetic mean of the shift time is between \qty{80}{\milli\second} and \qty{92}{\milli\second} (Table~\ref{tab:shifttimes}). To simplify the following calculation, a shift time of $t_{\mathrm{shift}}=\qty{100}{\milli\second}$ is assumed. For the design of the actuator motor, the difference between the mean values is of minor importance. If the project is successful, the time difference between upshifting and downshifting can be taken into account for optimization. From these data, the mean angular velocity is determined with the motor rotation angle from Eq.~\eqref{eq:4} and, for illustration, represented as a rotational speed:
\[ {\omega }_{\mathrm{mean}}=\frac{{\varphi }_{\mathrm{motor}}}{{t}_{\mathrm{shift}}}=\qty{314}{\radian\per\second}\quad\Rightarrow\quad
   {n}_{\mathrm{mean}}=\frac{{\omega }_{\mathrm{mean}}\cdot 60}{2\cdot \pi }=\qty{3000}{\per\minute} \]
Including motor run-up and braking, an average motor speed of \qty{3000}{\per\minute} is required to meet the time target of \qty{100}{\milli\second} for the gear changes. The maximum speed is considerably higher.

\textbf{Maximum motor speed during the gear change:} To estimate the required motor parameters, a trapezoidal speed profile is defined. As a simplification, the motor torque during the acceleration and braking phases can be assumed to be constant.

\begin{wrapfigure}{l}{0.58\textwidth}
\vspace*{-\intextsep}\vspace*{8.9pt}%
\centering
\begin{tikzpicture}[>={Stealth[length=1.6mm]}]
\begin{axis}[
    thesisplot, grid=none,
    width=0.58\textwidth, height=6.0cm,
    xmin=0, xmax=104, ymin=0, ymax=800,%
    xtick={0,20,...,100}, ytick={0,200,...,600},
    xlabel={$t$ in \unit{\milli\second}}, ylabel={$\omega$ in \unit{\radian\per\second}},
    clip=false,
  ]
  \addplot[plotline, fill=plothighlight] coordinates {(0,0) (40,524) (60,524) (100,0)} -- cycle;
  \draw[<->, thin] (axis cs:0,660) -- node[above, font=\scriptsize] {$t_{\mathrm{acc}}=\qty{40}{\milli\second}$} (axis cs:40,660);
  \draw[<->, thin] (axis cs:40,660) -- node[above, font=\scriptsize] {\qty{20}{\milli\second}} (axis cs:60,660);
  \draw[<->, thin] (axis cs:60,660) -- node[above, font=\scriptsize] {$t_{\mathrm{brake}}=\qty{40}{\milli\second}$} (axis cs:100,660);
  \draw[densely dotted] (axis cs:40,524) -- (axis cs:40,680) (axis cs:60,524) -- (axis cs:60,680)
                        (axis cs:0,0) -- (axis cs:0,680) (axis cs:100,0) -- (axis cs:100,680);
  \node[above, font=\scriptsize] at (axis cs:50,526) {$\omega_{\mathrm{max}}=\qty{524}{\radian\per\second}$};
  \path (axis cs:0,0) -- node[sloped, above, pos=0.42, font=\scriptsize]
    {$\alpha=\qty{13.1e3}{\radian\per\second\squared}$} (axis cs:40,524);
  \path (axis cs:60,524) -- node[sloped, above, pos=0.58, font=\scriptsize]
    {$\alpha=\qty{-13.1e3}{\radian\per\second\squared}$} (axis cs:100,0);
  \node[font=\scriptsize] at (axis cs:50,210) {$\int\!\omega\,\mathrm{d}t=\qty{31.4}{\radian}$};
\end{axis}
\end{tikzpicture}\par
\caption{Trapezoidal speed profile of the actuator motor\nocc}
\label{fig:8}
\end{wrapfigure}
For the calculation, the speed profile is divided into three sections: run-up, constant speed and braking (Figure~\ref{fig:8}). In this way, the operating points can be considered separately and independently of each other. For the series application, the exact speed profile will later be determined by means of simulation.

The calculation uses the times $t_{\mathrm{acc}}=\qty{40}{\milli\second}$, $t_{\mathrm{const}}=\qty{20}{\milli\second}$ and $t_{\mathrm{brake}}=\qty{40}{\milli\second}$. Three individual functions define the speed profile mathematically. The first partial function $f1_{\mathrm{acc}}(x)$ for the time interval $0<x\leq t_{\mathrm{acc}}$ consists of a linear equation $y=k\cdot x+d$ with the slope $k=\omega_{\mathrm{max}}/t_{\mathrm{acc}}$ starting from the point $x=0$, $y=0$:
\[ {f1}_{\mathrm{acc}}\left( x \right) =\frac{{\omega }_{\mathrm{max}}}{{t}_{\mathrm{acc}}}\cdot x+0 \]
The second subfunction $f1_{\mathrm{const}}(x)$ defines the speed profile for the time interval $t_{\mathrm{acc}}<x\leq t_{\mathrm{acc}}+t_{\mathrm{const}}$ consisting of a straight-line equation $y=d$ starting from the point $x=t_{\mathrm{acc}}$, $y=\omega_{\mathrm{max}}$:
\[ {f1}_{\mathrm{const}}\left( x \right) ={\omega }_{\mathrm{max}} \]
The third subfunction $f1_{\mathrm{brake}}(x)$ for the time interval $t_{\mathrm{acc}}+t_{\mathrm{const}}<x\leq t_{\mathrm{shift}}$ consists of a straight-line equation $y=k\cdot x+d$ with the slope $k=-\omega_{\mathrm{max}}/t_{\mathrm{acc}}$. The value $d$ is calculated starting from the point $x=t_{\mathrm{acc}}+t_{\mathrm{const}}$, $y=\omega_{\mathrm{max}}$:
\[ {f1}_{\mathrm{brake}}\left( x \right) =\frac{-{\omega }_{\mathrm{max}}}{{t}_{\mathrm{acc}}}\cdot x+{\omega }_{\mathrm{max}}-\frac{-{\omega }_{\mathrm{max}}}{{t}_{\mathrm{acc}}}\cdot \left( {t}_{\mathrm{acc}}+{t}_{\mathrm{const}} \right) \]
The real speed profile will only approximately correspond to the assumed one. The linear relationships were created to simplify the calculation.

In total, all three functions yield the desired speed profile for the time interval $0<x\leq t_{\mathrm{shift}}$. Each individual function is defined for its specific time interval within the overall function:
\[ f1\left( x \right) =\left\{ \begin{array}{ll}{f1}_{\mathrm{acc}}\left( x \right) , & 0<x\leq {t}_{\mathrm{acc}} \\ {f1}_{\mathrm{const}}\left( x \right) , & {t}_{\mathrm{acc}}<x\leq {t}_{\mathrm{acc}}+{t}_{\mathrm{const}} \\ {f1}_{\mathrm{brake}}\left( x \right) , & {t}_{\mathrm{acc}}+{t}_{\mathrm{const}}<x\leq {t}_{\mathrm{shift}}\end{array} \right. \]
From the knowledge of the mean angular velocity and the function $f1(x)$, the maximum angular velocity $\omega_{\mathrm{max}}$ can now be determined:
\[ {\omega }_{\mathrm{mean}}=\frac{1}{{t}_{\mathrm{shift}}}\cdot \int_{0}^{{t}_{\mathrm{shift}}} f1\left( x \right) \,\mathrm{d}x\quad \Rightarrow \quad {\omega }_{\mathrm{max}}=\qty{524}{\radian\per\second} \]
For simpler graphical representation in the CAS system, the function $f2(x)$ is created as a copy of the function $f1(x)$ and the result of $\omega_{\mathrm{max}}$ is inserted:
\[ f2\left( x \right) ={\left. f1\left( x \right) \right| }_{{\omega }_{\mathrm{max}}=\qty{524}{\radian\per\second}} \]
From the angular velocity in \unit{\radian\per\second}, the speed in \unit{\per\second} is now calculated and, for better illustration, converted to \unit{\per\minute}:
\[ {n}_{\mathrm{max}}=\frac{{\omega }_{\mathrm{max}}}{2\cdot \pi }=\qty{83.4}{\per\second}\quad\Rightarrow\quad
   {n}_{\mathrm{max}}\cdot 60=\qty{5000}{\per\minute} \]
Due to the assumed acceleration and braking behavior, a maximum motor speed of \qty{5000}{\per\minute} is required to meet the time target of \qty{100}{\milli\second}.

\textbf{Angular acceleration $\boldsymbol{0<x\leq t_{\mathrm{acc}}}$:} The motor torque during the acceleration process is calculated via the angular acceleration $\alpha$:
\[ {\alpha }_{\mathrm{motor}}=\frac{{\omega }_{\mathrm{max}}}{{t}_{\mathrm{acc}}}=\qty{13.1e3}{\radian\per\second\squared} \]

\textbf{Reduction of the inertia to the shift drum:} To reduce the inertia, the shift drum is regarded as a spindle with a defined thread pitch. The shift travel of the shift fork is $s_{\mathrm{fork}}=\qty{5}{\milli\metre}$; for this travel, the shift drum rotates by $\Delta\varphi_{\mathrm{drum}}=\ang{27}$. The shift fork with the largest mass has $m_{\mathrm{fork}}=\qty{0.063}{\kilogram}$, the gear wheel with the largest mass $m_{\text{gear wheel}}=\qty{0.236}{\kilogram}$ (all values from Table~\ref{tab:1}). The linearly moved mass is the sum of gear wheel and shift fork, and the ``thread pitch'' of the considered spindle follows from the data on the shift travel:
\[ {m}_{\mathrm{linear}}={m}_{\mathrm{fork}}+{m}_{\text{gear wheel}}=\qty{0.299}{\kilogram},\qquad
   {p}_{\mathrm{spindle}}=\frac{{s}_{\mathrm{fork}}}{\Delta {\varphi }_{\mathrm{drum}}}\cdot \ang{360}=\qty{66.7}{\milli\metre} \]
The force for moving the shift fork is not transmitted without losses. For the considered spindle, an efficiency of $\eta_{\mathrm{spindle}}=0.75$ is therefore assumed. The mass of the shift fork and the gear wheel is now reduced to the shift drum as rotational inertia:
\[ {J}_{\text{linear,red}}=\frac{{m}_{\mathrm{linear}}}{{\eta }_{\mathrm{spindle}}}\cdot \frac{{p}_{\mathrm{spindle}}^{2}}{4\cdot {\pi }^{2}}=\qty{44.9e-6}{\kilogram\metre\squared} \]
The largest inertia occurs when shifting gears 1, 3 and 4. Due to the gear ratio of the worm gear, however, it plays a minor role in the overall system.

\textbf{Calculation of the inertia acting on the motor incl. motor shaft:} The moment of inertia of the shift drum is $J_{\mathrm{drum}}=\qty{120e-6}{\kilogram\metre\squared}$ and the efficiency of the worm gear is $\eta_{\mathrm{worm}}=0.42$ (Table~\ref{tab:1}). According to the design documents~\cite[p.~1]{Ron19}, the moment of inertia of the gear wheel incl. shaft, spacer sleeve and screw is $J_{\text{worm wheel}}=\qty{139e-6}{\kilogram\metre\squared}$ and that of the worm is $J_{\mathrm{worm}}=\qty{15e-6}{\kilogram\metre\squared}$. The moment of inertia of the motor shaft (rotor) is $J_{\text{motor shaft}}=\qty{7.76e-6}{\kilogram\metre\squared}$ according to the data sheet~\cite[p.~1]{B7A19}. From these data, the inertia acting on the motor is calculated:
\begin{equation}
{J}_{\text{motor,red}}=\frac{{J}_{\text{linear,red}}+{J}_{\mathrm{drum}}+{J}_{\text{worm wheel}}}{{i}_{\mathrm{worm}}^{2}\cdot {\eta }_{\mathrm{worm}}}+{J}_{\mathrm{worm}}+{J}_{\text{motor shaft}}=\qty{23.6e-6}{\kilogram\metre\squared}
\label{eq:6}
\end{equation}
\textbf{Calculation of the reduced friction torque at the motor shaft:} For simplification, the friction torque is assumed to be independent of speed. The calculation is based on the largest measured friction torque of the shift drum when moving the shift forks, $M_{\text{fric,drum,red}}=\qty{0.13}{\newton\metre}$ (Table~\ref{tab:5}, guide~C, shift track~1/3). The friction torque at the motor shaft is calculated neglecting the bearing friction of the motor shaft:
\begin{equation}
{M}_{\text{fric,shaft,red}}=\frac{{M}_{\text{fric,drum,red}}}{{i}_{\mathrm{worm}}\cdot {\eta }_{\mathrm{worm}}}=\qty{0.0103}{\newton\metre}
\label{eq:7}
\end{equation}
From this, the required motor torque for acceleration and the required peak power at $\omega_{\mathrm{max}}$ are calculated:
\begin{gather*}
{M}_{\text{motor,run-up}}={J}_{\text{motor,red}}\cdot {\alpha }_{\mathrm{motor}}+{M}_{\text{fric,shaft,red}}=\qty{0.319}{\newton\metre} \\
{P}_{\text{motor,max}}={\omega }_{\mathrm{max}}\cdot {M}_{\text{motor,run-up}}=\qty{167}{\watt}
\end{gather*}
\phantomsection\label{bm:Ref14442146}\phantomsection\label{bm:Ref14442136}%
This is only a theoretical short-term value at the end of the acceleration phase. Under real conditions, the value may change.

\subsection{Consideration of the Service Life}\label{sec:22}

The actuator motor is operated under extreme short-time load. The rated duty type according to DIN EN 60034-1 is:

``S5 -- Intermittent periodic duty with electric braking''~\cite[p.~398, Sec.~7.5]{Han03}

For the design of the actuator motor, the number of gear changes is decisive; the total operating time matters little due to the short gear changes of \qty{100}{\milli\second}. The number of gear changes is determined from the rated service life of the gearbox, $s_{\mathrm{gearbox}}=\qty{60000}{\kilo\metre}$ (Table~\ref{tab:1}). As an average value in urban operation, one gear change every \qty{0.2}{\kilo\metre} can be assumed~\cite[p.~1]{Ron19}. With $t_{\mathrm{shift}}=\qty{100}{\milli\second}$ per gear change (Section~\ref{sec:8}), the number of gear changes within the rated service life and the actual operating time of the actuator motor result as follows:
\[ {N}_{\mathrm{shifts}}=\frac{{s}_{\mathrm{gearbox}}}{0.2\,\mathrm{km}}=300\,000\quad\Rightarrow\quad
   {t}_{\mathrm{motor}}={N}_{\mathrm{shifts}}\cdot {t}_{\mathrm{shift}}=\qty{30000}{\second} \]
The required service life is \qty{30000}{\second} or \qty{8.3}{\hour}.

\chapter{Shift Controller Hardware}\label{sec:23}

The shift controller controls the automatic gear changes on the gearbox test rig. In total, two prototype circuit boards were manufactured. In the course of development, several possibilities for improvement emerged, which were implemented in version 2 shown here. Above all, a simple and cost-effective realization of the electronics was the priority. In a further development, the controller is to be adapted as an auxiliary control unit for use in the vehicle. The shift controller consists of a custom-made control board in combination with a microcontroller for the intelligent handling of the control tasks. A cost-effective development of the shift controller takes place on the basis of the Arduino Zero. A critical look at the Arduino hardware shows some possibilities for improvement with regard to circuit and layout design. Nevertheless, its reliable function is confirmed by the worldwide distribution of the hardware. Above all, the large community is important. In forums, possible solutions to specific tasks are documented in detail.

\begin{wrapfigure}{r}{0.52\textwidth}
\vspace*{-\intextsep}\vspace*{-0.9pt}%
\centering
\includegraphics[angle=90,width=0.5\textwidth]{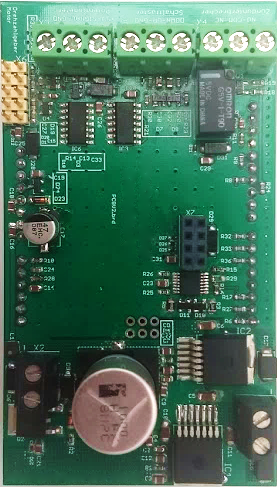}
\caption{Circuit board of the shift controller}
\label{fig:9}
\end{wrapfigure}
The Arduino IDE offers great features for beginners and hobbyists. Should a project go beyond the experimental scope, however, its limits quickly become apparent. The lack of an ANSI C standard implementation and the poor overview of source and header files are a knockout criterion for larger projects. Therefore, neither the Arduino IDE nor the Arduino libraries are used in this project. As an alternative, Arduino has implemented the EDBG interface on its development boards~\cite[pp.~3\,ff.]{EDB19}. EDBG uses a single chip as the building block for a fully fledged debugging interface. Atmel Studio supports EDBG via USB as a direct programming and debugging interface without additional hardware. The Atmel Studio programming environment meets the requirements of complex project development. Thus, the full functionality of the microcontroller integrated in the Arduino Zero (SAMD21G18A) can be exploited.

For use on the vehicle, the shift controller must be supplemented, e.g., with safety routines. If required, the program functions are transferred to a universal development environment such as MATLAB/Simulink or ETAS. This ensures that the program can be used independently of the processor, provided the hardware is adapted accordingly.

Table~\ref{tab:6} lists all inputs and outputs of the shift controller. Signals that have a connection to the I/O pins of the microcontroller are specified in more detail in a further table (Table~\ref{tab:7}). The incremental encoder, debug and API interfaces serve programming and measurement purposes and are not required for everyday operation of the shift controller.

\begin{table}[tbp]
\centering
\caption{I/O overview of the shift controller}\label{tab:6}
{\small
\begin{tabular}{>{\raggedright\arraybackslash}p{\dimexpr 0.29\textwidth-2\tabcolsep\relax}>{\raggedright\arraybackslash}p{\dimexpr 0.19\textwidth-2\tabcolsep\relax}>{\raggedright\arraybackslash}p{\dimexpr 0.244\textwidth-2\tabcolsep\relax}>{\raggedright\arraybackslash}p{\dimexpr 0.276\textwidth-2\tabcolsep\relax}}
\toprule
\textbf{Interface} & \textbf{Description} & \textbf{Voltage} & \textbf{Pins} \\
\midrule
Supply & Supply input & 10 to \qty{15}{\volt} & GND, +\qty{12}{\volt} \\
Potentiometer & Analog input & 0 to +\qty{5}{\volt} & GND, Signal, +\qty{5}{\volt} \\
Push button Upshift & Digital input & +\qty{3.3}{\volt} & GND, Signal \\
Push button Downshift & Digital input & +\qty{3.3}{\volt} & GND, Signal \\
Incremental encoder & Digital input & +\qty{5}{\volt} & GND, +\qty{5}{\volt}, A, B \\
Ignition cut & Relay output & 0 to +\qty{12}{\volt} & COM, NO, NC \\
Debug & USART & +\qty{3.3}{\volt} & GND, TX, RX, +\qty{3.3}{\volt} \\
API & USB & +4.75 to +\qty{5.25}{\volt} & GND, D+, D-, \qty{5}{\volt} \\
Actuator motor & PWM output & \qty{-12}{\volt} to +\qty{12}{\volt} & Motor+, Motor- \\
\bottomrule
\end{tabular}}
\end{table}

\section{Electromagnetic Compatibility}\label{sec:24}

The requirements for device design with regard to EMC are constantly growing. If preventive measures are not taken into account in the development process of new electronics, this inevitably leads to a high cost factor toward the end of the development phase. The development of the shift controller in the course of the diploma thesis is understood as a ``proof of concept''. The electronics are therefore not subject to any EMC-relevant standards and requirements; nevertheless, it is sensible to take certain criteria into account. The electronics must function in all specified fields of application and, above all, must not be destroyed by electrostatic discharges during the development work. As a first measure, the power electronics are separated from the control electronics (circuit board in Figure~\ref{fig:9}). Due to the high pulse currents of up to approx. \qty{100}{\ampere}, the measurement signal of the potentiometer at the ADC (\qtyrange{0}{1}{\volt}) must not be subject to any impermissible influence. The smaller the signal voltage and the larger the currents, the higher the requirements for the hardware design. To avoid influence from voltage drops on the conductor tracks of the circuit board, the ground of the power electronics on the one hand and the ground of the control electronics on the other hand are connected starting from a central ground point at terminal block X2. Pulsed currents of up to \qty{100}{\ampere} cause a dip in the battery voltage. In the shift controller, both the power section and the control electronics are supplied by the same \qty{12}{\volt} battery. Filter circuits are necessary here to avoid interference. Both analog and digital signals occur within the control electronics. Here, particular attention must be paid to ensuring that digital switching operations do not influence analog measurement voltages. As a measure, each integrated circuit has a capacitor to stabilize the supply voltage; furthermore, the supply of the control electronics is implemented over a large area in a separate layer. An important topic is the coupling of signal pulses onto adjacent signal lines. To limit this crosstalk, a larger safety clearance is kept to critical, high-impedance signals; furthermore, signals with a large interference factor are additionally separated by a ground plane. If interference still occurs despite these measures, it is still quite easy to make appropriate changes at this early stage of the development process. These requirements led to a 4-layer design of the circuit board.

In addition to the circuit-internal sources of interference, external influences must also be taken into account. Control units in the vehicle network must function satisfactorily in the electromagnetic environment without impermissibly influencing other devices. Electrostatic discharges in particular pose a danger. A person can be charged up to several thousand volts and discharge this energy into the electronics as ESD pulses. This impairs the function of the device, causes defects or leads to total failure.

\section{Supply Voltage Input Filter}\label{sec:25}

\begin{wrapfigure}{r}{0.5\textwidth}
\vspace*{-\intextsep}\vspace*{-0.9pt}%
\centering
\includegraphics[width=0.48\textwidth]{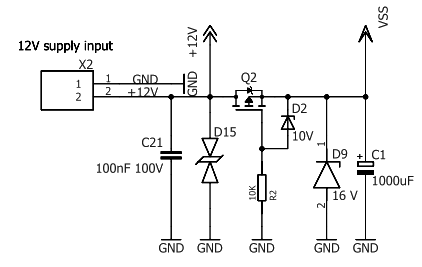}
\caption{Supply voltage}
\label{fig:10}
\end{wrapfigure}
The shift controller is supplied by a \qty{12}{\volt} lead-acid battery. Both the power electronics (supply VSS) of the electric motor and the control electronics (supply +\qty{12}{\volt}) are connected to this voltage source. Therefore, suitable filters for interference suppression are necessary in addition to the reverse polarity protection. Figure~\ref{fig:10} shows the input circuit. The capacitor C21 and the bidirectional TVS diode D15 suppress ESD pulses and prevent the destruction of the downstream electronics. The P-channel MOSFET in inverse operation provides the reverse polarity protection. When voltage is applied, positive current flows through the body diode. The voltage at the source terminal increases. The gate is connected to ground via R2. If $U_{\mathrm{Source}}$ increases, $U_{\text{Gate-Source}}$ increases equally. The MOSFET becomes conductive and short-circuits the body diode. The current now flows through the conductive channel of the MOSFET with correspondingly low power dissipation. D2 limits $U_{\text{Gate-Source}}$ at input voltages above \qty{10}{\volt}. C1 serves as a buffer capacitor for the high pulse currents of the actuator motor. When the rotation of the DC motor is braked, the motor operates in generator mode. To discharge impermissibly high voltages, the Zener diode D9 limits the voltage to \qty{16}{\volt}. The circuit of the input filter is designed for a rated current of \qty{13.5}{\ampere} and a short-term load (\qty{100}{\milli\second}) of \qty{100}{\ampere}. The limiting element here is the terminal connector X2.

\textbf{Dimensioning of Q2:} At the moment of switch-on, the actuator motor draws a maximum starting current of approx. \qty{70}{\ampere}; in braking operation, up to \qty{90}{\ampere} were measured (oscillogram in Appendix~\ref{app:2}). According to the data sheet, the inrush current is \qty{111}{\ampere}~\cite[p.~7]{B7A19}. Due to real resistances in the voltage supply (ohmic, inductive), the actual starting current does not correspond to the value given in the data sheet. The MOSFET must be dimensioned so that it can withstand pulse currents of up to \qty{90}{\ampere} without problems. The IPD90P04P4L-04 was selected as MOSFET Q2 (continuous drain current according to the data sheet~\cite[p.~1]{IPD11}).

\section{Incremental Encoder Signal Processing}\label{sec:26}

\begin{figure}[tbp]
\centering
\includegraphics[width=\textwidth,height=0.6\textheight,keepaspectratio]{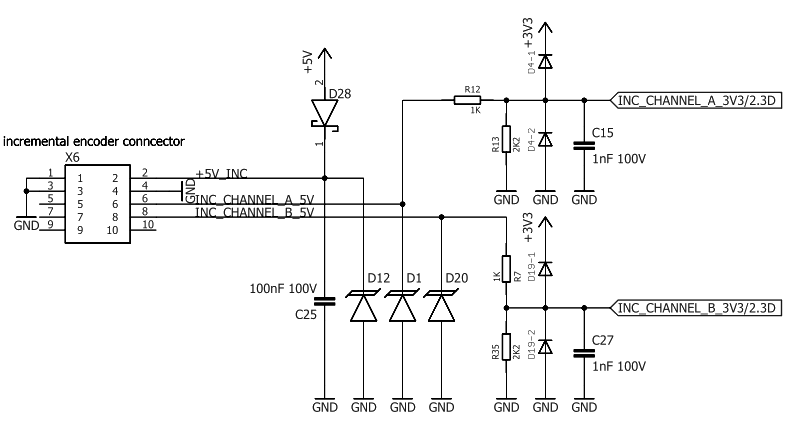}
\caption{Incremental encoder input filter}
\label{fig:11}
\end{figure}
The incremental encoder ENX16 EASY 256IMP supplies a Gray code with three channels as output: channels A, B and I (index)~\cite[p.~9]{B7A19}. 256 pulses are emitted per revolution. The encoder operates with \qty{5}{\volt}, the microcontroller used operates with \qty{3.3}{\volt} signals, thus a level adaptation is required.

The diode D28 protects the internal \qty{5}{\volt} supply against positive external voltages of up to \qty{20}{\volt}. Three TVS diodes D1, D12 and D20 as well as the capacitor C25 discharge ESD pulses. The voltage dividers R12-R13 and R7-R35 divide the \qty{5}{\volt} signals of the incremental encoder down to the voltage level of the microcontroller (\qty{3.3}{\volt}). D4-1 and D4-2 as well as D19-1 and D19-2 protect the downstream electronics against impermissible voltages in the positive and negative direction. The input circuit of channel A is constructed analogously to channel B (component designations as in Figure~\ref{fig:11}).

\begin{figure}[tbp]
\centering
\includegraphics[width=0.8\textwidth,height=0.6\textheight,keepaspectratio]{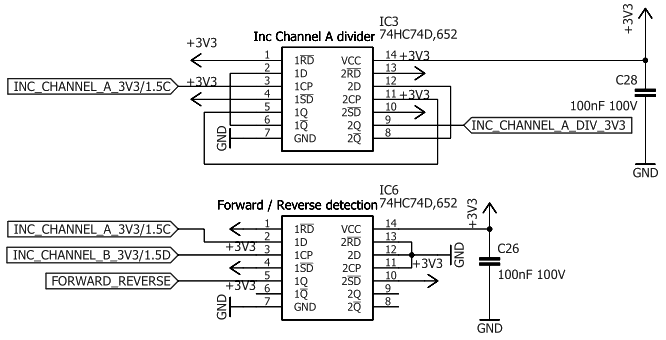}
\caption{Incremental encoder signal processing}
\label{fig:12}
\end{figure}

The two Gray code signals are passed on to two D flip-flops (Figure~\ref{fig:12}).

IC3 and IC6 are each dual D flip-flops. The flip-flop IC3 operates as a two-stage frequency divider with a division factor of 4. Instead of the original 256 pulses per revolution, 64 pulses per revolution are thus passed on to the microcontroller. The angular velocity of the motor can thereby be determined every \ang{5.63}. 64~values per revolution are sufficient for determining the dynamic behavior. The reduction of the resolution results in a reduction of the CPU load and computing time (component designations as in Figure~\ref{fig:12}).

\textbf{Calculation of the CPU load:} The maximum speed of the electric motor at rated voltage (no-load speed) is $n_{\text{no-load}}=\qty{7120}{\per\minute}=\qty{119}{\per\second}$ according to the data sheet~\cite[p.~7]{B7A19}.\phantomsection\label{bm:Ref10110661} The encoder delivers $N_{\text{pulse,orig}}=256$ pulses per revolution~\cite[p.~4]{B7A19}, and $N_{\text{pulse,div}}=64$ after the frequency divider (division factor 4). This gives the following pulse frequency and period before and after the frequency divider:
\begin{gather*}
{f}_{\mathrm{orig}}={n}_{\text{no-load}}\cdot {N}_{\text{pulse,orig}}=\qty{30.4}{\kilo\hertz}\quad\Rightarrow\quad {T}_{\mathrm{orig}}=\frac{1}{{f}_{\mathrm{orig}}}=\qty{32.9}{\micro\second} \\
{f}_{\mathrm{div}}={n}_{\text{no-load}}\cdot {N}_{\text{pulse,div}}=\qty{7.59}{\kilo\hertz}\quad\Rightarrow\quad {T}_{\mathrm{div}}=\frac{1}{{f}_{\mathrm{div}}}=\qty{132}{\micro\second}
\end{gather*}
To calculate the processor load caused by the interrupt of the speed signal, port A 09 is set at the beginning of the interrupt routine and reset at the end.\footnote{\label{fn:54}Port A 09 corresponds to Arduino pin D3.} The load can be determined from the duty cycle. From the voltage curve in the oscillogram, it can be read that the pulse width is \qty{4.4}{\micro\second} and the actual cycle time is \qty{133.8}{\micro\second} (Appendix~\ref{app:3}). From this, the load caused by the computing time in the interrupt is determined:
\[ {\rho }_{\mathrm{CPU}}=\frac{\qty{4.4}{\micro\second}}{\qty{133.8}{\micro\second}}\cdot \qty{100}{\percent}=\qty{3.29}{\percent} \]

\section{Potentiometer Signal Processing}\label{sec:27}

\begin{figure}[tbp]
\centering
\includegraphics[width=0.8\textwidth]{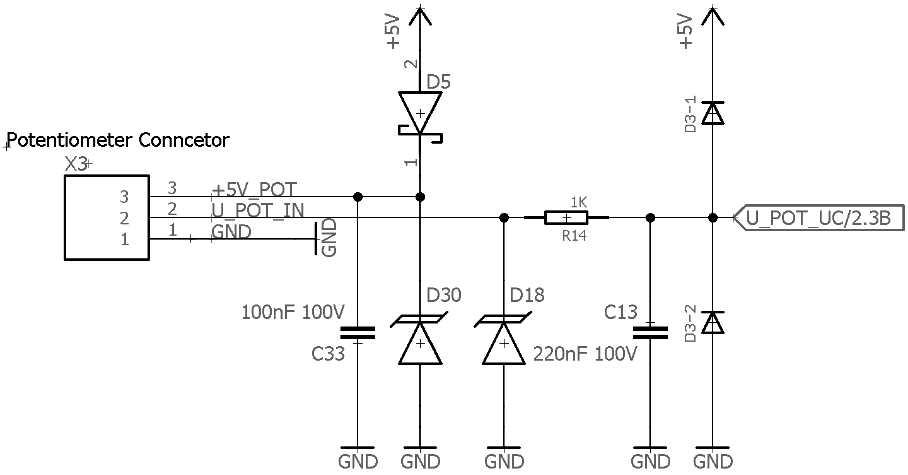}
\caption[Signal input of potentiometer 1]{Signal input of potentiometer 1 with protection diodes D18/D30 and RC filter R14/C13}
\label{fig:13}
\end{figure}
The Hall-effect potentiometer serves as position sensor and operates with a voltage of \qty{5}{\volt}. Depending on the gear position, the potentiometer supplies an output voltage between \qty{0.5}{\volt} and \qty{4.4}{\volt} (Table~\ref{tab:3}). The supply may vary between \qty{4.5}{\volt} and \qty{5.5}{\volt} without affecting the correct function. Thus, no precise reference voltage is required; the supply voltage (\qty{5}{\volt}) from the Arduino board is sufficient. Figure~\ref{fig:13} shows the signal input. The two TVS diodes D18 and D30 are important to mention, so that the downstream electronics are not damaged by ESD pulses. R14 and C13 filter high-frequency interference pulses by their low-pass effect. The diodes D3-1 and D3-2 discharge overvoltages. Subsequently, the signal is passed through a two-stage voltage follower (IC5): The first stage (IC5, input IN3+) decouples the signal and divides it by means of a voltage divider down to the correct voltage range of \qtyrange{0}{1}{\volt} for measurement by the ADC. The second stage (IC5, input IN4+) operates as an impedance converter and provides the signal to the ADC at low impedance. Since the op-amp has a very low low-level output voltage, signals close to \qty{0}{\volt} are also transmitted without distortion~\cite[p.~8]{Hig13}.\footnote{Op-amp type TSV734IPT. The component designations refer to the complete circuit diagram of the shift controller, which is not included in this publication.}

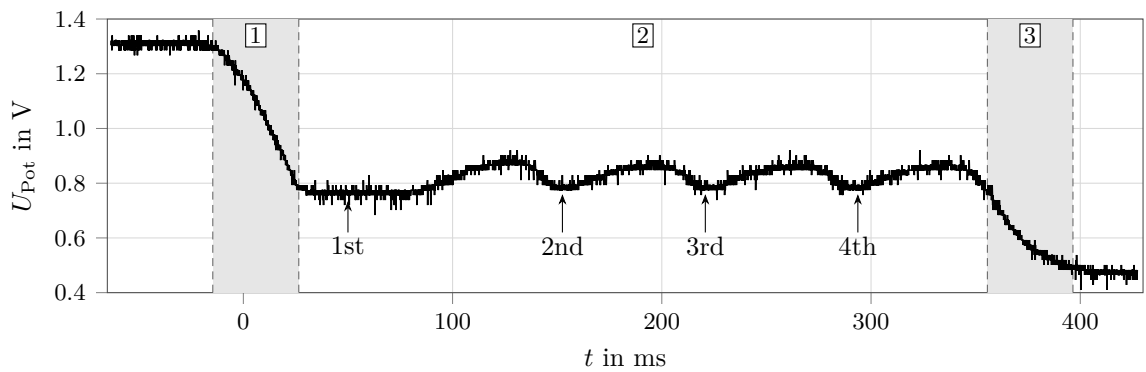
\begin{figure}[tbp]
\centering
\begin{tikzpicture}
\begin{axis}[
    thesisplot,
    width=0.97\textwidth, height=5.2cm,
    xmin=-65, xmax=430, xtick={0,100,...,400},
    ymin=0.4, ymax=1.4, ytick={0.4,0.6,...,1.4},
    yticklabel style={/pgf/number format/.cd, fixed, fixed zerofill, precision=1},
    xlabel={$t$ in \unit{\milli\second}}, ylabel={$U_{\mathrm{Pot}}$ in \unit{\volt}},
    clip=false,
  ]
  \fill[plothighlight] (axis cs:-14.6,0.4) rectangle (axis cs:26.5,1.4);
  \fill[plothighlight] (axis cs:355.6,0.4) rectangle (axis cs:396.5,1.4);
  \pgfplotsinvokeforeach{-14.6,26.5,355.6,396.5}{
    \draw[densely dashed, black!60] (axis cs:#1,0.4) -- (axis cs:#1,1.4);}
  \node[draw, inner sep=1.5pt, fill=white, font=\footnotesize] at (axis cs:5.95,1.34) {1};
  \node[draw, inner sep=1.5pt, fill=white, font=\footnotesize] at (axis cs:191,1.34) {2};
  \node[draw, inner sep=1.5pt, fill=white, font=\footnotesize] at (axis cs:376,1.34) {3};
  \draw[->, >={Stealth[length=1.4mm]}, thin] (axis cs:50,0.62) -- (axis cs:50,0.735);
  \node[below, font=\small, inner sep=1.5pt] at (axis cs:50,0.62) {1st};
  \draw[->, >={Stealth[length=1.4mm]}, thin] (axis cs:152.5,0.62) -- (axis cs:152.5,0.76);
  \node[below, font=\small, inner sep=1.5pt] at (axis cs:152.5,0.62) {2nd};
  \draw[->, >={Stealth[length=1.4mm]}, thin] (axis cs:220.8,0.62) -- (axis cs:220.8,0.76);
  \node[below, font=\small, inner sep=1.5pt] at (axis cs:220.8,0.62) {3rd};
  \draw[->, >={Stealth[length=1.4mm]}, thin] (axis cs:293.7,0.62) -- (axis cs:293.7,0.76);
  \node[below, font=\small, inner sep=1.5pt] at (axis cs:293.7,0.62) {4th};
  \addplot[plotline] table[x=t_ms, y=U_V] {figures/data/gear_change_5to6_potentiometer.dat};
\end{axis}
\end{tikzpicture}\par
\caption[Oscillogram of a gear change on the gearbox test rig]{Oscillogram of a gear change from 5th to 6th gear on the gearbox test rig. 1: shift movement out of 5th gear, 2: shift dogs tooth on tooth, 1st to 4th engagement attempt, 3: dogs slide into each other at the 5th attempt, 6th gear engaged\nocc}
\label{fig:14}
\end{figure}
\textbf{Calculation of the filter R14/C13:} The oscillogram (Figure~\ref{fig:14}) shows the voltage value at the potentiometer for the gear change from 5th gear to 6th~gear.

In this gear change, 6th gear does not engage at the first attempt. Range 1 is the shift movement out of 5th gear. Then the shift dogs of the sliding gear wheel meet the shift dogs of 6th gear in a tooth-on-tooth position and cannot slide into each other, so the shift drum is blocked (range 2). The software detects the blockage, releases the actuator motor after a short time and then pushes again. In the voltage curve, each attempt is a dip (marked in Figure~\ref{fig:14}), each release a hump. At the 5th attempt the dogs meet a gap, slide into each other, and the shift movement is completed (range 3). In the majority of gear changes, range 2 does not occur and the shift movement proceeds continuously.

The filter must be dimensioned in such a way that high-frequency interference signals are filtered, but the position signal of the potentiometer is passed on to the microcontroller almost undistorted. If the influence is impermissibly high, the end positions and angles of rotation no longer match the specification. Dead times or signal propagation delays in a control loop cause instabilities. The sum of the deviation in amplitude and phase must at no time exceed \qty{25}{\milli\volt} of the actual signal.\footnote{Corresponds to approx. \ang{2} at the shift drum.} The position signal during the gear changes represents an aperiodic signal. Gear changes can follow one another quickly, but can also occur only after several minutes. The well-known transfer function of the low-pass filter always requires knowledge of a frequency or of a periodic behavior:
\[ \frac{{U}_{\mathrm{out}}}{{U}_{\mathrm{in}}}=\frac{1}{\sqrt{1+{\left( 2\cdot \pi \cdot f\cdot R\cdot C \right) }^{2}}} \]
\phantomsection\label{bm:Ref14804289}\phantomsection\label{bm:Ref14804283}\phantomsection\label{bm:Ref14804299}

Three possible approaches are explained in the following:

\textbf{Fourier transform:} It is possible to carry out a Fourier transform by means of a limit consideration for $T\rightarrow\infty$~\cite[pp.~7\,f., Ch.~7]{ASV09}. The result is a continuous frequency spectrum of the signal. The cut-off frequency of the filter is chosen so that only those frequency components are cut off whose absence causes a total deviation of the signal of less than \qty{25}{\milli\volt}. This can vary depending on the application and the signal. Compared to other methods, this method requires a high computational effort.

\begin{wrapfigure}{r}{0.5\textwidth}
\vspace*{-\intextsep}\vspace*{9.1pt}%
\centering
\includegraphics[width=0.48\textwidth]{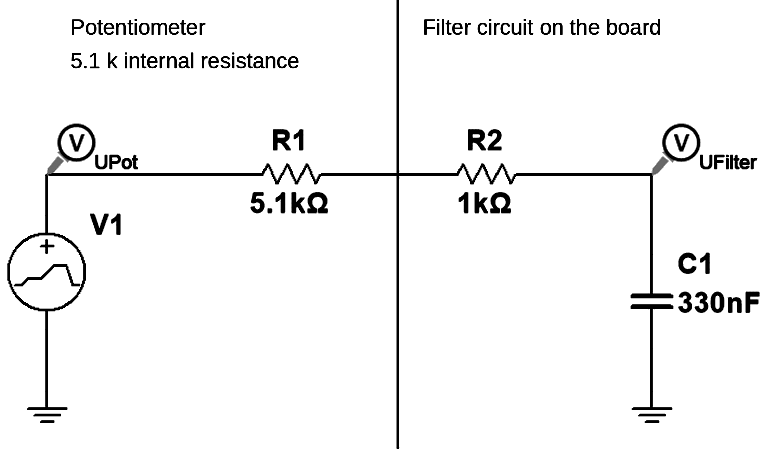}
\caption{Circuit in NI Multisim}
\label{fig:15}
\end{wrapfigure}

\textbf{Estimation:} The signal change with the largest voltage change per unit of time is to be considered. In the present case, this results in a difference of $\Delta U=\qty{506}{\milli\volt}$ in a time of $\Delta t=\qty{40.81}{\milli\second}$ (Figure~\ref{fig:14}, range~3).
A linearly rising voltage at an RC low-pass filter results in a constant offset between the input and output voltage. The slope of the input signal equals the slope of the output signal. The constant offset arises from the voltage drop across the resistor. A larger capacitance requires a higher current through the resistor so that the voltage at the output can follow with the same slope. Since the conditions are approximately linear, the following formula can be applied:
\begin{equation}
I\cdot t=C\cdot U
\label{eq:8}
\end{equation}
The sum of the internal resistance of the potentiometer (\qty{5.1}{\kilo\ohm}) and the filter resistance (\qty{1}{\kilo\ohm}) is \qty{6.1}{\kilo\ohm} (Figure~\ref{fig:15}).

According to Ohm's law, a voltage drop of \qty{25}{\milli\volt} results in a current of $I=\qty{4.1}{\micro\ampere}$. The capacitance value can now be calculated with Eq.~\eqref{eq:8}:
\[ C=I\cdot \frac{\Delta t}{\Delta U}=\qty{330.7}{\nano\farad}\quad\Rightarrow\quad C\leq\qty{330}{\nano\farad} \]

\textbf{Simulation:} A third, illustrative and practice-oriented option is offered by simulation. The electrical behavior of the potentiometer and the filter is modeled in the software Multisim from National Instruments. In the simulation, the ``PIECEWISE\_LINEAR\_VOLTAGE'' is used as the voltage source. This voltage source can represent a voltage curve specified by measurement points. In the circuit shown, V1 models the behavior of the real potentiometer. The internal resistance of the Hall-effect potentiometer is \qty{5.1}{\kilo\ohm}. The resistor R2 of the passive RC filter on the circuit board has \qty{1}{\kilo\ohm}. The capacitance of the capacitor is determined. The result of the estimation is the starting point for the simulation (component designations as in Figure~\ref{fig:15}).

The capacitance is determined with the parameter sweep analysis. C1 is used as the variable with \qty{200}{\nano\farad}, \qty{250}{\nano\farad}, \qty{300}{\nano\farad} and \qty{350}{\nano\farad}. The output is plotted over time as the difference $V(U_{\mathrm{Filter}})-V(U_{\mathrm{Pot}})$. Capacitance values that cause a deviation of less than \qty{25}{\milli\volt} are permissible. If the cut-off frequency is chosen too high, the signal remains almost unaffected. However, the filter then fails to have an effect on higher-frequency interference signals. To improve the filter effect, the cut-off frequency of the filter must be chosen as low as possible. The capacitor is dimensioned so that the deviation of the output signal is just still tolerable.

Figure~\ref{fig:16} shows the voltage drop across the filter. The curve for \qty{200}{\nano\farad} (lightest grey) has sufficient margin to the maximum deviation of \qty{25}{\milli\volt}. The curve for \qty{250}{\nano\farad} shows that the actual capacitance value must be lower than the estimation yielded. Capacitance values below \qty{250}{\nano\farad} are permissible. \qty{220}{\nano\farad} were chosen.

\par\medskip\noindent\begin{minipage}{\textwidth}
\centering
\begin{tikzpicture}
\begin{axis}[
    thesisplot,
    width=0.97\textwidth, height=4.4cm,%
    xmin=0, xmax=400, xtick={0,50,...,400},
    ymin=-12, ymax=50, ytick={-10,0,...,40},%
    xlabel={$t$ in \unit{\milli\second}}, ylabel={$\Delta U$ in \unit{\milli\volt}},
    legend style={draw=none, fill=white, fill opacity=0.85, text opacity=1, font=\footnotesize,
                  at={(0.99,0.97)}, anchor=north east, legend columns=4},
    legend cell align=left,
    every axis plot/.append style={line width=0.5pt},
  ]
  \addplot[black!30] table[x=t_ms, y=C200] {figures/data/filter_simulation_deviation.dat};
  \addplot[black!50] table[x=t_ms, y=C250] {figures/data/filter_simulation_deviation.dat};
  \addplot[black!75] table[x=t_ms, y=C300] {figures/data/filter_simulation_deviation.dat};
  \addplot[black]    table[x=t_ms, y=C350] {figures/data/filter_simulation_deviation.dat};
  \legend{\qty{200}{\nano\farad}, \qty{250}{\nano\farad}, \qty{300}{\nano\farad}, \qty{350}{\nano\farad}}
  \draw[densely dashed, black!60] (axis cs:0,25) -- (axis cs:400,25)
    node[pos=0.02, above right, font=\scriptsize, inner sep=1pt] {limit \qty{25}{\milli\volt}};
\end{axis}
\end{tikzpicture}\par

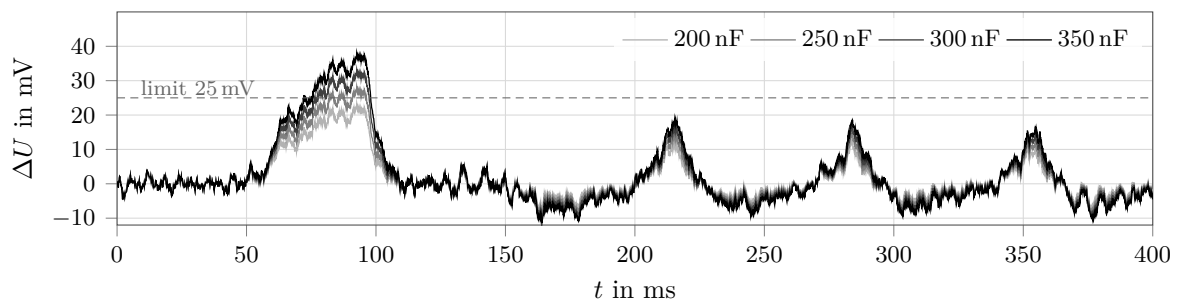
\captionof{figure}[Simulated deviation of the filter output for different capacitances]{Simulated deviation $\Delta U=U_{\mathrm{Filter}}-U_{\mathrm{Pot}}$ of the filter output for different capacitances\nocc}
\label{fig:16}
\end{minipage}

\section{Serial Interface}\label{sec:28}

The hardware-implemented USART of the microcontroller is used as the debug interface. A level converter connects the data lines TX and RX to the PC. The microcontroller operates with a +\qty{3.3}{\volt} logic level, the serial interface of the PC operates with \qty{\pm 12}{\volt}. The MAX3232 converts the levels in both signal directions for TX and RX. Breakout boards with the MAX3232 chip and the associated circuitry are available as complete elements. A separate circuit design is not necessary. The signals RX, TX, +\qty{3.3}{\volt} and GND are routed out on the circuit board so that the breakout board can be connected easily. For the shift controller, the RS232/TTL converter from the company Pollin is used.\footnote{Product EAN code: 4049702039495.}

\begin{figure}[hbp]%
\centering
\includegraphics[width=0.8\textwidth,height=0.6\textheight,keepaspectratio]{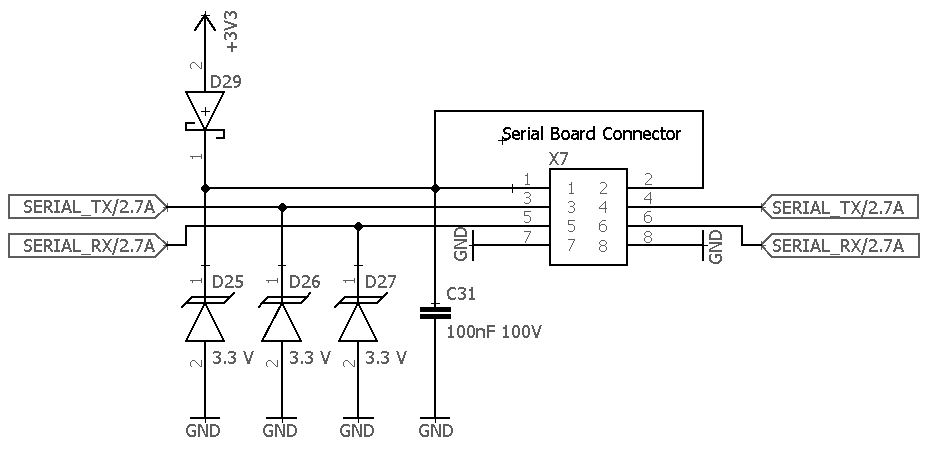}
\caption{Circuit diagram of the USART}
\label{fig:17}
\end{figure}
Opposite signals on the Serial Board Connector X7 are connected to each other: 1-2, 3-4, 5-6, and 7-8. The signals SERIAL\_TX and SERIAL\_RX lead directly to the corresponding pins on the microcontroller. D29 protects the internal supply voltage against external voltage. The capacitor C31 stabilizes the supply voltage during digital switching pulses and, together with the TVS diodes D25, D26 and D27, discharges occurring ESD pulses (component designations as in Figure~\ref{fig:17}).

\section{Ignition Cut}\label{sec:29}

For optimal gear changes, the gearbox must be free of load. In manual gearboxes, this is ensured by the clutch. In automated gear changes, this is done by cutting the ignition and/or the fuel supply. Cutting the ignition alone causes unburned fuel to enter the exhaust. This increases the pollutant emissions, which is why this method may no longer be used in vehicles with the current emission standard. In the gearbox test rig used, the load-free state is achieved by switching off the gearbox drive motor (relay in Figure~\ref{fig:2}).

\begin{figure}[tbp]
\centering
\includegraphics[width=\textwidth,height=0.6\textheight,keepaspectratio]{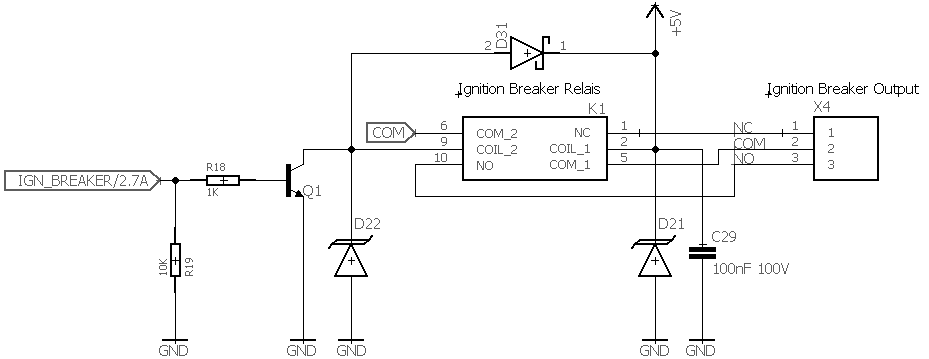}
\caption{Ignition cut circuit}
\label{fig:18}
\end{figure}

The relay K1 cannot be driven directly by the microcontroller because of the limited output current of the microcontroller and the different operating voltage. The maximum output current of the microcontroller pin is \qty{7}{\milli\ampere}~\cite[p.~1109]{ATD18}. The relay K1 has a coil current of \qty{30}{\milli\ampere}~\cite[p.~1]{G5V19} and is therefore driven via the transistor Q1. R19 is required to establish a defined low level at the moment of switching on. The TVS diodes D22 and D21 serve to discharge occurring ESD pulses. D31 discharges the self-induction voltage of the coil at the moment of switching off. C29 stabilizes the supply voltage (component designations as in Figure~\ref{fig:18}).

\textbf{Calculation of R18:} The coil current of the relay is $I_{\mathrm{relay}}=\qty{30}{\milli\ampere}$~\cite[p.~1]{G5V19}, and the DC current gain of the transistor is $B_{\mathrm{Q1}}=100$ according to the data sheet~\cite[p.~3]{Gen07}. With the supply voltage of the microcontroller $U_{\upmu\mathrm{C}}=\qty{3.3}{\volt}$, the base-emitter voltage drop $U_{\mathrm{BE}}=\qty{0.7}{\volt}$ and an overdrive factor of $B_{\mathrm{od}}=10$, the series resistor at the base is:
\[ \mathrm{R18}=\frac{{U}_{\upmu \mathrm{C}}-{U}_{\mathrm{BE}}}{\frac{{I}_{\mathrm{relay}}}{{B}_{\mathrm{Q1}}}\cdot {B}_{\mathrm{od}}}=\qty{867}{\ohm} \]
A resistor R18~=~\qty{1}{\kilo\ohm} is selected.

\section{Motor Driver}\label{sec:30}

\begin{figure}[tbp]
\centering
\includegraphics[width=\textwidth,height=0.6\textheight,keepaspectratio]{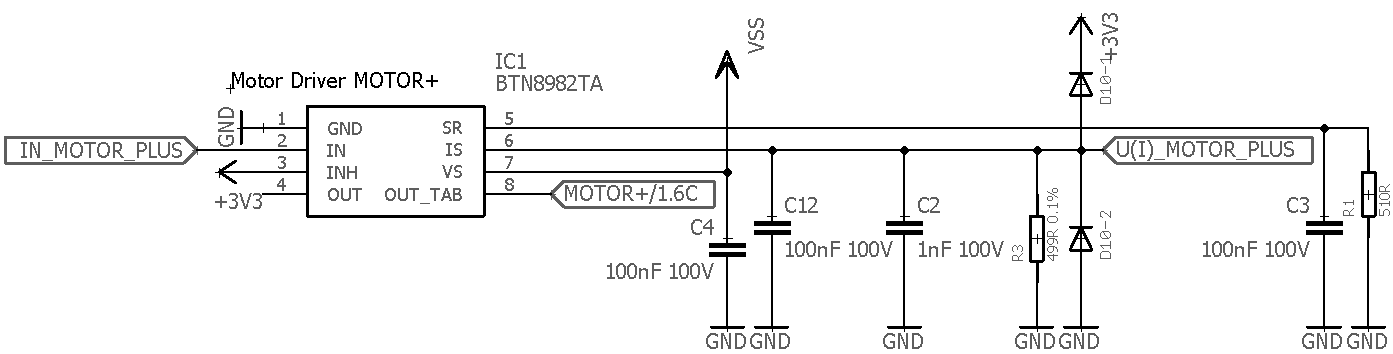}
\caption{Motor driver MOTOR+}
\label{fig:19}
\end{figure}

The chip BTN8982TA was selected primarily because of its simple control option and the high possible current. For better clarity, only the driver for the MOTOR+ terminal of the actuator motor is shown in Figure~\ref{fig:19}. The control of the motor terminal MOTOR+ behaves analogously to the control of MOTOR-. The functional description is therefore given only for the motor driver at the MOTOR+ terminal.\footnote{\label{fn:75}The complete circuit diagram of the shift controller is not included in this publication.}

The actuator motor for driving the shift drum can be controlled in four-quadrant operation: driving and braking (generator operation) are possible in both directions. The power electronics contain a PWM-controllable half bridge for each motor terminal (IC1, IC2).\footnote{Motor driver type BTN8982TA.} The chip used also contains an output for analog current measurement. So that the analog voltage signals are not distorted by leakage currents of the ADC, an impedance converter is connected upstream of the inputs of the microcontroller (IC5).\footref{fn:75}

Depending on the logic level, the signal IN\_MOTOR\_PLUS closes either the high-side or the low-side MOSFET of IC1. Opening both switches is not provided for by the manufacturer. At a low level, the pin INH activates the power-saving function of IC1. In the electronics of the shift controller, the IC is permanently activated, INH at high level. The slew rate for reducing EMC emissions is set with the pin SR. A sense current proportional to the load current flows via the pin IS. If no load current flows, a defined offset current is output at IS. The value $\mathrm{dk}_{\mathrm{ILIS}}$ gives the ratio of motor current to sense current. In the data sheet, large spreads are specified for $\mathrm{dk}_{\mathrm{ILIS}}$ and $I_{\mathrm{IS(Offset)}}$~\cite[p.~20, items~5.4.6 and 5.4.9]{Inf13}. The values are therefore optimized by measuring the actual quantities.

\textbf{Calculation of R5/R3:} The aim of the current measurement is to realize an overload protection for the electronics as well as for the electric motor. The current and time behavior $I^{2}\cdot t$ is used as the evaluation criterion for the load situation. The maximum starting current of the electric motor is \qty{111}{\ampere}~\cite[p.~7]{B7A19}. The actuator motor has a continuous load current (rated current) of \qty{6}{\ampere}~\cite[p.~7]{B7A19}. Designing the measuring range for the highest possible motor current carries the risk that small but permanent exceedances of the rated data are not detected due to measurement inaccuracies. To increase the resolution and accuracy at low currents, the measuring range is dimensioned to \qtyrange{0}{35}{\ampere}. For currents above \qty{35}{\ampere}, the maximum possible motor current is assumed. \qty{100}{\ampere} are possible for \qty{100}{\milli\second}. The reference voltage of the ADC is $U_{\text{ref,ADC}}=\qty{1}{\volt}$. For the resistor calculation, the typical values $\mathrm{dk}_{\text{ILIS,typ}}=\num{19.5e3}$ and $I_{\text{IS(Offset),typ}}=\qty{170}{\micro\ampere}$ are assumed according to the data sheet~\cite[p.~20, items~5.4.6 and 5.4.9]{Inf13}. The sense current $I_{\mathrm{IS}}$ at the upper end of the measuring range, $I_{\text{L,range}}=\qty{35}{\ampere}$, is calculated according to the data sheet~\cite[p.~17, Eq.~(1)]{Inf13}:
\[ {I}_{\mathrm{IS}}=\frac{1}{{\mathrm{dk}}_{\text{ILIS,typ}}}\cdot {I}_{\text{L,range}}+{I}_{\text{IS(Offset),typ}}=\qty{1.965}{\milli\ampere} \]
The resistance value follows from the reference voltage of the ADC; the value \qty{499}{\ohm} is selected for both R3 and R5:
\[ \frac{{U}_{\text{ref,ADC}}}{{I}_{\mathrm{IS}}}=\qty{508.9}{\ohm}\quad\Rightarrow\quad
   \mathrm{R3}=\mathrm{R5}=\qty{499}{\ohm} \]

\textbf{Determination of $\boldsymbol{I_{\mathrm{IS(Offset)}}}$ of IC1:} The high-side switch of the half bridge is closed and the actuator motor is not connected. No motor current flows, $I_{\mathrm{L}}=\qty{0}{\ampere}$. The voltage measured across R3 with the multimeter is $U_{\text{R3,offset}}=\qty{89}{\milli\volt}$. From this, the actual offset current of IC1 is calculated:
\[ {I}_{\text{IS(Offset),IC1}}=\frac{{U}_{\text{R3,offset}}}{\mathrm{R3}}=\frac{\qty{89}{\milli\volt}}{\qty{499}{\ohm}}=\qty{178}{\micro\ampere} \]

\textbf{Determination of $\boldsymbol{\mathrm{dk}_{\mathrm{ILIS}}}$ of IC1:} The high-side switch of the half bridge is closed and a load current of \qty{5}{\ampere} or \qty{10}{\ampere} is set via a load resistor. The sense current is determined from the voltage across R3. In this case, a sense current of $I_{\text{IS1,IC1}}=\qty{419.5}{\micro\ampere}$ results for a load current of $I_{\text{L1,IC1}}=\qty{5}{\ampere}$ and a sense current of $I_{\text{IS2,IC1}}=\qty{661}{\micro\ampere}$ for a load current of $I_{\text{L2,IC1}}=\qty{10}{\ampere}$. The actual value of $\mathrm{dk}_{\mathrm{ILIS}}$ is calculated according to the equation given in the data sheet~\cite[p.~17, Eq.~(3)]{Inf13}:
\[ {\mathrm{dk}}_{\text{ILIS,IC1}}=\frac{{I}_{\text{L2,IC1}}-{I}_{\text{L1,IC1}}}{{I}_{\text{IS2,IC1}}-{I}_{\text{IS1,IC1}}}=\num{20.7e3} \]
Since $\mathrm{dk}_{\mathrm{ILIS}}$ is calculated from the difference between two measurement points according to Equation (3) of the data sheet, the offset current cancels out.

\textbf{Determination of $\boldsymbol{I_{\mathrm{IS(Offset)}}}$ of IC2:} The high-side switch of the half bridge is closed, the actuator motor is not connected. No motor current flows, $I_{\mathrm{L}}=\qty{0}{\ampere}$. The voltage measured across R5 with the multimeter is $U_{\text{R5,offset}}=\qty{87}{\milli\volt}$. From this, the actual offset current of IC2 is calculated:
\[ {I}_{\text{IS(Offset),IC2}}=\frac{{U}_{\text{R5,offset}}}{\mathrm{R5}}=\frac{\qty{87}{\milli\volt}}{\qty{499}{\ohm}}=\qty{174}{\micro\ampere} \]

\textbf{Determination of $\boldsymbol{\mathrm{dk}_{\mathrm{ILIS}}}$ of IC2:} The high-side switch of the half bridge is closed and a load current of \qty{5}{\ampere} or \qty{10}{\ampere} is set via a load resistor. The sense current is determined from the voltage across R5. In this case, a sense current of $I_{\text{IS1,IC2}}=\qty{434}{\micro\ampere}$ results for a load current of $I_{\text{L1,IC2}}=\qty{5}{\ampere}$ and a sense current of $I_{\text{IS2,IC2}}=\qty{694}{\micro\ampere}$ for a load current of $I_{\text{L2,IC2}}=\qty{10}{\ampere}$. The actual value of $\mathrm{dk}_{\mathrm{ILIS}}$ is calculated according to the equation given in the data sheet~\cite[p.~17, Eq.~(3)]{Inf13}:
\[ {\mathrm{dk}}_{\text{ILIS,IC2}}=\frac{{I}_{\text{L2,IC2}}-{I}_{\text{L1,IC2}}}{{I}_{\text{IS2,IC2}}-{I}_{\text{IS1,IC2}}}=\num{19.2e3} \]

\textbf{Explanation of correct current measurement:} The voltage at the terminals of the actuator motor is controlled via two motor drivers by means of PWM. The motor current must be determined to assess the load behavior and to implement the overload protection. In the future, it should also be possible to control the torque of the DC motor. The special feature is that the motor controller used only delivers a usable sense current for certain switch states. In the following, different ways of measuring the current are therefore presented.

The IS pin of the BTN8982TA (IC1, IC2) delivers a proportional sense current for positive load currents through the high-side switch~\cite[pp.~17\,f., Sec.~5.4.4]{Inf13}.\footnote{The component designations refer to the complete circuit diagram of the shift controller, which is not included in this publication.} Negative currents or currents flowing through the low-side switch cannot be detected. The operating strategy of the PWM must be selected with this property in mind. To simplify the explanation, only one direction of rotation of the electric motor is treated. The reversal of the direction of rotation results from the complementary control of the switches.

\textbf{Operating strategy 1:} During ``PWM OFF'', both low-side switches are closed. In the ``PWM ON'' state, S3 and S2 are closed. The rising current through S3 is converted into a positive sense current and can be processed by the ADC. When the PWM changes to the ``PWM OFF'' state, S3 is opened and S4 is closed. If the PWM frequency is selected sufficiently high, the inductance of the electric motor maintains the current flow in the same direction. The motor terminals are short-circuited via GND. The current now flows in the negative direction through S4 and in the positive direction through S2. The inductance of the motor winding results in a decaying current. The motor torque still acts in the same direction and drives the load. The load current flows through the low-side switches and is no longer converted into a proportional sense current by the motor controller IC. The value of the decaying current during the ``PWM OFF'' phase cannot be processed by the microcontroller. A well-functioning load monitoring of the DC drive is therefore only possible to a limited extent.

\textbf{Operating strategy 2:} During ``PWM OFF'', both high-side switches are closed. As in operating strategy 1 (Figure~\ref{fig:20}\,a), S3 and S2 are closed in the ``PWM ON'' state. In contrast to operating strategy 1, the current now flows through S3 and S1 during ``PWM OFF''. The positive current direction through S3 is maintained. With this operating strategy, the load current in motor operation can be evaluated continuously during ``PWM ON'' and ``PWM OFF''. A true RMS current measurement is necessary for the exact determination of the motor current. A gapless evaluation of the load current in controlled generator operation is nevertheless not possible. If the current direction reverses during the ``PWM ON'' phase, current flows in the negative direction through S3. Due to the limitation of the motor controller, this current cannot be measured. However, the current value can be calculated with the aid of a model from the known current in the ``PWM OFF'' state.

\newcommand{\hbswitch}[4]{%
  \ifx#3c\draw[very thick] (#1,#2) -- ++(0,0.6);
  \else\draw[thick] (#1,#2) -- ++(#4*0.28,0.55);\fi}
\newcommand{\hbridge}[7]{%
\begin{tikzpicture}[font=\scriptsize, >={Stealth[length=1.8mm]}]
  \foreach \x in {0,2.6} {
    \draw (\x,2.65) -- (\x,2.35) (\x,1.75) -- (\x,0.75) (\x,0.15) -- (\x,-0.1);
    \draw[->] (\x,2.65) -- (\x,2.95) node[above, inner sep=1pt] {$V_{\mathrm{CC}}$};
    \draw (\x-0.15,-0.1) -- (\x+0.15,-0.1) node[pos=0.5, below, inner sep=2pt] {GND};
  }
  \hbswitch{0}{1.75}{#2}{1}\hbswitch{0}{0.15}{#3}{1}\hbswitch{2.6}{1.75}{#4}{-1}\hbswitch{2.6}{0.15}{#5}{-1}
  \node[left] at (0,2.05) {S1}; \node[left] at (0,0.45) {S2};
  \node[right] at (2.6,2.05) {S3}; \node[right] at (2.6,0.45) {S4};
  \draw (0,1.6) -- (2.6,1.6);
  \fill (0,1.6) circle (1.2pt) (2.6,1.6) circle (1.2pt);
  \draw[thick, fill=white] (1.55,1.6) circle (0.24) node {M};
  \ifx#7s\draw[thick, fill=white] (0.4,1.5) rectangle (0.95,1.7);
    \node[above, inner sep=1pt] at (0.675,1.7) {Shunt};\fi
  \def\pathtype{#6}\def\on{on}\def\lowfree{lowfree}\def\highfree{highfree}
  \begin{scope}[very thick, black!55, ->]
    \ifx\pathtype\on
      \draw (2.42,2.75) -- (2.42,1.3) -- (0.18,1.3) -- (0.18,0.0);
      \node[black, below] at (1.3,1.3) {Current flow};
    \fi
    \ifx\pathtype\lowfree
      \draw (2.42,0.0) -- (2.42,1.3) -- (0.18,1.3) -- (0.18,0.0);
      \node[black, below] at (1.3,1.3) {Current flow};
    \fi
    \ifx\pathtype\highfree
      \draw (2.42,2.75) -- (2.42,1.92) -- (0.18,1.92) -- (0.18,2.75);
      \node[black, above] at (1.3,1.92) {Current flow};
    \fi
  \end{scope}
  \node[font=\small] at (1.3,3.7) {#1};%
\end{tikzpicture}}

\begin{figure}[tbp]
\begin{minipage}[t]{0.32\textwidth}
\textbf{a)} Operating strategy 1\par\medskip
\centering
\hbridge{``PWM ON''}{o}{c}{c}{o}{on}{}\par\bigskip
\hbridge{``PWM OFF''}{o}{c}{o}{c}{lowfree}{}
\end{minipage}\hfill
\begin{minipage}[t]{0.32\textwidth}
\textbf{b)} Operating strategy 2\par\medskip
\centering
\hbridge{``PWM ON''}{o}{c}{c}{o}{on}{}\par\bigskip
\hbridge{``PWM OFF''}{c}{o}{c}{o}{highfree}{}
\end{minipage}\hfill
\begin{minipage}[t]{0.32\textwidth}
\textbf{c)} Shunt measurement\par\medskip
\centering
\hbridge{``PWM ON''}{o}{c}{c}{o}{on}{s}\par\bigskip
\hbridge{``PWM OFF''}{o}{c}{o}{c}{lowfree}{s}
\end{minipage}
\caption[Switching states of the motor driver (H-bridge)]{Switching states of the motor driver (H-bridge). a) PWM operating strategy 1,
  b) PWM operating strategy 2, c) current measurement via shunt resistor; grey arrows: current flow\nocc}
\label{fig:20}
\end{figure}
Another option is to measure the current of the supply voltage $U_{\text{PWM-IN}}$ of the motor driver. Control by means of PWM can be regarded as a current or voltage conversion. If the losses of the electronics are neglected, the following equations can be set up:
\begin{gather*}
{P}_{\text{PWM-IN}}={P}_{\text{PWM-OUT}} \\
{U}_{\text{PWM-IN}}\cdot {I}_{\text{PWM-IN}}={U}_{\text{PWM-OUT}}\cdot {I}_{\text{PWM-OUT}}
\end{gather*}
The supply current therefore does not correspond to the motor current. Depending on the operating point, the output voltage of the motor driver can change. An actuating range between \qty{0}{\volt} and $U_{\text{PWM-IN}}$ is possible. In a motor vehicle, the battery voltage also varies depending on load and state of charge, and with it the supply voltage $U_{\text{PWM-IN}}$ of the motor driver. If the supply current is measured to determine the motor current, the supply voltage and the output voltage of the motor driver must also be measured in order to obtain information about the motor current. For exact current measurement, measurement of the true RMS is also necessary.

\textbf{Current measurement via shunt resistor:} A further option is current measurement at the motor terminal or in the motor lead by means of a shunt: The voltage drop across the shunt represents a proportional quantity for determining the motor current. Depending on the operating point of the motor or the state of the switches S1 and S2, a varying voltage component is superimposed on the voltage drop. The differential voltage across the resistor can have GND or VCC as its reference point. To evaluate the actual differential voltage, additional circuitry is required that subtracts the varying reference point from the differential voltage. In addition, as with all methods, the current must be measured using the true RMS method for correct acquisition.

In summary, it can be seen that motor current measurement outside the capabilities of the driver ICs is not easily possible. In some cases, considerable additional effort is required. For the shift controller, the capabilities of the ICs used are therefore relied upon. By means of operating strategy 2 (Figure~\ref{fig:20}\,b), the current can be fully acquired in motor operation. The problem of current measurement in generator operation can be solved cost-effectively by means of software using various methods. Among others, calculation, modeling, or avoidance of the operating state come into consideration. If the motor current profile can be acquired with sufficient accuracy by averaging, the likewise complex true RMS measurement can be avoided.

\textbf{PWM frequency and current measurement with ADC:} To determine the required PWM frequency, the shift forks are removed so that the shift drum can rotate freely. The actuator motor can thus be operated at a certain voltage with constant speed. The torque applied by the motor is determined by the bearing friction of the mechanical components. The aim is to select the PWM frequency such that no discontinuous current occurs in motor operation. The current thus does not reach negative values. Under this condition, the measurement can be carried out by the motor driver with operating strategy 2 (Figure~\ref{fig:20}\,b). The smallest possible duty cycle in compliance with the component specification is \qty{25}{\percent}.

\begin{figure}[hbp]%
\centering
\begin{tikzpicture}
\begin{groupplot}[
    thesisplot,
    group style={group size=1 by 2, vertical sep=0.35cm, x descriptions at=edge bottom},
    width=0.97\textwidth, height=3.6cm,%
    xmin=28, xmax=150, xtick={40,60,...,140},
    xlabel={$t$ in \unit{\micro\second}},
    ylabel style={align=center},
  ]
  \nextgroupplot[ymin=0, ymax=250, ytick={0,100,200}, ylabel={$U_{\mathrm{IS}}$\\in \unit{\milli\volt}}]
    \addplot[plotline] table[x=t_us, y=U_IS_mV] {figures/data/motor_current_25pwm_voltage.dat};
  \nextgroupplot[ymin=0, ymax=3, ytick={0,1,2,3}, ylabel={$I_{\mathrm{IC1}}$\\in \unit{\ampere}}]
    \addplot[plotline] table[x=t_us, y=I_A] {figures/data/motor_current_25pwm_current.dat};
\end{groupplot}
\end{tikzpicture}\par
\caption[Motor current profile at \qty{25}{\percent} PWM]{Motor current profile at \qty{25}{\percent} PWM: voltage at pin IS (top) and current through IC1 (bottom)\nocc}
\label{fig:23}
\end{figure}
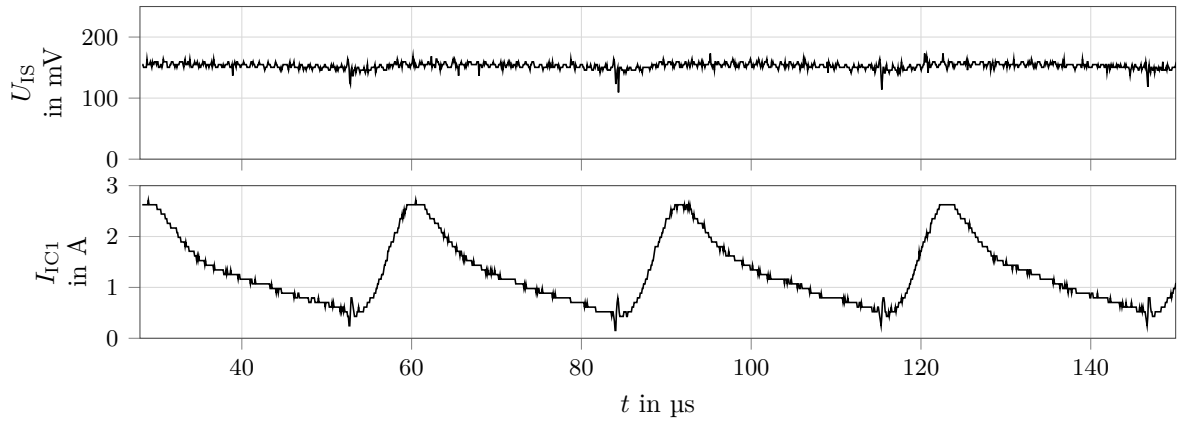
\begin{wrapfigure}{r}{0.5\textwidth}
\vspace*{-\intextsep}\vspace*{-0.9pt}%
\centering
\includegraphics[width=0.48\textwidth]{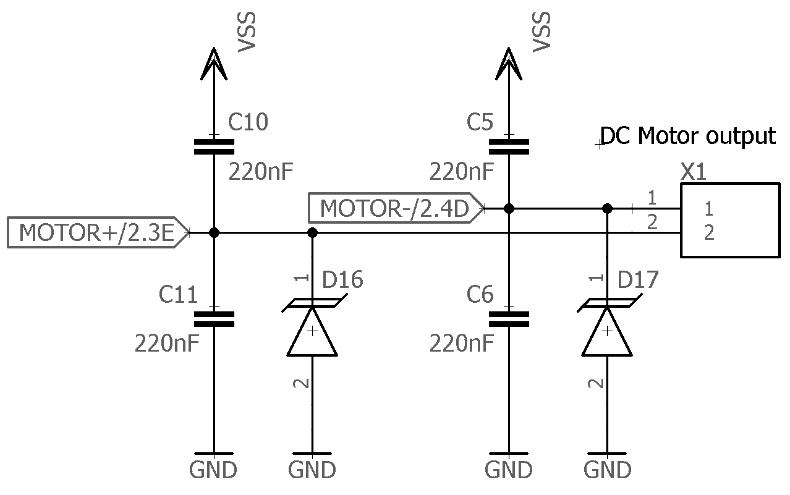}
\caption{Motor driver output}
\label{fig:24}
\end{wrapfigure}
The PWM frequency was set to \qty{32}{\kilo\hertz}. The upper plot in Figure~\ref{fig:23} shows the voltage profile at pin IS of the motor driver. The distinct averaging in the voltage profile is clearly visible. If measurement errors due to the missing true RMS measurement are tolerable, the voltage can be measured by means of the ADC and converted into a proportional current value. No information on the frequency response or the low-pass effect of the current measurement can be found in the data sheet of the chip manufacturer~\cite{Inf13}. Information on accuracy and linearity is also undocumented. For series use of the electronics, this type of current measurement is therefore only suitable to a limited extent and requires additional tuning. For the development of the shift controller, the sufficient accuracy of the current measurement was confirmed by measurement tests. Circuits that contain analog circuit sections and at the same time control PWM-controlled power drives require increased attention with regard to EMC. Short switching edges and high currents favor above all three types of interference with adjacent circuits: galvanic coupling, magnetic coupling, and capacitive coupling. Stray inductances and voltage differences between GND pins must be avoided. The power supply of the motor drivers is therefore supported by a low-ESR electrolytic capacitor (C1 in Figure~\ref{fig:10}). To reduce mutual interference, it is above all important to attenuate possible disturbances already at the input of the circuit board. At each motor terminal, two \qty{220}{\nano\farad} capacitors (C11, C6) are used to reduce voltage spikes from commutation. Additional ESD protection is provided by the unidirectional TVS diodes (D16, D17); the component designations refer to Figure~\ref{fig:24}. Further along the circuit, additional filtering measures are appropriate for signals that lead to input pins of the microcontroller (Table~\ref{tab:7}).

\chapter{Shift Controller Software}\label{sec:31}

The software development is carried out in the Atmel Studio 7 development environment. The program code is written in ANSI C. The Arduino libraries and the Arduino IDE programming environment are not used. The EDBG interface is used in Atmel Studio 7 as the programming and debugging interface. Compared to the Atmel-ICE, this interface offers a low-cost debug interface with a reduced range of functions. For smaller development projects, the functionality is completely sufficient; for larger projects, the Atmel-ICE should be used due to its more solid implementation. The SAMD21G18 microcontroller offers a large number of timers, many of which can be configured as 8-, 16- or 32-bit variants. Above all, useful timer functions are already implemented in hardware. Thus, the ``Timer/Counter for Control Applications'' offers fully implemented features for controlling H-bridges or motor controllers. The CPU is relieved, and the available computing power is thus available to a greater extent for the implementation of time-critical control processes. The microcontroller offers further important functions in hardware, e.g. a 12-bit ADC with 350~ksps and an integrated serial interface. For use in motor vehicles, the manufacturer offers an AEC-Q100 qualified variant of the SAMD21 with the SAM DA1 series.

\section{Initialization}\label{sec:32}

\begin{wrapfigure}{r}{0.4\textwidth}
\vspace*{-\intextsep}\vspace*{-1.3pt}%
\centering
\begin{tikzpicture}[font=\footnotesize,
    box/.style={minimum height=0.75cm, text width=2.35cm, align=center, inner xsep=2pt,
                font=\footnotesize\ttfamily},
    used/.style={box, draw, thick, fill=white},
    dbg/.style={box, draw, dashed, fill=black!8}]
  \node[used] at (0,1.0)   {IO\_init()};
  \node[used] at (2.75,1.0) {Clock\_init()};
  \node[used] at (0,0)     {ADC\_init()};
  \node[used] at (2.75,0)   {Timer\_init()};
  \node[dbg] at (0,-1.3)   {Serial\_init()};
  \node[dbg] at (2.75,-1.3) {EVSYS\_init()};
  \node[dbg] at (0,-2.3)   {EIC\_init()};
  \draw[thick, fill=white] (-1.3,-3.38) rectangle ++(0.45,0.26);
  \node[anchor=west] at (-0.8,-3.25) {used in the main program};
  \draw[dashed, fill=black!8] (-1.3,-3.88) rectangle ++(0.45,0.26);
  \node[anchor=west] at (-0.8,-3.75) {basic investigation and debugging};
\end{tikzpicture}\par
\caption{Software initialization\nocc}
\label{fig:25}
\end{wrapfigure}
In order to be able to use the microcontroller for the intended purpose, the hardware must first be initialized with all required registers (Figure~\ref{fig:25}). The SAMD21G18 uses the 32-bit ARM Cortex-M0+ processor as its CPU, thus the registers required for initialization are 32 bits wide. When changing register contents, particular care must be taken not to violate any of the register properties explained below:

\textbf{Write-Protected:} The write protection of certain registers can be activated and deactivated in the ``Peripheral Access Controller'' with the registers WPSET and WPCLR. For register accesses by the debugger, the write protection is ignored~\cite[pp.~53\,f., Sec.~11.6]{ATD18}.

\textbf{Enable-Protected:} Bits in registers with this property can only be changed if the respective peripheral is disabled. Which bit prevents access to the register is defined separately for each peripheral.

\textbf{Read/Write-Synchronized:} Due to several clock sources in the system, the CPU clock and the peripheral clock may not be synchronous. This requires a synchronization process when accessing the respective register. The registers to which this property applies are explained separately for each register in the register description. Peripheral modules with registers of this property have their own status register with the corresponding bit SYNCBUSY. For the duration of the synchronization process, SYNCBUSY=1, and no further access to the register is permitted during this time. After completion of the synchronization, the bit is reset automatically.

In principle, there are two ways of working with registers. The simplest method is to use the predefined register addresses of the implemented Atmel libraries. The names are assigned very intuitively and are easy to remember. In the following example, Port A 2 is defined as an output.

All register definitions always begin with ``REG'', followed by the name of the peripheral, e.g., ``PORT'', and finally the name of the register is appended, in this case the register ``DIR''. If pins of port group A are configured, a ``0'' must also be appended. For pins of port group B, a ``1'' is added. The complete command is thus:

\begin{lstlisting}
REG_PORT_DIR0 |= (1 << 2);                                // Port A 2 as output
\end{lstlisting}

If the pin is to be used as an input again, the corresponding bit must be reset. This is best done with the command:

\begin{lstlisting}
REG_PORT_DIR0 &= ~(1 << 2);                               // Port A 2 as input
\end{lstlisting}

However, not all register addresses are defined in the Atmel libraries. As a remedy, nested structures (structures within structures) are used. If the commands are to be executed in this way, the instructions are as follows:

\begin{lstlisting}
PORT->Group[0].DIR.reg |= (1 << 2);                       // Port A 2 as output
PORT->Group[0].DIR.reg &= ~(1 << 2);                      // Port A 2 as input
\end{lstlisting}

Particular attention must be paid to the fact that some registers require a ``single 32-bit write''. If this is overlooked and the bits are nevertheless set individually, this can lead to strange behavior and problems that are difficult to diagnose.

The individual initialization measures are explained only for modules of the completed software development. Serial\_init(), EVSYS\_init() and EIC\_init() serve for experimental testing and for debugging measures and are not required for operation.

\subsection{IO\_init()}\label{sec:33}

The central element of the I/O functionality of the microcontroller is the I/O pin controller. The individual ports are organized in groups of 32 pins each. Ports are identified by letters, e.g., Port A, Port B, etc.; the pins are numbered 0-31. For example, Port A 0 designates Port A, pin number 0. The individual pins can either be managed as standard I/O pins via the PORT registers or be connected directly to internal peripherals~\cite[pp.~28\,f., Table~7-1]{ATD18}. Each pin can only be connected to one peripheral; multiple assignments are not possible.

It is important not to lose track of the pins already in use. Incorrect assignment can quickly destroy the internal logic of the pins, above all if an input is mistakenly defined as an output. Table~\ref{tab:7} gives the complete list of the I/O used.

\begin{table}[tbp]
\centering
\caption{I/O overview of the microcontroller}\label{tab:7}
{\small
\begin{tabular}{>{\centering\arraybackslash}p{\dimexpr 0.188\textwidth-2\tabcolsep\relax}>{\centering\arraybackslash}p{\dimexpr 0.161\textwidth-2\tabcolsep\relax}>{\raggedright\arraybackslash}p{\dimexpr 0.307\textwidth-2\tabcolsep\relax}>{\raggedright\arraybackslash}p{\dimexpr 0.320\textwidth-2\tabcolsep\relax}}
\toprule
\textbf{Arduino pin} & \textbf{Atmel pin} & \textbf{Signal} & \textbf{Peripheral signal} \\
\midrule
D0 & Port A 11 & Serial RX & SERCOM0/PAD[3] \\
D1 & Port A 10 & Serial TX & SERCOM0/PAD[2] \\
D2 & Port A 14 & INC\_CHANNEL\_A\_DIV & EIC/EXTINT[14] \\
D5 & Port A 15 & Forward / Reverse & General Purpose Input \\
D6 & Port A 20 & Switch Upshift & General Purpose Input \\
D7 & Port A 21 & Switch Downshift & General Purpose Input \\
D10 & Port A 18 & Ignition Breaker & General Purpose Output \\
D11 & Port A 16 & IN\_MOTOR\_MINUS & TCC2/WO[0] \\
D13 & Port A 17 & IN\_MOTOR\_PLUS & TCC2/WO[1] \\
A0 & Port A 02 & U(I) Motor+ & AIN[0] \\
AREF & Port A 03 & U(I) Motor- & AIN[1] \\
A1 & Port B 08 & U(Pos) shift drum & AIN[2] \\
\bottomrule
\end{tabular}}
\end{table}
The initialization of the standard I/O pins is carried out as described above (Section~\ref{sec:32}). For the configuration of the peripheral signals, the use of the PORT.WRCONFIG register has proven to be advantageous. The PORT.WRCONFIG register must always be written with a ``single 32-bit write''. In the following, the individual bits of the register are explained~\cite[pp.~430\,f., Sec.~23.8.11]{ATD18}:

\textbf{HWSEL:} Half-Word Select. If the bit is set, pins 16 to 31 are configured; if the bit is set to 0, pins 0 to 15.

\textbf{WRPINCFG:} ``Write Pin Configuration'' If the bit is set, the configuration of WRCONFIG.DRVSTR, WRCONFIG.PULLEN and WRCONFIG.INEN is applied to the pins selected in WRCONFIG.PINMASK.

\textbf{WRPMUX:} ``Write Peripheral Multiplexer'' If the bit is set, the configuration of PMUX is applied to the pins selected in WRCONFIG.PINMASK.

\textbf{PMUX:} ``Peripheral Multiplexing'' The respective bits establish the connection from the pin to the peripheral (assignment table in~\cite[pp.~28\,f., Table~7-1]{ATD18}). The configuration is written for the pins selected in WRCONFIG.PINMASK.

\textbf{PULLEN:} ``Pull Enable'' This bit activates the corresponding pull-up resistor for the pins selected in WRCONFIG.PINMASK.

\textbf{INEN:} Input Enable. This bit activates the input buffer for the pins selected in WRCONFIG.PINMASK.

\textbf{PMUXEN:} ``Peripheral Multiplexer Enable'' If the bit is set, the pin is connected by the multiplexer to the peripheral selected in WRCONFIG.PMUX. If the bit is not set, the pin operates as a general purpose I/O. The pins are selected by WRCONFIG.PINMASK.

\textbf{PINMASK:} ``Pin Mask for Multiple Pin Configuration'' 16-bit register for the selection of the pins. Each bit defines one pin. Depending on the state of WRCONFIG.HWSEL, pins 0 to 15 or 16 to 31 can be selected. If, e.g., the state of WRCONFIG.HWSEL=0, pin 2 is configured by setting bit number 2.

\subsection{Clock\_init()}\label{sec:34}

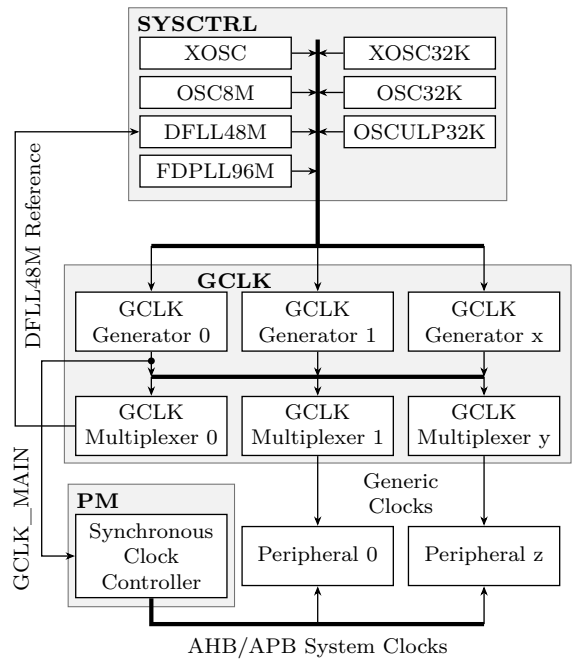
\begin{wrapfigure}[21]{r}{0.52\textwidth}
\vspace*{-\intextsep}\vspace*{-1.1pt}%
\centering
\begin{tikzpicture}[font=\scriptsize, >={Stealth[length=3.6pt, width=2.8pt]},
    blk/.style={draw, fill=white, align=center, inner xsep=2pt, inner ysep=2.5pt,
                minimum height=0.42cm, line width=0.5pt},
    src/.style={blk, minimum width=2.0cm},
    gen/.style={blk, minimum width=2.0cm, minimum height=0.78cm},
    per/.style={blk, minimum width=2.0cm, minimum height=0.78cm},
    module/.style={draw=black!55, fill=black!5, line width=0.4pt},
    title/.style={anchor=north west, inner sep=3pt, font=\scriptsize\bfseries},
    bus/.style={line width=1.6pt},
    sig/.style={->, line width=0.5pt}]

  \draw[module] (0.85,0) rectangle (5.85,-2.55);
  \node[title] at (0.85,0) {SYSCTRL};
  \node[src] (xosc)   at (2.0,-0.6)  {XOSC};
  \node[src] (osc8m)  at (2.0,-1.12) {OSC8M};
  \node[src] (dfll)   at (2.0,-1.64) {DFLL48M};
  \node[src] (fdpll)  at (2.0,-2.16) {FDPLL96M};
  \node[src] (xosc32) at (4.7,-0.6)  {XOSC32K};
  \node[src] (osc32)  at (4.7,-1.12) {OSC32K};
  \node[src] (osculp) at (4.7,-1.64) {OSCULP32K};
  \draw[bus] (3.35,-0.42) -- (3.35,-3.15);
  \draw[bus] (1.15,-3.15) -- (5.55,-3.15);
  \foreach \n in {xosc,osc8m,dfll,fdpll}  \draw[sig] (\n.east) -- (\n.east -| 3.35,0);
  \foreach \n in {xosc32,osc32,osculp}    \draw[sig] (\n.west) -- (\n.west -| 3.35,0);

  \draw[module] (0,-3.4) rectangle (6.7,-6.02);
  \node[title, anchor=north] at (2.25,-3.4) {GCLK};%
  \node[gen] (g0) at (1.15,-4.15) {GCLK\\Generator 0};
  \node[gen] (g1) at (3.35,-4.15) {GCLK\\Generator 1};
  \node[gen] (gx) at (5.55,-4.15) {GCLK\\Generator x};
  \draw[bus] (1.15,-4.88) -- (5.55,-4.88);
  \node[gen] (m0) at (1.15,-5.53) {GCLK\\Multiplexer 0};
  \node[gen] (m1) at (3.35,-5.53) {GCLK\\Multiplexer 1};
  \node[gen] (my) at (5.55,-5.53) {GCLK\\Multiplexer y};
  \foreach \g in {g0,g1,gx} {
    \draw[sig] (\g.north |- 0,-3.15) -- (\g.north);
    \draw[sig] (\g.south) -- (\g.south |- 0,-4.88);
  }
  \foreach \m in {m0,m1,my} \draw[sig] (\m.north |- 0,-4.88) -- (\m.north);

  \draw[module] (0.05,-6.3) rectangle (2.25,-7.92);
  \node[title] at (0.05,-6.3) {PM};
  \node[per, minimum height=1.1cm] (scc) at (1.15,-7.25) {Synchronous\\Clock\\Controller};
  \node[per] (p0)  at (3.35,-7.25) {Peripheral 0};
  \node[per] (pz)  at (5.55,-7.25) {Peripheral z};

  \draw[sig] (m1.south) -- (p0.north);
  \draw[sig] (my.south) -- (pz.north);
  \node[align=center] at (4.45,-6.42) {Generic\\Clocks};

  \draw[bus] (scc.south) -- (1.15,-8.15) -- (5.55,-8.15);
  \draw[sig] (3.35,-8.15) -- (p0.south);
  \draw[sig] (5.55,-8.15) -- (pz.south);
  \node[anchor=north] at (3.35,-8.18) {AHB/APB System Clocks};

  \fill (1.15,-4.67) circle (1.3pt);
  \draw[sig] (1.15,-4.67) -- (-0.3,-4.67) -- (-0.3,-7.25) -- (scc.west);
  \node[rotate=90, anchor=south] at (-0.3,-6.75) {GCLK\_MAIN};

  \draw[sig] (m0.west) -- (-0.65,-5.53) -- (-0.65,-1.64) -- (dfll.west);
  \node[rotate=90, anchor=north] at (-0.65,-3.15) {DFLL48M Reference};
\end{tikzpicture}\par
\caption{SAMD21G18 clock system (based on~\protect\cite[p.~109, Fig.~14-1]{ATD18})\nocc}
\label{fig:27}
\end{wrapfigure}
The function initializes the clock system of the SAMD21G18 microcontroller and activates the external crystal for the clock supply. The external crystal operates at a resonance frequency of \qty{32.768}{\kilo\hertz} and is connected to pins 1 (XIN32) and 2 (XOUT32) of the microcontroller. The aim is to generate the system clock GCLK\_MAIN (\qty{48}{\mega\hertz}) from the external oscillator XOSC32K (Figure~\ref{fig:27}). To achieve this, various parts of the clock system must be initialized~\cite[p.~109, Ch.~14; p.~117, Ch.~15]{ATD18}:

\textbf{Clock Sources:} The microcontroller has various clock sources such as DFLL48M and OSC8M, including external clock inputs such as XOSC32K. The basic function of the DFLL48M can be understood as a precise clock multiplier~\cite[pp.~180\,f.]{ATD18}. The clock sources are managed by the system controller SYSCTRL.

\textbf{Generic Clock Generators:} These are connected to the clock sources via multiplexers and operate as adjustable prescalers.

\textbf{Generic Clocks:} These are the clock supply of the peripherals, e.g. timers, event system, USART. A multiplexer is used to set which Generic Clock Generator serves as the source for the various ``Generic Clocks'' supplying the peripherals.

The following steps are necessary to set up the clock supply:

\begin{itemize}
\item Setting the flash wait states for~\qty{48}{\mega\hertz}
\begin{lstlisting}
REG_NVMCTRL_CTRLB |= (1 << NVMCTRL_CTRLB_RWS_Pos);
\end{lstlisting}
\item Enabling the XOSC32K as clock source for the DFLL48M
\begin{lstlisting}
REG_SYSCTRL_XOSC32K |=  (0x3 << SYSCTRL_XOSC32K_STARTUP_Pos) | (!SYSCTRL_XOSC32K_ONDEMAND) | (!SYSCTRL_XOSC32K_AAMPEN) | SYSCTRL_XOSC32K_EN32K | SYSCTRL_XOSC32K_XTALEN | SYSCTRL_XOSC32K_ENABLE;
\end{lstlisting}
\item Defining the XOSC32K as source for Generic Clock Generator 1
\begin{lstlisting}
REG_GCLK_GENCTRL = (uint32_t) GCLK_GENCTRL_RUNSTDBY | (!GCLK_GENCTRL_DIVSEL) | !GCLK_GENCTRL_OE | !GCLK_GENCTRL_OOV | GCLK_GENCTRL_IDC | GCLK_GENCTRL_GENEN | (0x5 << GCLK_GENCTRL_SRC_Pos) | (0x1 << GCLK_GENCTRL_ID_Pos);
while(REG_GCLK_STATUS & GCLK_STATUS_SYNCBUSY);            // Wait for sync.
\end{lstlisting}
\item Connecting the output of Clock Generator 1 to the input of the DFLL48M
\begin{lstlisting}
REG_GCLK_CLKCTRL = (uint16_t) (!GCLK_CLKCTRL_WRTLOCK) | GCLK_CLKCTRL_CLKEN | (0x1 << GCLK_CLKCTRL_GEN_Pos) | (0 << GCLK_CLKCTRL_ID_Pos);
\end{lstlisting}
\item Initializing the DFLL48M in closed-loop mode
\begin{lstlisting}
REG_SYSCTRL_DFLLCTRL = SYSCTRL_DFLLCTRL_MODE | SYSCTRL_DFLLCTRL_ENABLE;
while(!(REG_SYSCTRL_PCLKSR & SYSCTRL_PCLKSR_DFLLRDY));    // Wait for sync.
\end{lstlisting}
\item Setting the DFLL48M multiplication factor
\begin{lstlisting}
REG_SYSCTRL_DFLLMUL |= (31 << SYSCTRL_DFLLMUL_CSTEP_Pos) | (511 << SYSCTRL_DFLLMUL_FSTEP_Pos) | (1465 << SYSCTRL_DFLLMUL_MUL_Pos);
\end{lstlisting}
\item Setting the DFLL48M as clock source for Generic Clock Generator 0
\begin{lstlisting}
REG_GCLK_GENCTRL = (uint32_t) GCLK_GENCTRL_RUNSTDBY | !GCLK_GENCTRL_DIVSEL | !GCLK_GENCTRL_OE | !GCLK_GENCTRL_OOV | GCLK_GENCTRL_IDC | GCLK_GENCTRL_GENEN | (7 << GCLK_GENCTRL_SRC_Pos) | (0 << GCLK_GENCTRL_ID_Pos);
while(REG_GCLK_STATUS & GCLK_STATUS_SYNCBUSY);            // Wait for sync.
\end{lstlisting}
\end{itemize}

\subsection{ADC\_init()}\label{sec:35}

\begin{wraptable}{r}{0.36\textwidth}
\vspace*{-\intextsep}\vspace*{-2.5pt}%
\centering
\caption[Prescaler]{Prescaler (source:~\protect\cite[p.~910, Sec.~33.8.5]{ATD18})\nocc}\label{tab:8}
{\small
\begin{tabular}{lr}%
\toprule
\textbf{Prescaler} & \multicolumn{1}{c}{\textbf{GCLK\_ADC}} \\
\midrule
DIV4 & \qty{12000}{\kilo\hertz} \\
DIV8 & \qty{6000}{\kilo\hertz} \\
DIV16 & \qty{3000}{\kilo\hertz} \\
DIV32 & \qty{1500}{\kilo\hertz} \\
DIV64 & \qty{750}{\kilo\hertz} \\
DIV128 & \qty{375}{\kilo\hertz} \\
DIV256 & \qty{188}{\kilo\hertz} \\
DIV512 & \qty{94}{\kilo\hertz} \\
\bottomrule
\end{tabular}}
\end{wraptable}
The function initializes the ADC of the microcontroller. According to the ADC specifications, the ADC can be operated at a clock rate between \qty{30}{\kilo\hertz} and \qty{2100}{\kilo\hertz}~\cite[p.~1115, Table~40-24]{ATD18}. The clock supplying the ADC (GCLK\_ADC) is derived from Clock Generator 0, which is already configured to \qty{48}{\mega\hertz}. Only a limited selection of divider values is available as prescaler in the ADC.CTRLB register, so the ADC frequency cannot be set exactly to the maximum possible frequency of \qty{2100}{\kilo\hertz} with a base clock of \qty{48}{\mega\hertz}. The divider values listed in Table~\ref{tab:8} are possible. In order not to operate the ADC outside its specification, at most the prescaler DIV32 can be used. This results in an ADC clock frequency of \qty{1.5}{\mega\hertz}. To operate the ADC above \qty{1.5}{\mega\hertz}, a separate clock must be generated with Clock Generator 2, since the prescaler DIV16 yields a frequency of \qty{3}{\mega\hertz} and the ADC would thus be operated outside its specification. However, it must be noted that the sample-and-hold time decreases with increasing ADC frequency. This leads to reduced accuracy and a larger spread of the measurement results. An optimal value for the sample-and-hold time was determined by test measurements. With the value SAMPLEN=15 in the ADC.SAMPCTRL register, the sample-and-hold time is set to \qty{5.33}{\micro\second}. Calculation of the time using the equation according to the manufacturer~\cite[p.~909, Sec.~33.8.4]{ATD18}:
\begin{gather*}
{f}_{\text{CLK,ADC}}=\qty{1.5}{\mega\hertz},\qquad {T}_{\text{CLK,ADC}}=\frac{1}{{f}_{\text{CLK,ADC}}}=\qty{666.7}{\nano\second},\qquad \mathrm{SAMPLEN}=15 \\
{t}_{\mathrm{sample}}=\left( \mathrm{SAMPLEN}+1 \right) \cdot \frac{{T}_{\text{CLK,ADC}}}{2}=\qty{5.33}{\micro\second}
\end{gather*}
In the present application, three measured values have to be read within the \qty{50}{\micro\second} sampling time of the position control: signal U(I)\_MOTOR\_PLUS\_ADC at pin AIN[0], signal U(I)\_MOTOR\_MINUS\_ADC at pin AIN[1] and signal U(POS)\_DRUM\_ADC at pin AIN[2] (Table~\ref{tab:7}). All signals are in the voltage range \qtyrange{0}{1}{\volt}, so the single-ended mode is used. For reading several signals one after the other, the free-running mode in combination with the input scan function is best suited. In summary, the following settings are made on the ADC:

\textbf{Free-running mode:} If the ADC is enabled, a new conversion is started automatically after a conversion has been completed. The results are stored in the ADC.RESULT register until the following conversion and are then overwritten automatically.

\textbf{Input scan AIN[0] to AIN[2]:} The inputs AIN[0], AIN[1], and AIN[2] are converted one after the other autonomously. AIN[2] is followed by input AIN[0] again.

\textbf{Internal reference voltage INT1V:} The internal reference voltage is used. The signals must not exceed +\qty{1}{\volt} with respect to ground.

\textbf{Single-ended mode:} All voltage signals are referenced to a common potential, in this case ground. The conversion of signals in single-ended mode is considerably faster than in differential mode.

The conversion time of one measurement is 6.5 clock cycles + sampling time~\cite[p.~898, Fig.~33-7]{ATD18}. The calculation of the time for measuring all signals is therefore as follows:
\[ {t}_{\mathrm{conv}}=3\cdot \left( 6.5\cdot {T}_{\text{CLK,ADC}}+{t}_{\mathrm{sample}} \right) =\qty{29}{\micro\second} \]
The register initialization of the ADC comprises the following steps:

\begin{itemize}
\item Generic Clock Generator 0 as source for GCLK\_ADC
\begin{lstlisting}
REG_GCLK_CLKCTRL = (uint16_t) GCLK_CLKCTRL_CLKEN | (0 << GCLK_CLKCTRL_GEN_Pos) | (0x1E << GCLK_CLKCTRL_ID_Pos);
\end{lstlisting}
\item Setting the reference voltage
\begin{lstlisting}
REG_ADC_REFCTRL |= (0x00 << ADC_REFCTRL_REFSEL_Pos);
\end{lstlisting}
\item Configuring the prescaler, setting the digital correction values and setting the free-running mode
\begin{lstlisting}
REG_ADC_CTRLB |= (0x03 << ADC_CTRLB_PRESCALER_Pos) | ADC_CTRLB_CORREN |
    ADC_CTRLB_FREERUN;
while (REG_ADC_STATUS & ADC_STATUS_SYNCBUSY);             // Wait for sync.
\end{lstlisting}
\item Configuring the inputs and setting the input scan
\begin{lstlisting}
REG_ADC_INPUTCTRL |= (0x18 << ADC_INPUTCTRL_MUXNEG_Pos) | (0x02 << ADC_INPUTCTRL_MUXPOS_Pos) | (0x0 << ADC_INPUTCTRL_INPUTSCAN_Pos) | (0 << ADC_INPUTCTRL_INPUTOFFSET_Pos);
while (REG_ADC_STATUS & ADC_STATUS_SYNCBUSY);             // Wait for sync.
\end{lstlisting}
\item Enabling the result-ready interrupt
\begin{lstlisting}
REG_ADC_INTENSET |= ADC_INTENSET_RESRDY;
\end{lstlisting}
\item Enabling the interrupts in the NVIC
\begin{lstlisting}
NVIC_EnableIRQ(ADC_IRQn);
\end{lstlisting}
\item Reading the manufacturer correction values from the NVRAM
\begin{lstlisting}
uint8_t adc_linearity = ( (*(uint32_t*)NVMCTRL_OTP4) & NVM_ADC_LINEARITY_Msk ) >> NVM_ADC_LINEARITY_Pos;
uint8_t adc_biascal = ( (*(uint32_t*)NVMCTRL_OTP4) & NVM_ADC_BIASCAL_Msk ) >> NVM_ADC_BIASCAL_Pos;
\end{lstlisting}
\item Setting the manufacturer correction values in the ADC.CALIB register
\begin{lstlisting}
REG_ADC_CALIB |= (adc_biascal << ADC_CALIB_BIAS_CAL_Pos) | (adc_linearity << ADC_CALIB_LINEARITY_CAL_Pos);
\end{lstlisting}
\item Setting the manual correction values
\begin{lstlisting}
REG_ADC_GAINCORR = 0b10000000010;                         // Set manual gain correction
REG_ADC_OFFSETCORR = 13;                                  // Set manual offset value
\end{lstlisting}
\item Setting the sample time
\begin{lstlisting}
ADC->SAMPCTRL.bit.SAMPLEN = 15;                           // Sample time 5.33 µs
\end{lstlisting}
\item Enabling the ADC
\begin{lstlisting}
REG_ADC_CTRLA |= ADC_CTRLA_ENABLE;
while (REG_ADC_STATUS & ADC_STATUS_SYNCBUSY);             // Wait for sync.
\end{lstlisting}
\end{itemize}

\subsection{Timer\_init()}\label{sec:36}

The microcontroller SAMD21G18A has a large number of timers with various functions~\cite[pp.~662\,ff., Ch.~30; pp.~704\,ff., Ch.~31]{ATD18}. With the extensive capabilities of the component, the configuration effort for the initialization also increases. The advantage of the hardware-based variety of functions lies above all in a reduced CPU load. For the implementation of the software components, two timers in particular are used as core components; their initialization is explained in the following.

\textbf{TCC -- Timer/Counter for Control Applications:} This timer module offers three different instances with several independent signal outputs~\cite[pp.~704\,ff., Ch.~31]{ATD18}. The signal generation is specifically tailored to the requirements of external peripherals. Special features for motor controllers, switched-mode power supplies and H-bridges are integrated directly in hardware.

To control the motor drivers, one PWM signal is required for each half-bridge. If a low level is applied to the ``IN'' pin of a half-bridge, the low-side MOSFET is closed. A high-level signal closes the high-side MOSFET. The timer instance is selected according to the pin assignment on the microcontroller. IN\_MOTOR\_PLUS is connected to the signal TCC2/WO[1] and IN\_MOTOR\_MINUS is connected to the signal TCC2/WO[0] (Table~\ref{tab:7}). TCC2 is therefore used for signal generation. The switches are controlled according to operating strategy~2 (Figure~\ref{fig:20}\,b). For this purpose, the timer mode ``Single-Slope PWM Operation'' is used (Figure~\ref{fig:28}).

\par\medskip\noindent\begin{minipage}{\textwidth}
\centering
\begin{tikzpicture}[x=\linewidth/13.4, y=1cm, font=\scriptsize, >={Stealth[length=3.6pt, width=2.8pt]},
    lvl/.style={densely dotted, line width=0.4pt},
    cnt/.style={line width=0.9pt},
    cc/.style={dashed, line width=0.6pt},
    wo/.style={line width=0.8pt},
    ext/.style={densely dotted, line width=0.4pt},
    rst/.style={circle, fill=black, inner sep=0pt, minimum size=4pt},
    mtc/.style={cross out, draw, line width=0.7pt, inner sep=0pt, minimum size=5pt}]
  \def\T{3.4}\def\P{2.2}
  \draw[lvl] (0,\P) -- (10.4,\P);  \node[anchor=east] at (-0.1,\P) {TCC2.PER};
  \draw[lvl] (0,0)  -- (10.4,0);   \node[anchor=east] at (-0.1,0)  {0};
  \node[anchor=east, align=right, font=\scriptsize\bfseries] at (-0.1,1.1) {TCC2.COUNT};
  \foreach \i/\c in {0/0.6, 1/0.3, 2/0.85} {
    \pgfmathsetmacro{\xa}{\i*\T}
    \pgfmathsetmacro{\xb}{(\i+1)*\T}
    \pgfmathsetmacro{\xm}{\xa+\c*\T}
    \draw[cnt] (\xa,0) -- (\xb,\P);
    \draw[cnt] (\xb,\P) -- (\xb,0);
    \draw[cc] (\xa,\c*\P) -- (\xb,\c*\P);
    \node[rst] at (\xa,0) {};
    \node[mtc] at (\xm,\c*\P) {};
    \draw[wo] (\xa,-1.45) -- (\xa,-0.95) -- (\xm,-0.95) -- (\xm,-1.45) -- (\xb,-1.45);
    \draw[ext] (\xa,0) -- (\xa,-0.95);
    \draw[ext] (\xm,\c*\P) -- (\xm,-0.95);
  }
  \node[rst] at (3*\T,0) {};
  \draw[wo] (3*\T,-1.45) -- (3*\T,-0.95) -- (10.4,-0.95);
  \draw[ext] (3*\T,0) -- (3*\T,-0.95);
  \node[anchor=south] at (1.72*\T,0.3*\P) {TCC2.CC0};
  \node[anchor=east, font=\scriptsize\bfseries] at (-0.1,-1.2) {WO[0]};
  \draw[ext] (0,\P) -- (0,\P+0.55);  \draw[ext] (\T,\P) -- (\T,\P+0.55);
  \draw[<->, line width=0.5pt] (0,\P+0.45) -- node[above] {$T = 1/f_{\mathrm{PWM,TCC2}}$} (\T,\P+0.45);
  \node[rst] at (7.05,\P+0.85) {};
  \node[anchor=west] at (7.15,\P+0.85) {COUNT restarts, WO[0] set};
  \node[mtc] at (7.05,\P+0.5) {};
  \node[anchor=west] at (7.15,\P+0.5) {match with CC0, WO[0] reset};
\end{tikzpicture}\par
\captionof{figure}{Single-slope PWM of TCC2 (based on~\protect\cite[p.~715, Fig.~31-6]{ATD18})\nocc}
\label{fig:28}
\end{minipage}\par\medskip
Several registers are relevant for the signal configuration. The period T is set using the register TCC2.PER: With each clock cycle, the counter value in the register TCC2.COUNT is incremented by 1. When the counter value reaches the value of the register TCC2.PER, the counter value is reset to 0. At the same time, the selected signal outputs WO[x] are set. The duty cycle is controlled by the associated registers TCC2.CC0 for WO[0] and TCC2.CC1 for WO[1]. When the counter value reaches the register value in TCC2.CC0, the signal output WO[0] is reset; however, the counter value in the register TCC2.COUNT continues to be incremented until the period is reached. The same happens for the signal output WO[1]. This mode allows a simple setting of the duty cycle of the individual signals. The period is the same for both signal outputs. The PWM frequency was set to $f_{\text{PWM,TCC2}}=\qty{32}{\kilo\hertz}$ (period of \qty{31.25}{\micro\second} in Figure~\ref{fig:23}). For this, the register values for signal generation are calculated. The peripheral clock supplying the timer is $f_{\text{GCLK,TCC2}}=\qty{48}{\mega\hertz}$; it corresponds to the CPU clock, no prescaler is used ($N_{\text{presc,TCC2}}=1$). The register value TCC2.PER ``TOP'' is calculated as follows~\cite[p.~716, Sec.~31.6.2.5.5]{ATD18}:
\begin{equation}
{\mathrm{TOP}}_{\mathrm{TCC2}}=\frac{{f}_{\text{GCLK,TCC2}}}{{f}_{\text{PWM,TCC2}}\cdot {N}_{\text{presc,TCC2}}}-1=1499
\label{eq:9}
\end{equation}
For a frequency of \qty{32}{\kilo\hertz}, the register TCC2.PER must be initialized to the value 1499. The values of the registers TCC2.CC1 and TCC2.CC0 are initialized with 1499 to close the high-side MOSFETs. During operation, these can be varied between 0 (\qty{0}{\percent} duty cycle) and 1499 (\qty{100}{\percent} duty cycle) depending on the desired motor voltage.

The resolution of the duty cycle can be calculated with the given equations~\cite[p.~716, Sec.~31.6.2.5.5]{ATD18}:
\[ {b}_{\text{PWM,TCC2}}=\frac{{\log}_{10}\,\left( {\mathrm{TOP}}_{\mathrm{TCC2}}+1 \right) }{{\log}_{10}\,\left( 2 \right) }=10.55\,\mathrm{bit} \]

The following steps are necessary to initialize the timer:

\begin{itemize}
\item Enable the APBC bus clock to supply the clock logic
\begin{lstlisting}
REG_PM_APBCMASK |= PM_APBCMASK_TCC2;
\end{lstlisting}
\item Define Generic Clock Generator 0 as the peripheral clock source
\begin{lstlisting}
REG_GCLK_CLKCTRL = (uint16_t) GCLK_CLKCTRL_CLKEN | (0 << GCLK_CLKCTRL_GEN_Pos) | (0x1B << GCLK_CLKCTRL_ID_Pos);
\end{lstlisting}
\item Put TCC2 into single-slope mode
\begin{lstlisting}
REG_TCC2_WAVE |= (0x02 << TCC_WAVE_WAVEGEN_Pos);
while(REG_TCC2_SYNCBUSY & TCC_SYNCBUSY_WAVE);             // Wait for sync.
\end{lstlisting}
\item Initialize the PWM period
\begin{lstlisting}
while(REG_TCC2_SYNCBUSY & TCC_SYNCBUSY_PER);              // Wait for sync.
REG_TCC2_PER = 1499;                                      // Period = 32 kHz
while(REG_TCC2_SYNCBUSY & TCC_SYNCBUSY_PER);              // Wait for sync.
\end{lstlisting}
\item Initialize the duty cycle.
\begin{lstlisting}
REG_TCC2_CC1 = 1499;                                      // Duty cycle MOTOR+ pin 100 %
REG_TCC2_CC0 = 1499;                                      // Duty cycle MOTOR- pin 100 %
\end{lstlisting}
\item Enable the timer
\begin{lstlisting}
REG_TCC2_CTRLA |= TCC_CTRLA_ENABLE;
\end{lstlisting}
\end{itemize}
The duty cycle defined in TCC2.CC1 applies to positive directions of rotation, the value in TCC2.CC0 to negative directions of rotation.

\textbf{TC -- Timer/Counter:} Timer with functions for generating repeatable events~\cite[pp.~662\,ff., Ch.~30]{ATD18}. Additional functions such as signal generation (PWM) exist; the variety of functions is smaller compared to the TCC module. In the software, the timer is used to generate the sampling time for the control of the actuator motor. This is \qty{50}{\micro\second}. For a periodic interrupt, the timer is operated in the Waveform/Match-Frequency mode. The module TC3 is used. The goal is to reset the running counter of the timer at a specific value and to trigger an interrupt. In the Match-Frequency mode, the register TC3.CC0 is compared with the running counter value TC3.COUNT for this purpose. When the counter value reaches the value of TC3.CC0, the counter value is reset. To obtain a periodic interrupt, the compare-match interrupt CC0 is enabled (Figure~\ref{fig:29}).

The peripheral clock supplying the timer is $f_{\text{GCLK,TC3}}=\qty{48}{\mega\hertz}$, and the sampling time is defined as $T_{\mathrm{ctrl}}=\qty{50}{\micro\second}$. From the period duration of the clock, the maximum counter value ``TOP'' in the register TC3.CC0 is calculated:
\[ {T}_{\text{GCLK,TC3}}=\frac{1}{{f}_{\text{GCLK,TC3}}}=\qty{20.83}{\nano\second}\quad\Rightarrow\quad
   {\mathrm{TOP}}_{\mathrm{TC3}}=\frac{{T}_{\mathrm{ctrl}}}{{T}_{\text{GCLK,TC3}}}=2400 \]
For a periodic counter overflow of \qty{50}{\micro\second}, 2400 clock cycles are thus necessary. Since the counter counts from 0 up to and including TC3.CC0, the value of the register TC3.CC0 must be 2399.

\begin{figure}[ht]
\centering
\begin{tikzpicture}[x=\linewidth/13.4, y=0.85cm, font=\scriptsize, >={Stealth[length=3.6pt, width=2.8pt]},
    lvl/.style={densely dotted, line width=0.4pt},
    cnt/.style={line width=0.9pt},
    ext/.style={densely dotted, line width=0.4pt},
    mtc/.style={circle, draw, fill=white, line width=0.6pt, inner sep=0pt, minimum size=5pt}]
  \def\T{2.55}\def\P{1.8}
  \draw[lvl] (0,\P) -- (10.4,\P);  \node[anchor=east] at (-0.1,\P) {TC3.CC0};
  \draw[lvl] (0,0)  -- (10.4,0);   \node[anchor=east] at (-0.1,0)  {0};
  \node[anchor=east, font=\scriptsize\bfseries] at (-0.1,0.9) {TC3.COUNT};
  \foreach \i in {0,1,2,3} {
    \pgfmathsetmacro{\xa}{\i*\T}
    \pgfmathsetmacro{\xb}{(\i+1)*\T}
    \draw[cnt] (\xa,0) -- (\xb,\P) -- (\xb,0);
    \node[mtc] at (\xb,\P) {};
    \draw[ext] (\xb,\P) -- (\xb,-0.75);
    \draw[->, line width=0.7pt] (\xb,-1.35) -- (\xb,-0.8);
  }
  \node[anchor=east, font=\scriptsize\bfseries] at (-0.1,-1.07) {Interrupt};
  \draw[ext] (\T,\P) -- (\T,\P+0.55);  \draw[ext] (2*\T,\P) -- (2*\T,\P+0.55);
  \draw[<->, line width=0.5pt] (\T,\P+0.45) -- node[above] {$T_{\mathrm{ctrl}} = \qty{50}{\micro\second}$ (2400 clock cycles)} (2*\T,\P+0.45);
  \node[mtc] at (7.05,\P+0.5) {};
  \node[anchor=west] at (7.15,\P+0.5) {match with CC0: COUNT restarts};
\end{tikzpicture}\par
\caption{Match-frequency mode of TC3 (based on~\protect\cite[p.~670, Fig.~30-5]{ATD18})\nocc}
\label{fig:29}
\end{figure}

The following steps are necessary to initialize the timer:

\begin{itemize}
\item Enabling the APBC bus clock to supply the clock logic
\begin{lstlisting}
REG_PM_APBCMASK |= PM_APBCMASK_TC3;
\end{lstlisting}
\item Defining Generic Clock Generator 0 as peripheral clock source
\begin{lstlisting}
REG_GCLK_CLKCTRL = (uint16_t) GCLK_CLKCTRL_CLKEN | (0 << GCLK_CLKCTRL_GEN_Pos) | GCLK_CLKCTRL_ID_TCC2_TC3;
\end{lstlisting}
\item Initialize TC3 in waveform/match frequency mode
\begin{lstlisting}
REG_TC3_CTRLA |= (1 << TC_CTRLA_WAVEGEN_Pos);
while(REG_TC3_STATUS & TC_STATUS_SYNCBUSY);               // Wait for sync.
\end{lstlisting}
\item Set the period duration.
\begin{lstlisting}
REG_TC3_COUNT16_CC0 = 2399;                               // Period = 50 µs
while(REG_TC3_STATUS & TC_STATUS_SYNCBUSY);               // Wait for sync.
\end{lstlisting}
\item Enabling the compare match interrupt CC0
\begin{lstlisting}
REG_TC3_INTENSET |= TC_INTENSET_MC0;
\end{lstlisting}
\item Enable the interrupt in the NVIC
\begin{lstlisting}
NVIC_EnableIRQ(TC3_IRQn);
\end{lstlisting}
\item Enable the timer.
\begin{lstlisting}
REG_TC3_CTRLA |= TC_CTRLA_ENABLE;
\end{lstlisting}
\end{itemize}

\section{Main Program}\label{sec:37}

Like the initialization, the main program is divided into individual functions. The arrows in Figure~\ref{fig:30} show the signal flow.

\par\medskip\noindent\begin{minipage}{\textwidth}
\centering
\begin{tikzpicture}[
    font=\footnotesize,
    >={Stealth[length=2mm]},
    fn/.style={draw, thick, fill=white, font=\scriptsize\ttfamily, minimum height=5mm,
               inner xsep=2.5pt, inner ysep=0pt},
    blob/.style={draw, thick, fill=white, circle, align=center, inner sep=2pt},
    sig/.style={->, semithick},
    bisig/.style={<->, semithick},
    layer/.style={draw, dashed, thin},
    file/.style={font=\scriptsize\ttfamily, anchor=south east, inner sep=3pt},
  ]
  \def\gap{0.55}%
  \def\wl{4.85}%
  \def\wr{5.6}%
  \def\fl{-5.55}%
  \def\fr{6.3}%
  \def\drop{2.6}%

  \node[fn, minimum width=\wl cm, anchor=east] (motor) at (-\gap,0)     {MOTOR\_set\_rpm(long su)};
  \node[fn, minimum width=\wl cm, anchor=east] (dpos)  at (-\gap,-0.65) {GET\_drum\_pos(uint16\_t drum\_val)};
  \node[fn, minimum width=\wr cm, anchor=west] (port)  at (\gap,0)      {PORT\_read\_switch(int group, int pin)};
  \node[fn, minimum width=\wr cm, anchor=west] (dval)  at (\gap,-0.65)  {GET\_drum\_val(uint8\_t drum\_pos)};
  \node[fn, minimum width=3.4cm, minimum height=7.5mm] (shift) at (0,-1.75) {SHIFT\_application()};

  \node[blob] (timer)  at (0,-3.85)    {Software\\Timer};
  \node[blob] (global) at (-4.35,-3.85) {Global\\Variables \&\\Constants};

  \node[fn, minimum width=2.9cm] (tc3) at (-3.1,-5.55) {void TC3\_Handler()};
  \node[fn, minimum width=2.9cm] (rtc) at (0,-5.55)    {void RTC\_Handler()};
  \node[fn, minimum width=2.9cm] (adc) at (3.1,-5.55)  {void ADC\_Handler()};

  \node[blob, minimum size=1.5cm, inner sep=0.5pt] (tcc2) at (\fl-0.95,0) {TCC2\\Register};
  \node[blob, minimum size=1.5cm, inner sep=0.5pt] (preg) at (\fr+0.95,0) {PORT\\Register};

  \draw[layer] (\fl,0.45)  rectangle (\fr,-2.55);
  \node[file] at (\fr,-2.55) {apps.c};
  \draw[layer] (\fl,-2.7)  rectangle (\fr,-5.0);
  \node[file] at (\fr,-5.0)  {global.c, global.h};
  \draw[layer] (\fl,-5.15) rectangle (\fr,-6.3);
  \node[file] at (\fr,-6.3)  {Interrupts.c};

  \draw[sig]   (motor.west) -- (tcc2.east);
  \draw[sig]   (preg.west) -- (port.east);
  \draw[sig]   (shift.north -| -0.2,0) |- (motor.east);
  \draw[bisig] (shift.north -| 0.2,0) |- (port.west);
  \draw[bisig] (dpos.south -| -\drop,0) |- (shift.west);
  \draw[bisig] (dval.south -| \drop,0) |- (shift.east);
  \draw[bisig] (shift.south) -- (timer.north);
  \draw[bisig] (rtc.north) -- (timer.south);
  \draw[bisig] (tc3.north -| -\drop,0) |- (timer.west);
  \draw[sig]   (adc.north -| \drop,0) -- (\drop,-2.95) -| (shift.south -| 1.0,0);
\end{tikzpicture}\par
\captionof{figure}{Signal flow diagram of the main program\nocc}
\label{fig:30}
\end{minipage}

\subsection{apps.c}\label{sec:38}

This C file contains all functions required for open-loop and closed-loop control.

\textbf{MOTOR\_set\_rpm(long su):} To implement operating strategy 2 (Figure~\ref{fig:20}\,b) by means of TCC2, a separate function is created for driving the switching transistors. The speed request $s_{\mathrm{u}}$ of the P position controller serves as the transfer parameter.

\begin{lstlisting}
void MOTOR_set_rpm(long su);                              // Set motor rpm (-1499 to 1499)
\end{lstlisting}

Within this function, the registers TCC2.CC0 and TCC2.CC1 are written with the actuating variable according to the direction of rotation and the speed. The CCx registers can have a value range from 0 to 1499 (Figure~\ref{fig:28} and Eq.~\eqref{eq:9}). Therefore, the variable $s_{\mathrm{u}}$ is limited to +1499 and -1499, respectively, within the function. Operating strategy 2 (Figure~\ref{fig:20}\,b) is realized as follows: For a positive direction of rotation, the register TCC2.CC0 is set to the value 1499. The voltage at the motor is varied by means of TCC2.CC1. If both registers take on equally large values, the voltage at the motor terminals is zero. For a negative direction of rotation, the speed control is carried out equivalently with the register TCC2.CC0, while TCC2.CC1 contains the value 1499. The complete function is thus obtained as:

\begin{lstlisting}
void MOTOR_set_rpm(long su)                               // Set motor rpm (-1499 to 1499)
{
    if (su>1499) su=1499;                                 // Limit, positive direction
    else if (su<-1499) su=-1499;                          // Limit, negative direction
    if(su >= 0) {                                         // Positive direction requested
        REG_TCC2_CC0 = 1499;                              // High-side MOSFET closed
        REG_TCC2_CC1 = 1499 - (uint16_t) su;              // Set the duty cycle
    }
    if(su < 0) {                                          // Negative direction requested
        su=su*(-1);
        REG_TCC2_CC0 = 1499 -(uint16_t) su;               // Set the duty cycle
        REG_TCC2_CC1 = 1499;                              // High-side MOSFET closed
    }
}                                                         //end Motor_set_rpm
\end{lstlisting}

\textbf{PORT\_read\_switch(int group, int pin):} This function can be applied to any inputs of the microcontroller. The aim of the function is to debounce the two push buttons ``Upshift'' and ``Downshift''. When the push button is pressed, the state of the input is a low level; when the push button is not pressed, the input has a high level. The port group (0 or 1) and the number of the pin serve as transfer parameters. The function is called in the control cycle every \qty{50}{\micro\second}.

Determination of the current input level:

\begin{lstlisting}
if(PORT->Group[group].IN.reg & (1 << pin)) {              // Read pin state
buffer=HIGH;                                              // Write high-state
} else buffer=LOW;                                        // Write low-state
return_var=UNDEF;                                         // First, set pin-state to undefined
\end{lstlisting}

As the initial value, the variable buffer takes on the signal state of the pin. Due to push button bounce, alternating logic levels can occur. The current push button position is not yet known at this point, so the variable return\_var receives the value 2 (undefined). In the next step, it is queried whether the input state has changed:

\begin{lstlisting}
if(i<255) i++;                                            // Time base 255*50 µs
else {
return_var=buffer;                                        // Set buffer as return value
    switch_var=0;                                         // Start again
}
/* Compare pin-state changes */
if( ((PORT->Group[group].IN.reg & (1 << pin)) >> pin) ^ buffer ) switch_var=0;
\end{lstlisting}

If the state of the pin has not changed within 255 calls, the switch state can be regarded as permanent. The variable return\_var takes on this switch state as the return value. 255 calls form a time duration of \qty{12.75}{\milli\second}. A new query starts if switch bounce occurs and the switch state thus changes during this time duration.

\textbf{GET\_drum\_pos(uint16\_t drum\_val):} Based on the digital position value of the shift drum, this function returns a defined position value drum\_pos (Figure~\ref{fig:32}). Based on the position value drum\_pos and the commands ``upshift'' or ``downshift'', the sequence control decides which position value is requested by the operator.

\textbf{GET\_drum\_val(uint8\_t drum\_pos):} For the possible gear positions, defined by the position values 20, 30, 40, 50, 60, 70 and 80 of the variable drum\_pos, this function returns the exact digital position value drum\_val (Figure~\ref{fig:32}). The sequence control specifies a position value drum\_pos for the control. From this, the setpoint for the control is determined as a digital position value drum\_val with the help of this function.

\subsection{Interrupts.c}\label{sec:39}

This file contains all interrupt calls in the program.

\textbf{ADC\_Handler():} The position of the shift drum (actual value) is acquired with the help of the ADC input AIN[2]. Due to the mounting position in the gearbox test rig, the rotation angle of the potentiometer is negative (Section~\ref{sec:18}). This leads to a falling voltage value during upshifting. For better comprehensibility, the ADC value is mirrored in the interrupt routine:

\begin{lstlisting}
void ADC_Handler(){                                       // ADC Result-Ready Interrupt
    ADC_buffer[DRUM_POT_PIN] = 4095 - (ADC->RESULT.reg);  // Correct turning direction
    REG_ADC_INTFLAG |= ADC_INTFLAG_RESRDY;                // Reset intflag
}                                                         //end ADC_Handler
\end{lstlisting}

\textbf{TC3\_Handler():} The sampling time of the control is based on a periodic interrupt every \qty{50}{\micro\second}. At this frequency, all functions required for the control are executed: debouncing of the push buttons, calculations of the control algorithm, sequence control, etc. The ADC conversions are controlled in such a way that all three measured values are available in time with the sampling frequency.

\subsection{Control}\label{sec:40}

The control of the gear changes is developed with a view to series production that is as cost-effective as possible. Particular attention must therefore be paid to ensuring that a small number of signals is required for position control, while the software nevertheless works in a resource-saving manner. Usually, the position control loop of a DC machine is built as a cascade control consisting of torque, speed and position controllers. By means of speed and torque feedforward control, this can be implemented in a very robust and high-performing way. In series application, however, some disadvantages arise. The actual speed of the actuator motor is difficult to determine in series application. Speed sensors and incremental encoders as well as the required signal processing are additional cost factors in production. If possible, the speed controller is to be omitted. As an option, the speed can be determined with limitations from the derivative of the position value. The torque of the motor can be determined mathematically from the current. For simplification, it is also to be tested here whether the current controller can be omitted. Taking the requirements into account, the position control loop is built without typical current and speed control (Section~\ref{sec:8}). The basic structure of the position control loop is implemented as a P controller.

\begin{figure}[tbp]
\centering
\begin{tikzpicture}[font=\scriptsize, line width=0.6pt,
    >={Stealth[length=4.5pt,width=3.2pt]},
    block/.style={draw, fill=white, minimum height=0.75cm, minimum width=0.7cm,
                  inner sep=2pt, align=center},
    sum/.style={draw, fill=white, circle, minimum size=0.34cm, inner sep=0pt},
    dot/.style={circle, fill, inner sep=0pt, minimum size=3pt},
    frame/.style={draw=black!55, dashed, line width=0.5pt},
    annot/.style={->, draw=black!45, line width=0.5pt},
    lbl/.style={align=center, inner sep=1pt}]

  \def\yA{0}       %
  \def\yT{-1.28}   %
  \def\yG{-2.02}   %
  \def\yB{-3.60}   %
  \def\yE{-4.45}   %
  \def\yC{-5.17}   %
  \def\yN{1.48}    %
  \def\yI{2.42}   %
  \def\yM{-2.63}   %
  \def\xD{0.8}     %

  \node[lbl, anchor=east] at (1.0,\yA) {Position\\setpoint};
  \node[sum] (s1) at (1.6,\yA) {};
  \node[block, right=0.42cm of s1] (kr) {$K_{\mathrm{R}}$};
  \node[block, right=0.75cm of kr, minimum width=0.9cm, minimum height=0.8cm] (lim) {};
  \draw[line width=0.3pt] ($(lim)+(-0.32,0)$) -- ($(lim)+(0.32,0)$)
                          ($(lim)+(0,-0.28)$) -- ($(lim)+(0,0.28)$);
  \draw ($(lim)+(-0.32,-0.2)$) -- ($(lim)+(-0.17,-0.2)$)
        -- ($(lim)+(0.17,0.2)$) -- ($(lim)+(0.32,0.2)$);
  \node[block, minimum width=0.45cm, minimum height=0.45cm] (sd)
    at ($(lim.west)+(-0.36,-0.72)$) {$s$};

  \draw[->] (1.0,\yA) -- node[above, pos=0.8] {$w$} (s1.west);
  \draw[->] (s1.east) -- node[above] {$e$} (kr.west);
  \draw[->] (kr.east) -- (lim.west);
  \draw[->] (sd.north) |- ($(lim.west)+(0,-0.22)$);

  \draw[->] (0,\yC) -- (0,\yT) -| (sd.south);
  \node[dot] at (s1 |- 0,\yT) {};
  \node[dot] at (kr |- 0,\yT) {};
  \draw[->] (s1 |- 0,\yT) -- (s1.south) node[pos=0.86, right=1pt] {$x$}
                                         node[pos=0.97, left=1pt] {$-$};
  \draw[->] (kr |- 0,\yT) -- (kr.south);
  \node[lbl, anchor=north west] at (0.04,{\yT-0.04}) {Position actual value};

  \node[lbl, above=1pt] at (kr.north) {Software\\position\\control};
  \node[lbl, below=2pt] at ($(lim.south)+(0.08,0)$) {Overload\\limitation};

  \draw[frame] (-0.18,-1.64) rectangle ($(lim.east |- 0,1.40)+(0.28,0)$);
  \node[lbl, anchor=north west, text=black!65] at (0.04,-1.67) {Discretization};

  \node[block, anchor=west] (drv) at (\xD,\yB) {$e^{-s\tau_{\mathrm{PWM}}}$};
  \node[sum, right=1.1cm of drv] (s2) {};
  \node[block, right=0.35cm of s2] (ra) {$\dfrac{1}{R_{\mathrm{A}}(1+s\tau_{\mathrm{A}})}$};
  \node[block, right=0.75cm of ra] (km) {$K_{\mathrm{M}}\Phi_{\mathrm{M}}$};
  \node[sum, right=0.68cm of km] (s3) {};
  \node[block, right=0.35cm of s3] (tm)
    {$\dfrac{R_{\mathrm{A}}}{(K_{\mathrm{M}}\Phi_{\mathrm{M}})^{2}\,s\tau_{\mathrm{M}}}$};
  \node[block, right=0.8cm of tm, minimum width=0.8cm] (kg) {$K_{\mathrm{G}}$};
  \node[block, right=0.45cm of kg] (mech) {$\dfrac{1}{1+s\tau_{\mathrm{G}}}$};
  \node[block, minimum height=0.6cm] (emf) at (s3 |- 0,\yE) {$K_{\mathrm{M}}\Phi_{\mathrm{M}}$};

  \coordinate (iA) at ($(ra.east)+(0.42,0)$);
  \coordinate (wM) at ($(tm.east)+(0.46,0)$);
  \coordinate (wS) at ($(mech.east)+(0.46,0)$);
  \node[dot] at (iA) {};
  \node[dot] at (wM) {};
  \node[dot] at (wS) {};

  \draw[->] (drv.east) -- (s2.west);
  \node[lbl, anchor=north west] at ($(drv.east)+(0.03,-0.06)$) {$U_{\mathrm{PWM}}$};
  \draw[->] (s2.east) -- (ra.west);
  \draw (ra.east) -- (iA);
  \node[lbl, anchor=north west] at ($(ra.east)+(0.03,-0.06)$) {$I_{\mathrm{A}}$};
  \draw[->] (iA) -- (km.west);
  \draw[->] (km.east) -- (s3.west);
  \node[lbl, anchor=north west] at ($(km.east)+(0.03,-0.06)$) {$M_{\mathrm{M}}$};
  \draw[->] (s3.east) -- (tm.west);
  \draw (tm.east) -- (wM);
  \node[lbl, anchor=north west] at ($(tm.east)+(0.03,-0.06)$) {$\omega_{\mathrm{M}}$};
  \draw[->] (wM) -- (kg.west);
  \draw[->] (kg.east) -- (mech.west);
  \draw (mech.east) -- (wS);
  \node[lbl, anchor=north west] at ($(mech.east)+(0.03,-0.06)$) {$\omega_{\mathrm{S}}$};
  \draw[->] (wS) -- ++(0.36,0);

  \draw[->] ($(s3)+(0,0.82)$) node[right=1pt, yshift=-1pt] {$M_{\text{fric,shaft,red}}$}
    -- (s3.north) node[pos=0.75, right=0pt] {$-$};

  \draw[->] (wM) |- (emf.east);
  \draw[->, densely dotted] (emf.west) -| (s2.south)
    node[pos=0.97, right=0pt] {$-$} node[pos=0.72, right=0pt] {$U_{\mathrm{q}}$};

  \node[lbl, above=1pt] at (drv.north) {Motor driver};
  \node[lbl, above=1pt] at ($(kg.north)+(0.12,0)$) {Gearbox};
  \node[lbl, above=1pt] at (mech.north) {Mechanics};

  \coordinate (mfL) at ($(s2.west)+(-0.2,0)$);
  \coordinate (mfR) at ($(wM)+(0.14,0)$);
  \draw[frame] (mfL |- 0,{\yE-0.42}) rectangle (mfR |- 0,{\yB+0.92});
  \node[lbl, anchor=south west, text=black!65] at ($(mfL |- 0,{\yE-0.42})+(0.04,0.03)$) {DC motor};

  \draw[->] (iA) -- (iA |- 0,\yI) -| (lim.north);
  \coordinate (wrap) at ($(lim.east)+(0.9,0)$);
  \draw[->] (lim.east) -- (wrap) -- (wrap |- 0,\yG) -- (0.3,\yG) -- (0.3,\yB) -- (drv.west);

  \node[block, minimum width=0.5cm] (int) at ($(wS |- 0,\yC)+(-0.65,0)$) {$\dfrac{1}{s}$};
  \node[block, left=0.3cm of int] (kl) {$\dfrac{K_{\mathrm{L}}}{1+s\tau_{\mathrm{L}}}$};
  \draw[->] (wS) |- (int.east);
  \draw[->] (int.west) -- (kl.east);
  \draw (kl.west) -- (0,\yC);
  \node[lbl, anchor=south east] at ($(wS |- int.north)+(-0.06,0.06)$) {Position sensor\\Hall potentiometer};

  \draw[annot] (1.12,\yN) -- (1.12,{\yA+0.04});
  \node[lbl, above] at (1.12,\yN) {12-bit\\digital value};
  \path ($(kr.east)!0.45!(lim.west)$) coordinate (a2);
  \draw[annot] (a2 |- 0,\yN) -- ($(a2)+(0,0.04)$);
  \node[lbl, above] at (a2 |- 0,\yN) {12-bit\\digital value};
  \draw[annot] (0.75,{\yT+0.3}) -- (0.75,{\yT+0.04});
  \node[lbl, above] at (0.75,{\yT+0.3}) {12-bit\\digital value};
  \path ($(lim.east)+(0.4,0)$) coordinate (a4);
  \draw[annot] (a4 |- 0,\yN) -- ($(a4)+(0,0.04)$);
  \node[lbl, above] at (a4 |- 0,\yN) {PWM\\\qtyrange{25}{100}{\percent}};
  \path ($(drv.east)!0.72!(s2.west)$) coordinate (a5);
  \draw[annot] (a5 |- 0,\yM) -- ($(a5)+(0,0.04)$);
  \node[lbl, above] at (a5 |- 0,\yM) {PWM \qtyrange{25}{100}{\percent}};
  \path ($(iA)+(0.22,0)$) coordinate (a6);
  \draw[annot] (a6 |- 0,\yM) -- ($(a6)+(0,0.04)$);
  \node[lbl, anchor=south west] at ($(iA |- 0,\yM)+(0.08,0)$) {\qty{\pm 111}{\ampere}};
  \path ($(km.east)!0.75!(s3.west)$) coordinate (a7);
  \draw[annot] (a7 |- 0,\yM) -- ($(a7)+(0,0.04)$);
  \node[lbl, above] at (a7 |- 0,\yM) {\qty{\pm 1.7}{\newton\metre}};
  \path ($(kg.west)+(-0.15,0)$) coordinate (a8);
  \draw[annot] (a8 |- 0,\yM) -- ($(a8)+(0,0.04)$);
  \node[lbl, above] at (a8 |- 0,\yM) {\qty{\pm 7120}{\per\minute}};
  \path ($(wS)+(0.18,0)$) coordinate (a9);
  \draw[annot] (a9 |- 0,\yM) -- ($(a9)+(0,0.04)$);
  \node[lbl, anchor=south east] at ($(a9 |- 0,\yM)+(0.18,0)$) {\qty{\pm 237}{\per\minute}};
  \draw[annot] (1.05,{\yC+0.32}) -- (1.05,{\yC+0.04});
  \node[lbl, above] at (1.05,{\yC+0.32}) {\qtyrange{0.4}{4.6}{\volt}};
\end{tikzpicture}\par
\caption{Complete position control loop (based on~\protect\cite[p.~42]{Pro11})\nocc}
\label{fig:31}
\end{figure}
Figure~\ref{fig:31} shows the entire position control loop in analog form. The control engineering equivalent circuit diagram of the mechanics is shown in abbreviated form. The mechanics are represented by the gearbox, the moment of inertia $J_{\text{motor,red}}$ reduced to the motor and the friction torque $M_{\text{fric,shaft,red}}$, which is assumed to be constant. The controlled system contains the transfer functions of the Hall potentiometer, the actuator motor and the motor driver. The basic structure of the position controller is shown as a P controller. To protect the electric motor, an overload limitation is carried out taking speed, torque and manipulated variable into account. Table~\ref{tab:loopparams} lists the parameters of the individual transfer functions.

\begin{table}[ht]
\centering
\caption{Parameters of the position control loop}\label{tab:loopparams}
{\small\renewcommand{\arraystretch}{1.15}%
\begin{tabular}{@{}llcl@{}}
\toprule
\textbf{Block} & \textbf{Parameter} & \textbf{Value} & \textbf{Determination} \\
\midrule
Position controller & $K_{\mathrm{R}}$ & 15 & experimental \\
Motor driver & $\tau_{\mathrm{PWM}}$ & \qty{31.25}{\micro\second} & period of $f_{\text{PWM,TCC2}}=\qty{32}{\kilo\hertz}$ \\
DC motor & $L_{\mathrm{A}}$ & \qty{33.6}{\micro\henry} & data sheet~\cite[p.~7]{B7A19} \\
 & $R_{\mathrm{A}}$ & \qty{0.108}{\ohm} & data sheet~\cite[p.~7]{B7A19} \\
 & $\tau_{\mathrm{A}}=L_{\mathrm{A}}/R_{\mathrm{A}}$ & \qty{311.1}{\micro\second} & calculated \\
 & $K_{\mathrm{M}}\Phi_{\mathrm{M}}$ & \qty{0.0156}{\newton\metre\per\ampere} & data sheet, torque constant~\cite[p.~7]{B7A19} \\
 & $\tau_{\mathrm{M}}$ & \qty{10.5}{\milli\second} & Eq.~\eqref{eq:tauM} \\
 & $M_{\text{fric,shaft,red}}$ & \qty{0.0103}{\newton\metre} & Eq.~\eqref{eq:7} \\
Gearbox & $K_{\mathrm{G}}=1/i_{\mathrm{worm}}$ & 0.0333 & Eq.~\eqref{eq:3} \\
Position sensor & $\tau_{\mathrm{L}}$ & \qty{1.3}{\milli\second} & potentiometer \qty{5.1}{\kilo\ohm}, filter \qty{1}{\kilo\ohm}/\qty{220}{\nano\farad} \\
Mechanics & $\tau_{\mathrm{G}}$ & -- & to be identified \\
\bottomrule
\end{tabular}}
\end{table}

The motor driver is modelled as a dead time element whose dead time corresponds to the PWM period. The mechanical time constant follows from $J_{\text{motor,red}}$ (Eq.~\eqref{eq:6})~\cite{Pro18}:
\begin{equation}
\tau_{\mathrm{M}}=J_{\text{motor,red}}\cdot\frac{R_{\mathrm{A}}}{\left(K_{\mathrm{M}}\Phi_{\mathrm{M}}\right)^{2}}=\qty{10.5}{\milli\second}
\label{eq:tauM}
\end{equation}
The position sensor could be described more precisely by a second-order lag element. However, the data sheet gives no information on its mechanical time behavior, so only the electrical behavior of the potentiometer and the filter is approximated as a PT1 element. The mechanics are also approximated as a PT1 element; in further development, the transfer function is to be determined with the ``System Identification Toolbox'' in Matlab/Simulink. To discretize the P controller, the dependencies $(t)$ are replaced by discrete dependencies $(k)$:
\[ u\left( t \right) ={K}_{\mathrm{R}}\cdot e\left( t \right) \quad\Rightarrow\quad u\left( k \right) ={K}_{\mathrm{R}}\cdot e\left( k \right) \]

The sampling time of the discretized control loop is \qty{50}{\micro\second}. To realize correct position control, it is not sufficient to merely specify the setpoint for the next gear. Various events must additionally be taken into account in the control software:

\begin{itemize}
\item Checking that the gearbox is free of load
\item Reducing the actuating forces to protect the mechanics when there is a risk of shift dogs colliding
\item Preventing the shift dogs from being permanently pressed against each other
\end{itemize}
The integrated overload limitation not only provides direct protection of the electric motor, mechanical defects such as a blocked motor shaft are also detected.

\subsection{Sequence Control}\label{sec:41}

For the sequence control, it is not decisive to know the exact position of the shift drum. Rather, it is important to store defined sequences for different position ranges. Sequences must also be stored for impermissible but nevertheless possible position values in order to reach certain gear positions again. In addition, it must be taken into account that a proportional controller behavior shows a fast response, but the exact end position is never reached without additional measures.

Figure~\ref{fig:32} shows the individual ranges; its rows are explained in the following.

\textbf{Range (drum\_pos):} The numbers 20, 30, 40, 50, 60, 70 and 80 define the control end position for the individual gears. If the ADC value drum\_val lies within the marked area, the gear is considered engaged. The threshold is \ang{\pm 5}. The numbers 15, 25, 35, 45, 55, 65, 75, 85 define intermediate positions or positions above and below the shift gates.

\textbf{Gear:} The gear as an actual number, where the positions X cannot be reached due to the missing shift gates. 0 marks neutral, the numbers 1-6 the gears.

\textbf{Drum angle:} The position of the shift drum starting from first gear in angular degrees. The reference point given here does not correspond to the angle specification in the data sheet of the potentiometer.

\textbf{ADC value (drum\_val):} The position of the shift drum as a 12-bit digital value. This value is calculated from the measurement result of the running ADC conversion. Now, starting from the individual values of the variable drum\_pos, the sequence can be defined (Table~\ref{tab:9}).

\par\medskip\noindent\begin{minipage}{\textwidth}
\centering
{\scriptsize\pgfmathsetlengthmacro{\gpunit}{(\linewidth - 0.5*width("3500"))/495}\global\let\gpunit\gpunit}
\begin{tikzpicture}[font=\scriptsize, x=\gpunit, every node/.style={inner xsep=0pt}]
  \foreach \v/\a/\b in {15/-30/-5, 20/-5/5, 25/5/25, 30/25/35, 35/35/55, 40/55/65,
                        45/65/115, 50/115/125, 55/125/175, 60/175/185, 65/185/235,
                        70/235/245, 75/245/295, 80/295/305, 85/305/330} {
    \draw (\a,0.22) -- (\a,0.38) -- (\b,0.38) -- (\b,0.22);
    \node[above] at ({(\a+\b)/2},0.4) {\v};
  }
  \draw[thick] (-30,0) -- (330,0);
  \foreach \c in {0,30,60,120,180,240,300}
    \draw[line width=3.5pt] (\c-5,0) -- (\c+5,0);
  \foreach \c in {-30,330} \draw[thick] (\c,-0.12) -- (\c,0.12);

  \foreach \c/\g/\val in {-30/X/{}, 0/1/700, 30/0/1000, 60/2/1280, 120/3/1840,
                          180/4/2480, 240/5/3100, 300/6/3700, 330/X/{}} {
    \node at (\c,-0.45) {\g};
    \node at (\c,-0.85) {\ang{\c}};
    \node at (\c,-1.25) {\val};
  }

  \foreach \y/\t in {0.62/{Range (\texttt{drum\_pos})}, -0.45/Gear, -0.85/Drum angle,
                     -1.25/{ADC value (\texttt{drum\_val})}}
    \node[anchor=west, inner xsep=0pt] at (-165,\y) {\t};
\end{tikzpicture}\par
\captionof{figure}[Gear positions and control ranges]{Gear positions and control ranges (bars: range in which the gear counts as engaged)\nocc}
\label{fig:32}
\bigskip
\captionof{table}{Table of values of the variable drum\_pos}\label{tab:9}
{\small\setlength{\tabcolsep}{3pt}%
\begin{tabular*}{\textwidth}{@{}l@{\extracolsep{\fill}}*{15}{c}@{}}
\toprule
\textbf{Current drum\_pos} & 15 & 20 & 25 & 30 & 35 & 40 & 45 & 50 & 55 & 60 & 65 & 70 & 75 & 80 & 85 \\
\midrule
Next, upshift button & 20 & 25 & 30 & 35 & 40 & 45 & 50 & 55 & 60 & 65 & 70 & 75 & 80 & 80 & 80 \\
Next, downshift button & 20 & 20 & 20 & 25 & 30 & 35 & 40 & 45 & 50 & 55 & 60 & 65 & 70 & 75 & 80 \\
\bottomrule
\end{tabular*}}
\end{minipage}

\chapter{Verification of the Gear Change}\label{sec:42}

This chapter verifies the developed system on the gearbox test rig. Section~\ref{sec:43} evaluates a measured gear change, Section~\ref{sec:44} compares the results with the requirements, Section~\ref{sec:45} discusses the influence of the worm gear ratio, and Section~\ref{sec:46} names the limitations of the investigation.

\section{Gear Change on the Gearbox Test Rig}\label{sec:43}

Figure~\ref{fig:33} shows a gear change on the gearbox test rig, using the downshift from 6th to 5th gear as an example.

\begin{figure}[ht]%
\centering
\begin{tikzpicture}[>={Stealth[length=1.6mm]},
    marker/.style={draw, circle, fill=white, inner sep=1pt, font=\scriptsize}]
\begin{groupplot}[
    thesisplot,
    group style={group size=1 by 2, vertical sep=0.4cm, x descriptions at=edge bottom},
    width=0.97\textwidth, height=4.3cm,
    xmin=-50, xmax=70, xtick={-40,-20,...,60},
    xlabel={$t$ in \unit{\milli\second}},
    ylabel style={align=center},
    clip=false,
  ]
  \nextgroupplot[ymin=0.4, ymax=1.85, ytick={0.4,0.8,...,1.6},%
    yticklabel style={/pgf/number format/.cd, fixed, fixed zerofill, precision=1},
    ylabel={$U_{\mathrm{Pot}}$\\in \unit{\volt}}]
    \addplot[plotline] table[x=t_ms, y=U_V] {figures/data/gear_change_rig_voltage.dat};
    \draw[densely dashed, black!60] (axis cs:-40,0.4) -- (axis cs:-40,1.7) (axis cs:37.5,0.4) -- (axis cs:37.5,1.7);
    \draw[<->, thin] (axis cs:-40,1.62) -- node[above, font=\scriptsize, inner sep=1.5pt]
      {$\Delta t<\qty{80}{\milli\second}$} (axis cs:37.5,1.62);
    \node[marker] at (axis cs:43.5,1.09) {5};
  \nextgroupplot[ymin=-90, ymax=70, ytick={-75,-50,...,50},
    ylabel={$I_{\mathrm{motor}}$\\in \unit{\ampere}}]
    \addplot[plotline] table[x=t_ms, y=I_A] {figures/data/gear_change_rig_current.dat};
    \node[marker] at (axis cs:-32.7,62) {1};
    \node[marker] at (axis cs:-12.4,-84) {2};
    \node[marker] at (axis cs:4.0,42) {3};
    \node[marker] at (axis cs:32.5,-61) {4};
\end{groupplot}
\end{tikzpicture}\par
\caption{Gear change on the gearbox test rig\nocc}
\label{fig:33}
\end{figure}

\begin{itemize}
\item Position 1: 6th gear is engaged \\
Push button ``Downshift'' pressed. The gearbox is load-free due to the drive motor being switched off. The shift drum is set in motion.
\item Position 2: Gear wheel is in the neutral position between 6th gear and 5th gear \\
Reduction of the actuating speed to relieve the mechanism in case the shift dogs collide.
\item Position 3: The shift dogs slide into one another \\
The actuating movement is accelerated again.
\item Position 4: End position almost reached \\
The actuating movement is decelerated.
\item Position 5: 5th gear is engaged \\
The gear change is completed in less than \qty{80}{\milli\second} without overshoot.
\end{itemize}

\section{Fulfillment of the Requirements}\label{sec:44}

Table~\ref{tab:10} compares the requirements from Section~\ref{sec:8} with the results achieved.

\begin{table}[ht]
\centering
\caption{Fulfillment of the requirements}\label{tab:10}
{\small
\begin{tabular}{>{\centering\arraybackslash}p{\dimexpr 0.066\textwidth-2\tabcolsep\relax}>{\raggedright\arraybackslash}p{\dimexpr 0.326\textwidth-2\tabcolsep\relax}>{\raggedright\arraybackslash}p{\dimexpr 0.424\textwidth-2\tabcolsep\relax}>{\centering\arraybackslash}p{\dimexpr 0.185\textwidth-2\tabcolsep\relax}}
\toprule
\textbf{No.} & \textbf{Requirement} & \textbf{Result} & \textbf{Assessment} \\
\midrule
1 & Weight of all components on the vehicle at most \qty{3}{\kilogram} & Not determined on the gearbox test rig with test adapters & open \\
2 & Shift movement by an electric motor & Actuator motor drives the shift drum directly through a worm gear & fulfilled \\
3 & Fewer mechanical components than with manual shift actuation & Foot shift mechanism is omitted, drive directly at the shift drum & fulfilled \\
4 & Average shift time of \qty{100}{\milli\second} & Single measurement below \qty{80}{\milli\second} (6th to 5th gear), no statistical evaluation & partially verified \\
5 & Control electronics also usable as an add-on control unit & Shift controller implemented as a prototype & partially fulfilled \\
6 & Maintenance-free within the service life of the vehicle & Required operating time of the actuator motor calculated as \qty{8.3}{\hour}, no endurance run & open \\
7 & Simplest control strategy with overload protection, without speed and current control & P position controller without subordinate control, overload limitation & fulfilled \\
\bottomrule
\end{tabular}}
\end{table}
The requirements for the actuators and the control strategy are fulfilled. For the weight and the maintenance-free operation, the experimental proof is missing. The shift time on the gearbox test rig is only verified by single measurements.

\section{Influence of the Gear Ratio of the Worm Gear}\label{sec:45}

During the design in Section~\ref{sec:21}, it became apparent that the required motor torque increases with increasing gear ratio. This behavior can be explained by inertia matching. The motor torque during acceleration is composed of three components: the acceleration of the motor shaft and worm, the acceleration of the masses on the shift drum side, and the friction. Referred to the angular acceleration of the shift drum $\alpha_{\mathrm{drum}}$, the following applies:
\[ {M}_{\mathrm{motor}}={\alpha }_{\mathrm{drum}}\cdot \left( {J}_{\mathrm{in}}\cdot i+\frac{{J}_{\mathrm{out}}}{i\cdot \eta } \right) +\frac{{M}_{\text{fric,drum}}}{i\cdot \eta } \]
Here, $J_{\mathrm{in}}=\qty{22.8e-6}{\kilogram\metre\squared}$ is the moment of inertia of the motor shaft and worm, and $J_{\mathrm{out}}=\qty{303.9e-6}{\kilogram\metre\squared}$ is the moment of inertia of the shift drum, worm wheel and reduced linear masses. The first term increases linearly with the gear ratio, the other two terms decrease with the gear ratio. Without friction, the minimum lies at $i_{\mathrm{opt}}=\sqrt{J_{\mathrm{out}}/(\eta\cdot J_{\mathrm{in}})}$~\cite{Pas84}. With the values from Section~\ref{sec:21}, $i_{\mathrm{opt}}\approx 5.6$ results.

Taking into account the friction and an angular acceleration of the shift drum of \qty{437}{\radian\per\second\squared}, the minimum lies at a gear ratio of about 8. The required motor torque thereby drops from \qty{319}{\milli\newton\metre} at $i=30$ to about \qty{158}{\milli\newton\metre}. At $i=30$, around \qty{93}{\percent} of the motor torque is spent on accelerating the motor shaft and worm. From a dynamic point of view, the selected gear ratio is therefore clearly too large.

A smaller gear ratio, however, is not readily possible. In worm gears, self-locking occurs only at small lead angles, and the efficiency is then below \qty{50}{\percent} \cite{Rol17}. The determined efficiency of \qty{42}{\percent} confirms this behavior. Small lead angles in turn require large gear ratios. There is thus a conflict of objectives between dynamics and self-locking. Two approaches for further development result from this: an actuator motor with a lower moment of inertia at the same gear ratio, or a smaller gear ratio with a different means of fixing the gear position.

\section{Limitations of the Investigation}\label{sec:46}

The results are to be considered under the following limitations:

\begin{itemize}
\item The gearbox test rig is operated without oil filling, and the rotating masses of the engine and vehicle are only reproduced in a simplified model. Load changes and engagement behavior therefore differ from the vehicle.
\item The shift time at the gearbox test rig is documented only by single measurements. A measurement series comparable with the test bench run is not available.
\item The friction torques were determined once at standstill. The influences of temperature, lubrication and wear are not taken into account.
\item The trapezoidal speed profile is an approximation. The design of the actuator motor is therefore an estimate.
\item The P controller exhibits a steady-state control error. The end position is therefore evaluated using a threshold of \ang{\pm 5}.
\item The service life of the actuator motor is considered only by calculation.
\end{itemize}

\chapter*{Summary}\addcontentsline{toc}{chapter}{Summary}\label{sec:47}

In the course of this diploma thesis, a functional gearbox test rig was created that fulfills the essential requirements. The verification of the weight and of the maintenance-free operation remains open (see Section~\ref{sec:44}). The idea of carrying out gear changes via a worm gear works. After initial difficulties, the necessary self-locking effect was also optimized. The results of various control concepts are documented both by means of the position curve at the potentiometer and by means of a slow-motion camera.

The essential results of the thesis are:

\begin{enumerate}
\item Manual gear changes on a large-displacement series-production motorcycle take \qty{80}{\milli\second} on average for upshifts and \qty{92}{\milli\second} for downshifts; \qty{83}{\percent} and \qty{69}{\percent}, respectively, are completed after \qty{100}{\milli\second} ($n=42$ per direction).
\item The direct drive of the shift drum via a worm gear shifts at the gearbox test rig with a simple P controller in less than \qty{80}{\milli\second} without overshoot in a single measurement.
\item At a gear ratio of 30, around \qty{93}{\percent} of the motor torque is spent on accelerating the motor shaft and worm. A gear ratio of about 8 reduces the required torque from \qty{319}{\milli\newton\metre} to around \qty{158}{\milli\newton\metre}.
\item Self-locking and low inertia are mutually exclusive in a worm gear: the gear is self-locking only at an efficiency below \qty{50}{\percent}, and the measured efficiency is \qty{42}{\percent}.
\item The friction torques of the shift tracks lie between \qty{55}{\milli\newton\metre} and \qty{130}{\milli\newton\metre}. For \num{300000} gear changes in \qty{60000}{\kilo\metre}, the actuator motor is in operation for only around \qty{8.3}{\hour}.
\end{enumerate}
The power, or rather the torque required to carry out the gear changes, was somewhat underestimated. Surprisingly, it emerged that the torque demand on the motor increases when the gear ratio is increased. As a logical consideration, the required motor torque should decrease, since the torque is amplified by the gear ratio. On closer inspection, it can be seen that the dynamic torque for accelerating the motor shaft and worm is many times higher than the torque actually required to rotate the shift drum. The efficiency of the worm gear must also be taken into account. The consideration in Section~\ref{sec:45} shows that, from a dynamic point of view, the gear ratio of 30 is too large by a factor of about four. At a gear ratio of about 8, the required motor torque would be halved. This is opposed by the required self-locking of the worm gear.

In the further course of development, it is possible to optimize the gear ratio of the worm gear and the moment of inertia of the motor shaft. In addition, it has turned out that the gearbox test rig corresponds to the conditions in the actual vehicle only in a simplified model. The moving masses at the gearbox input and output are too small. When the shift dogs meet, the gearbox drive is slowed down to a not inconsiderable extent, which has a negative effect above all when downshifting. In all likelihood, this behavior does not occur on the actual vehicle.

After adapting the gear ratio of the worm gear, a gearbox prototype is manufactured for use in the vehicle.

\begingroup
\let\clearpage\relax\let\cleardoublepage\relax
\RedeclareSectionCommand[beforeskip=2.2\baselineskip, afterskip=1.2\baselineskip]{chapter}
\chapter*{Outlook}\addcontentsline{toc}{chapter}{Outlook}\label{sec:48}
\endgroup

A major motorcycle manufacturer has already shown interest at this stage of development in using this system in a production motorcycle, provided it functions correctly. The advantages of the system persist in the long term if motorcycle manufacturers are obliged to increase the efficiency of their vehicles. With regard to emission regulations, the Euro 5 standard, mandatory from 2021, is regarded as pioneering for new exhaust aftertreatment concepts for two-wheelers.

As the next step, the suggested improvements mentioned are implemented on the gearbox test rig, followed by the design and prototype development. Only then will it become apparent how the system behaves in actual road operation. In the long term, safety tests on the test bench follow, with subsequent test rides in road traffic. In the end, all necessary legal regulations must be complied with in order to obtain domestic as well as foreign approvals as a component in the vehicle. The planned development time is three years.

Some points could not be addressed within the scope of this thesis. These include a statistical evaluation of the shift times on the gearbox test rig, the verification of the system weight, endurance tests on the service life of the actuator motor, and tests with oil filling and real vehicle masses. Furthermore, the influence of the automated gearbox on fuel consumption and emissions has so far only been justified qualitatively and should be quantified in driving cycles. In addition, the inertia matching gives rise to the task of jointly optimizing the actuator motor, gear ratio and fixing of the gear position.

It is pleasing to see that the foundation for a successful new development has been laid in the course of this diploma thesis. Where the strengths and weaknesses of the system lie will be shown by the subsequent development.

\clearpage
\begingroup
\let\clearpage\relax\let\cleardoublepage\relax
\RedeclareSectionCommand[beforeskip=2.2\baselineskip, afterskip=1.2\baselineskip]{chapter}
\chapter*{Software, Measuring Equipment and Tools}\addcontentsline{toc}{chapter}{Software, Measuring Equipment and Tools}\label{sec:49}

{\small
\begin{tabular*}{\textwidth}{@{}l@{\hspace{1em}}l@{\extracolsep{\fill}}l@{\hspace{1em}}l@{}}
\textbf{Measuring} & & \textbf{Software} & \\
Multimeter & Fluke 179 & Programming environment & Atmel Studio 7 \\
Current clamp & Pico Technology TA167 & Simulation & NI Multisim 14 \\
Oscilloscope & PicoScope 2205A & CAD & CadSoft EAGLE 7.7 \\
Calculator & CASIO fx-991DE X ClassWiz & Computer algebra & TI-Nspire CX CAS 5 \\
\end{tabular*}}

\chapter*{Appendices}\addcontentsline{toc}{chapter}{Appendices}\label{sec:50}

\begin{minipage}{\linewidth}
\centering
\captionof{appendixfloat}{Shift actuation FJR1300AS (description according to \protect\cite{yam19})}\label{app:1}
\end{minipage}
\nopagebreak

The shift actuation of the FJR1300AS model with YCC-S is documented as an exploded view in the manufacturer's spare parts catalog \cite{yam19}. In the following, the essential assemblies are described:

\textbf{Shift actuator:} An electric shift actuator is mounted on the engine housing. It consists of an electric motor with a reduction gear and is controlled by the MCU.

\textbf{Transmission linkage:} The shift actuator moves a lever connected to the shift shaft via a push rod with rod ends.

\textbf{Shift shaft and shift mechanism:} As in a conventional manual gearbox, the shift shaft transmits the movement to the shift drum via the shift mechanism. A return spring returns the shift shaft to the center position after each gear change.

\textbf{Foot shift lever:} The foot shift lever is not mechanically connected to the shift shaft. It serves exclusively as a control element by which the rider specifies the shift request; the actual shift actuation is carried out electromechanically by the shift actuator.

\par\medskip\noindent
\begin{minipage}{\linewidth}
\centering
\captionof{appendixfloat}{Inrush current profile of the actuator motor\nocc}\label{app:2}
\begin{tikzpicture}
\begin{axis}[
    thesisplot,
    width=0.97\textwidth, height=3.6cm,%
    xmin=-1.5, xmax=8.5, xtick={0,2,...,8},
    ymin=0, ymax=80, ytick={0,20,...,80},
    xlabel={$t$ in \unit{\milli\second}}, ylabel={$I_{\mathrm{motor}}$ in \unit{\ampere}},
  ]
  \addplot[plotline] table[x=t_ms, y=I_A] {figures/data/inrush_current.dat};
\end{axis}
\end{tikzpicture}

\end{minipage}
\par\bigskip\noindent
\begin{minipage}{\linewidth}
\centering
\captionof{appendixfloat}{Microcontroller load due to the speed interrupt (voltage at port A09 high during the interrupt routine)\nocc}\label{app:3}
\begin{tikzpicture}[>={Stealth[length=1.6mm]}]
\begin{axis}[
    thesisplot,
    width=0.97\textwidth, height=4.0cm,
    xmin=-12, xmax=188, xtick={0,25,...,175},
    ymin=-0.3, ymax=6.0, ytick={0,1,...,4},%
    xlabel={$t$ in \unit{\micro\second}}, ylabel={$U_{\mathrm{PA09}}$ in \unit{\volt}},
    clip=false,
  ]
  \addplot[plotline] table[x=t_us, y=U_V] {figures/data/cpu_load_port_a09.dat};
  \draw[<->, thin] (axis cs:0,4.55) -- node[above, font=\scriptsize] {\qty{133.8}{\micro\second}} (axis cs:133.9,4.55);
  \node[font=\scriptsize, anchor=west] at (axis cs:5.5,2.0) {\qty{4.4}{\micro\second}};
\end{axis}
\end{tikzpicture}
\end{minipage}
\par\bigskip\noindent
\begin{minipage}{\linewidth}
\centering
\captionof{appendixfloat}{Table of values of the shift time measurement (also available as a CSV data set~\protect\cite{Sto26})}\label{app:4}
{\small
\begin{tabular}{@{}ccc@{\hspace{2.5em}}ccc@{}}
\toprule
\textbf{No.} & $t_{\mathrm{up}}$ in \unit{\milli\second} & $t_{\mathrm{down}}$ in \unit{\milli\second} &
\textbf{No.} & $t_{\mathrm{up}}$ in \unit{\milli\second} & $t_{\mathrm{down}}$ in \unit{\milli\second} \\
\midrule
1 & 100 & 74 & 22 & 40 & 52 \\
2 & 91 & 52 & 23 & 47 & 84 \\
3 & 90 & 83 & 24 & 83 & 73 \\
4 & 92 & 71 & 25 & 63 & 84 \\
5 & 45 & 82 & 26 & 92 & 143 \\
6 & 32 & 88 & 27 & 87 & 63 \\
7 & 79 & 22 & 28 & 90 & 46 \\
8 & 40 & 42 & 29 & 94 & 30 \\
9 & 62 & 36 & 30 & 150 & 181 \\
10 & 55 & 25 & 31 & 31 & 145 \\
11 & 65 & 30 & 32 & 35 & 194 \\
12 & 54 & 62 & 33 & 61 & 148 \\
13 & 69 & 70 & 34 & 26 & 130 \\
14 & 54 & 76 & 35 & 129 & 132 \\
15 & 40 & 85 & 36 & 122 & 257 \\
16 & 42 & 244 & 37 & 300 & 81 \\
17 & 71 & 79 & 38 & 91 & 101 \\
18 & 54 & 90 & 39 & 129 & 76 \\
19 & 128 & 21 & 40 & 153 & 110 \\
20 & 72 & 51 & 41 & 96 & 107 \\
21 & 30 & 51 & 42 & 80 & 208 \\
\bottomrule
\end{tabular}}
\end{minipage}
\par\medskip

\endgroup
\clearpage
\renewcommand{\nocc}{}
\listoffigures
\listoftables

\chapter*{List of Abbreviations}\addcontentsline{toc}{chapter}{List of Abbreviations}
\begin{longtable}{@{}p{3cm}p{\dimexpr\textwidth-3cm-2\tabcolsep\relax}@{}}
ADC & Analog-to-Digital Converter \\
API & Application Programming Interface \\
CAS & Computer Algebra System \\
CPU & Central Processing Unit \\
CVT & Continuously Variable Transmission \\
DC & Direct Current \\
EDBG & Embedded Debugger \\
EMC & Electromagnetic Compatibility \\
ESD & Electrostatic Discharge \\
MCU & Motor Control Unit \\
MOSFET & Metal Oxide Semiconductor Field Effect Transistor \\
NVIC & Nested Vector Interrupt Controller \\
NVRAM & Nonvolatile Random-Access Memory \\
OEM & Original Equipment Manufacturer \\
op-amp & Operational Amplifier \\
PWM & Pulse Width Modulation \\
RMS & Root Mean Square \\
TC & Timer/Counter \\
TCC & Timer/Counter for Control Applications \\
TVS & Transient Voltage Suppressor \\
USART & Universal Synchronous/Asynchronous Receiver Transmitter \\
USB & Universal Serial Bus \\
YCC-S & Yamaha Chip Controlled-Shift \\
\end{longtable}

\chapter*{Symbols and Variables}\addcontentsline{toc}{chapter}{Symbols and Variables}
{\small
\begin{longtable}{>{\raggedright\arraybackslash}p{\dimexpr 0.218\textwidth-2\tabcolsep\relax}>{\centering\arraybackslash}p{\dimexpr 0.142\textwidth-2\tabcolsep\relax}>{\raggedright\arraybackslash}p{\dimexpr 0.640\textwidth-2\tabcolsep\relax}}
\toprule
\textbf{Symbol} & \textbf{Unit} & \textbf{Explanation} \\
\midrule
\endhead
$B_{\text{od}}$ & (1) & Overdrive factor of the transistor \\
$B_{\text{Q1}}$ & (1) & Current gain of the transistor \\
$b_{\text{PWM,TCC2}}$ & (bit) & Resolution of the duty cycle \\
$C$ & (F) & Capacitance \\
$\mathrm{dk}_{\text{ILIS,IC1}}$ & (1) & Ratio of load to sense current, motor driver IC1 \\
$\mathrm{dk}_{\text{ILIS,IC2}}$ & (1) & Ratio of load to sense current, motor driver IC2 \\
$\mathrm{dk}_{\text{ILIS,typ}}$ & (1) & Ratio of load to sense current, typical value \\
$e(k)$ & (1) & Control error of the position control loop, discrete-time \\
$e(t)$ & (1) & Control error of the position control loop \\
$f$ & (Hz) & Frequency \\
$f_{\text{c}}$ & (Hz) & Low-pass filter cut-off frequency \\
$f_{\text{CLK,ADC}}$ & (Hz) & Clock frequency of the ADC after the prescaler \\
$f_{\text{div}}$ & (Hz) & Encoder pulse frequency after the frequency divider \\
$f_{\text{GCLK,TC3}}$ & (Hz) & Peripheral clock frequency TC3 \\
$f_{\text{GCLK,TCC2}}$ & (Hz) & Peripheral clock frequency TCC2 \\
$f_{\text{orig}}$ & (Hz) & Encoder pulse frequency without frequency divider \\
$f_{\text{PWM,TCC2}}$ & (Hz) & PWM frequency in single-slope mode TCC2 \\
$f1(x)$ & (\unit{\radian\per\second}) & Trapezoidal speed profile \\
$f1_{\text{acc}}(x)$ & (\unit{\radian\per\second}) & Partial function acceleration of $f1(x)$ \\
$f1_{\text{brake}}(x)$ & (\unit{\radian\per\second}) & Partial function braking of $f1(x)$ \\
$f1_{\text{const}}(x)$ & (\unit{\radian\per\second}) & Partial function constant speed of $f1(x)$ \\
$f2(x)$ & (\unit{\radian\per\second}) & Copy of $f1(x)$ with defined $\omega_{\text{max}}$ \\
$I_{\text{IS}}$ & (A) & Sense current of the motor driver \\
$I_{\text{IS1,IC1}}$ & (A) & IC1 sense current IS1 at IL1 \\
$I_{\text{IS1,IC2}}$ & (A) & IC2 sense current IS1 at IL1 \\
$I_{\text{IS2,IC1}}$ & (A) & IC1 sense current IS2 at IL2 \\
$I_{\text{IS2,IC2}}$ & (A) & IC2 sense current IS2 at IL2 \\
$I_{\text{IS(Offset),IC1}}$ & (A) & Motor driver IC1 offset current \\
$I_{\text{IS(Offset),IC2}}$ & (A) & Motor driver IC2 offset current \\
$I_{\text{IS(Offset),typ}}$ & (A) & Typical offset current of the motor driver \\
$I_{\text{L1,IC1}}$ & (A) & IC1 load current at IS1 \\
$I_{\text{L1,IC2}}$ & (A) & IC2 load current at IS1 \\
$I_{\text{L2,IC1}}$ & (A) & IC1 load current at IS2 \\
$I_{\text{L2,IC2}}$ & (A) & IC2 load current at IS2 \\
$I_{\text{L,range}}$ & (A) & Measuring range for the load current of the motor driver \\
$I_{\text{PWM-IN}}$ & (A) & Supply current of the motor driver \\
$I_{\text{PWM-OUT}}$ & (A) & Output current of the motor driver \\
$I_{\text{relay}}$ & (A) & Coil current of the relay \\
$i_{\text{opt}}$ & (1) & Gear ratio with the lowest required motor torque \\
$i_{\text{worm}}$ & (1) & Gear ratio of the worm gear \\
$J_{\text{drum}}$ & (\unit{\kilogram\metre\squared}) & Inertia of the shift drum \\
$J_{\text{in}}$ & (\unit{\kilogram\metre\squared}) & Inertia of the motor shaft and the worm \\
$J_{\text{linear,red}}$ & (\unit{\kilogram\metre\squared}) & Shift fork and gear wheel reduced to the shift drum \\
$J_{\text{motor shaft}}$ & (\unit{\kilogram\metre\squared}) & Inertia of the motor shaft \\
$J_{\text{motor,red}}$ & (\unit{\kilogram\metre\squared}) & Inertia reduced to the motor, incl. motor shaft \\
$J_{\text{out}}$ & (\unit{\kilogram\metre\squared}) & Inertia of shift drum, worm wheel and linear masses \\
$J_{\text{worm}}$ & (\unit{\kilogram\metre\squared}) & Inertia of the worm \\
$J_{\text{worm wheel}}$ & (\unit{\kilogram\metre\squared}) & Inertia of the worm wheel incl. shaft, sleeve, screw \\
$K_{\text{G}}$ & (1) & Reciprocal of the gear ratio \\
$K_{\text{R}}$ & (1) & Controller gain of the P controller \\
$K_{\text{M}}\Phi_{\text{M}}$ & (\unit{\newton\metre\per\ampere}) & Torque constant of the actuator motor \\
$L_{\text{A}}$ & (H) & Armature inductance of the actuator motor \\
$M_{\text{fric,drum,red}}$ & (\unit{\newton\metre}) & Friction torque reduced to the shift drum \\
$M_{\text{fric,shaft,red}}$ & (\unit{\newton\metre}) & Friction torque reduced to the motor shaft \\
$M_{\text{motor,run-up}}$ & (\unit{\newton\metre}) & Required motor torque for the specified acceleration \\
$m_{\text{fork}}$ & (kg) & Mass of one shift fork \\
$m_{\text{gear wheel}}$ & (kg) & Mass of one gear wheel \\
$m_{\text{linear}}$ & (kg) & Linearly moved mass during the gear change \\
$N_{\text{presc,TCC2}}$ & (1) & Peripheral clock prescaler TCC2 \\
$N_{\text{pulse,div}}$ & (1) & Number of pulses per revolution after frequency divider \\
$N_{\text{pulse,orig}}$ & (1) & Number of pulses per revolution of the incremental encoder \\
$N_{\text{shifts}}$ & (1) & Number of gear changes within the rated service life \\
$n_{\text{max}}$ & (\unit{\per\second}) & Maximum motor speed \\
$n_{\text{mean}}$ & (\unit{\per\second}) & Mean motor speed during the gear change \\
$n_{\text{no-load}}$ & (\unit{\per\second}) & Motor speed at no load \\
$P_{\text{motor,max}}$ & (W) & Maximum motor power \\
$P_{\text{PWM-IN}}$ & (W) & Electrical input power of the motor driver \\
$P_{\text{PWM-OUT}}$ & (W) & Electrical output power of the motor driver \\
$p_{\text{spindle}}$ & (m) & Lead of the considered spindle \\
$R_{\text{A}}$ & (\unit{\ohm}) & Armature resistance of the actuator motor \\
$r_{\text{prim}}$ & (m) & Primary effective lever arm of the CVT transmission \\
$r_{\text{sec}}$ & (m) & Secondary effective lever arm of the CVT transmission \\
R3 & (\unit{\ohm}) & Sense resistor at motor driver IC1 \\
R5 & (\unit{\ohm}) & Sense resistor at motor driver IC2 \\
R18 & (\unit{\ohm}) & Base series resistor of transistor Q1 \\
$s_{\text{fork}}$ & (m) & Shift travel of the shift forks \\
$s_{\text{gearbox}}$ & (km) & Rated service life of the gearbox in kilometers \\
SAMPLEN & (1) & Value SAMPLEN in the register ADC.SAMPCTRL \\
$T$ & (s) & Period \\
$T_{\text{CLK,ADC}}$ & (s) & Period of the ADC clock \\
$T_{\text{ctrl}}$ & (s) & Sampling time of the position control \\
$T_{\text{div}}$ & (s) & Encoder signal period after the frequency divider \\
$T_{\text{GCLK,TC3}}$ & (s) & Period of the peripheral clock TC3 \\
$T_{\text{orig}}$ & (s) & Encoder signal period before the frequency divider \\
$t_{\text{acc}}$ & (s) & Time segment of the acceleration \\
$t_{\text{brake}}$ & (s) & Time segment of the braking process \\
$t_{\text{const}}$ & (s) & Time segment with constant motor speed \\
$t_{\text{conv}}$ & (s) & Duration of one ADC conversion \\
$t_{\text{down}}$ & (s) & List of measured values ``downshift'' \\
$t_{\text{motor}}$ & (s) & Operating time of the actuator motor in its service life \\
$t_{\text{sample}}$ & (s) & Sample-and-hold time of the ADC \\
$t_{\text{shift}}$ & (s) & Assumed shift time for the motor design \\
$t_{\text{up}}$ & (s) & List of measured values upshift \\
$\mathrm{TOP}_{\text{TC3}}$ & (1) & Value TOP in the register TC3.CC0 \\
$\mathrm{TOP}_{\text{TCC2}}$ & (1) & Value TOP in the register TCC2.PER \\
$U_{\text{BE}}$ & (V) & Base-emitter voltage at the transistor \\
$U_{\text{PWM-IN}}$ & (V) & Supply voltage of the motor drivers \\
$U_{\text{PWM-OUT}}$ & (V) & Output voltage of the motor driver \\
$U_{\text{R3,offset}}$ & (V) & Voltage drop across R3 at the offset current of IC1 \\
$U_{\text{R5,offset}}$ & (V) & Voltage drop across R5 at the offset current of IC2 \\
$U_{\text{ref,ADC}}$ & (V) & Reference voltage of the ADC \\
$U_{\upmu\mathrm{C}}$ & (V) & Supply voltage of the microcontroller \\
$u(k)$ & (1) & Control variable of the position loop, discrete-time \\
$u(t)$ & (1) & Control variable of the position control loop \\
$V(U_{\text{Filter}})$ & (V) & Simulated voltage curve at the filter output \\
$V(U_{\text{Pot}})$ & (V) & Measured voltage curve at the potentiometer \\
$\alpha_{\text{drum}}$ & (\unit{\radian\per\second\squared}) & Angular acceleration of the shift drum \\
$\alpha_{\text{motor}}$ & (\unit{\radian\per\second\squared}) & Angular acceleration of the motor shaft \\
$\eta_{\text{spindle}}$ & (1) & Efficiency of the considered spindle \\
$\eta_{\text{worm}}$ & (1) & Efficiency of the worm gear \\
$\rho_{\text{CPU}}$ & (\unit{\percent}) & Microcontroller load caused by the speed interrupt \\
$\tau_{\text{A}}$ & (s) & Armature time constant of the actuator motor \\
$\tau_{\text{G}}$ & (s) & Time constant of the mechanics \\
$\tau_{\text{L}}$ & (s) & Filter time constant of the low-pass filter \\
$\tau_{\text{M}}$ & (s) & Mechanical time constant of the actuator motor \\
$\tau_{\text{PWM}}$ & (s) & Period of the PWM signal \\
$\Delta\varphi_{\text{drum}}$ & (\unit{\degree}) & Shift drum angle for the shift travel $s_{\text{fork}}$ \\
$\varphi_{\text{gear}}$ & (rad) & Rotation angle of the shift drum during the gear change \\
$\varphi_{\text{motor}}$ & (rad) & Rotation angle of the motor shaft during the gear change \\
$\omega_{\text{max}}$ & (\unit{\radian\per\second}) & Maximum angular velocity of the motor shaft \\
$\omega_{\text{mean}}$ & (\unit{\radian\per\second}) & Mean angular velocity of the motor shaft \\
\bottomrule
\end{longtable}}

\end{document}